\documentclass[letter,11pt]{article}
\pdfoutput=1
\usepackage{jheppub}
\usepackage{xcolor} 
\usepackage{slashed}
\usepackage{titlesec}
\usepackage{comment}
\usepackage{hyperref}
\titleformat{\section}
{\normalfont\LARGE\bfseries}{Section~\thesection{:}}{1em}{}
\titleformat*{\subsection}{\Large\bfseries}
\titleformat*{\subsubsection}{\large\bfseries}
\titleformat*{\paragraph}{\large\bfseries}
\titleformat*{\subparagraph}{\large\bfseries}
\def\ltap{\ \raise.3ex\hbox{$<$\kern-.75em\lower1ex\hbox{$\sim$}}\ }
\def\gtap{\ \raise.3ex\hbox{$>$\kern-.75em\lower1ex\hbox{$\sim$}}\ }

\pdfoutput=1

\usepackage{chngcntr}
\counterwithin{figure}{section}
\counterwithin{table}{section}
\usepackage{float}
\floatstyle{plaintop}
\restylefloat{table}
\usepackage[tableposition=top]{caption}
\usepackage{afterpage}

\begin{document}

\title{Searches for Decays of New Particles in the DUNE Multi-Purpose Near Detector}

\author[1,2,3]{Jeffrey~M.~Berryman,}
\author[4]{Andr{\'e}~de~Gouv{\^e}a,}
\author[5]{Patrick~J.~Fox,}
\author[5]{Boris~J.~Kayser,}
\author[5]{Kevin~J.~Kelly,}
\author[5]{Jennifer~L.~Raaf }

\affiliation[1]{Center for Neutrino Physics, Department of Physics, Virginia Tech, Blacksburg, VA 24061, USA}
\affiliation[2]{Department of Physics and Astronomy, University of Kentucky, Lexington, KY 40506, USA}
\affiliation[3]{Department of Physics, University of California, Berkeley, CA 94720, USA}
\affiliation[4]{Northwestern University, Department of Physics \& Astronomy, 2145 Sheridan Road, Evanston, IL 60208, USA}
\affiliation[5]{Fermi National Accelerator Laboratory, Batavia, IL 60510, USA}

\emailAdd{jeffberryman@berkeley.edu}
\emailAdd{degouvea@northwestern.edu}
\emailAdd{pjfox@fnal.gov}
\emailAdd{boris@fnal.gov}
\emailAdd{kkelly12@fnal.gov}
\emailAdd{jlraaf@fnal.gov}

\abstract{One proposed component of the upcoming Deep Underground Neutrino Experiment (DUNE) near detector complex is a multi-purpose, magnetized, gaseous argon time projection chamber: the Multi-Purpose Detector (MPD). We explore the new-physics potential of the MPD, focusing on scenarios in which the MPD is significantly more sensitive to new physics than a liquid argon detector, specifically searches for semi-long-lived particles that are produced in/near the beam target and decay in the MPD. The specific physics possibilities studied are searches for dark vector bosons mixing kinetically with the Standard Model hypercharge group, leptophilic vector bosons, dark scalars mixing with the Standard Model Higgs boson, and heavy neutral leptons that mix with the Standard Model neutrinos. We demonstrate that the MPD can extend existing bounds in most of these scenarios. We illustrate how the ability of the MPD to measure the momentum and charge of the final state particles leads to these bounds.}

%\definecolor{kjkblue}{rgb}{0.39, 0.589, 0.6914}
%\newcommand{\KJK}[1]{{\color{kjkblue} \bf KJK: #1}}
%\definecolor{jmbred}{rgb}{0.926, 0.16, 0.223}
%\newcommand{\JMB}[1]{{\color{jmbred} \bf JMB: #1}}
%\definecolor{wildstrawberry}{rgb}{1.0, 0.26, 0.64}
%\newcommand{\pjf}[1]{{\color{wildstrawberry} \bf PJF: #1}}
%\definecolor{blue-violet}{rgb}{0.54, 0.17, 0.89}
%\newcommand{\JLR}[1]{{\color{blue-violet} \bf JLR: #1}}
%\definecolor{adggreen}{rgb}{0.17, 0.8, 0.3}
%\newcommand{\adg}[1]{{\color{adggreen} \bf AdG: #1}}

\preprint{FERMILAB-PUB-19-607-ND-T, NUHEP-TH/19-16}
\setcounter{tocdepth}{2}
\renewcommand*{\thefootnote}{\fnsymbol{footnote}}

\maketitle
\flushbottom

\afterpage{\clearpage}

%---------------------------
%Introduction Section
%---------------------------

\section{Introduction}\setcounter{footnote}{0}
\label{sec:Introduction}

The upcoming Deep Underground Neutrino Experiment (DUNE)~\cite{Abi:2018dnh} will use liquid argon time projection chambers (LArTPCs) at its far detector site.  The near detector facility is currently under design~\cite{LiquidSlides}, but is foreseen to include a modular LArTPC~\cite{LiquidSlides2} (ArgonCube) with a spectrometer immediately downstream, called the Multi-Purpose Detector~\cite{GasSlides,NewGasSlides} (MPD). The MPD will consist of a high-pressure gaseous argon time projection chamber (HPTPC) surrounded by an electromagnetic calorimeter, all situated within a magnet. The HPTPC has the same nuclear target as ArgonCube and the far detectors but has a much lower density, allowing this detector to more finely resolve the details of low-energy particles produced in neutrino interactions. The LArTPC component of the near detector, although not identical to the far detector, is critical to the DUNE mission of making precision measurements of neutrino oscillations. The MPD's main purpose is to track and momentum-analyze particles exiting the LArTPC, but it will also play an important role in reducing systematic uncertainties for the oscillation analysis. 

Beyond DUNE's nominal mission to perform precision measurements of neutrino properties and neutrino oscillations, both its near and far detectors will serve as powerful probes of new physics. Recently, interest in using neutrino detectors to search for new physics has grown, with proposals to search for dark matter/dark sectors~\cite{Batell:2009di, deNiverville:2011it,  deNiverville:2012ij, Dharmapalan:2012xp,Izaguirre:2013uxa, Batell:2014yra, Dobrescu:2014ita, Coloma:2015pih, deNiverville:2016rqh, deNiverville:2018dbu, Jordan:2018gcd, deGouvea:2018cfv,Jordan:2018qiy,DeRomeri:2019kic,Tsai:2019mtm,Batell:2019nwo}, millicharged particles~\cite{Magill:2018tbb,Harnik:2019zee,Acciarri:2019jly}, and exotic interactions among the Standard Model (SM) neutrinos~\cite{Berryman:2018ogk,Bertuzzo:2018itn,Ballett:2018ynz,Kelly:2019wow,Arguelles:2018mtc}, to name a few (see Ref.~\cite{Arguelles:2019xgp} for a more thorough summary of beyond-the-Standard-Model searches in neutrino experiments). Several of these scenarios have been investigated by the experimental collaborations, leading to stringent limits on the new physics scenarios of interest~\cite{Aguilar-Arevalo:2018wea,Acciarri:2019jly}. The next generation of neutrino experiments will undoubtedly be able to interrogate these and other new physics scenarios with increased precision.

In most new physics searches at neutrino facilities, the dominant backgrounds for the signals of interest arise from interactions of neutrinos from the beam with the detector.\footnote{The MiniBooNE-DM search avoided this issue by diverting the proton beam into a beam dump, allowing neutrino-related backgrounds to be significantly smaller than in beam-on searches. This strategy requires a dedicated experimental run for a new physics search. What we propose here is parasitic with the standard operation of DUNE.} These scale with the \emph{mass} of the detector. For new physics searches that rely on particles scattering in the detector, the signal rate is also proportional to the detector mass. However, signal rates in searches for novel particle decays scale as the active \emph{volume} of the detector. For the DUNE near detectors, the LArTPC and MPD will have roughly the same active volumes, but their different densities imply that their target masses will differ by a factor of $50$. In searches for decaying particles, the signal-to-background ratio in the MPD will consequently be, roughly, a factor of $50$ larger than in the LArTPC.

We propose using the DUNE MPD as a detector for the decays of new particles within its active volume for precisely this reason. Additionally, the excellent tracking ability and charge reconstruction, as well as the particle-identification capability, of the MPD will allow for these searches to be further optimized. For decays of new particles, the MPD, as a standalone detector, will be able to outperform the LArTPC.

As a guiding principle of which new physics searches to study, we focus on the three renormalizable portals to the SM~\cite{Foot:1991bp,Foot:1991kb,Patt:2006fw,Schabinger:2005ei,Cerdeno:2006ha,Espinosa:2007qk,MarchRussell:2008yu,Ahlers:2008qc,Batell:2009di}: the vector, scalar, and neutrino portals. Each portal consists of a new particle or mediator that couples to SM particles via a renormalizable operator constructed out of Lorentz- and gauge-invariant combinations of SM fields:
\begin{eqnarray}
F^{\mu\nu} F^\prime_{\mu\nu}\quad \quad &\mathrm{Vector\ Portal,\ Kinetic\ Mixing};  \nonumber \\
V^\mu J_\mu^\mathrm{SM}\quad \quad &\mathrm{Vector\ Portal,\ Gauge\ Coupling}; \nonumber\\
H^\dagger H \left\lvert S\right\rvert^2,\quad \mu^\prime H^\dagger H S \quad \quad &\mathrm{Scalar\ Portal}; \label{eq:portals} \\
\left(LH\right)N\quad\quad &\mathrm{Neutrino\ Portal}.\nonumber
\end{eqnarray}
Here, the new physics particles are represented by $F^\prime_{\mu\nu}$, $V^\mu$, $S$, or $N$ while $F_{\mu\nu}$ is the $U(1)$ hypercharge field strength, $J_{\mu}^{\rm SM}$ is a vector current made out of SM fermion fields, $H$ is the Higgs scalar doublet, and $L$ is the lepton doublet. The parameter $\mu^\prime$ is a coupling with mass dimension 1. These new particles are interesting in their own right, but they might also be connected to other new particles, such as an extended dark sector. In this manuscript, we will explore the capabilities of the DUNE MPD in searching for the decays of each of these types of new particles.

The remainder of this paper is organized as follows. In Section~\ref{sec:ExperimentalParameters}, we describe the DUNE Near Detector complex and the MPD. We also provide details regarding the measurement capabilities and expected background rates of interest for the MPD. Section~\ref{sec:Simulation} provides details of the simulation techniques we use throughout our studies. Sections~\ref{sec:DarkPhoton}-\ref{sec:HNL} contain the specific new physics searches for which we advocate: Section~\ref{sec:DarkPhoton} details a search for dark vector bosons that mix kinetically with the SM hypercharge group; Section~\ref{sec:Leptophilic} describes a search for vector bosons that are coupled to various combinations of lepton flavor number; Section~\ref{sec:DarkHiggs} explains a search for a dark scalar particle that mixes with the SM Higgs boson; and Section~\ref{sec:HNL} explores the capability to search for heavy neutral leptons (HNLs) that mix with the SM neutrinos. Also in Section~\ref{sec:HNL}, we explore the capability of the MPD to determine, if a heavy neutral lepton is discovered, whether lepton number is violated in nature and therefore whether neutrinos are Dirac or Majorana fermions.

In this work, we show that the DUNE MPD will have exceptional capacity to search for new physics in a variety of scenarios. With the DUNE Near Detector design not yet finalized, we use this aspect of the MPD to advocate for its inclusion in the final design for the Near Detector suite. While the LArTPC could search for these new physics scenarios, the added capabilities of the MPD allow for a considerable increase in sensitivity and the possibility of discovering new physics.

%---------------------------------------------
%Experimental Parameters Section
%---------------------------------------------

\section{DUNE Near Detector Complex}\label{sec:ExperimentalParameters}\setcounter{footnote}{0}

In this section, we outline our assumptions for the setup of the DUNE Near Detector complex, including the proton beam target, decay volume, and the layout of detectors in the experimental hall. We direct the reader to Refs.~\cite{Acciarri:2015uup,Acciarri:2016crz,Acciarri:2016ooe,Strait:2016mof} for more detail regarding the DUNE Experiment.

\begin{figure}[t]
    \centering
    \includegraphics[width=0.9\linewidth]{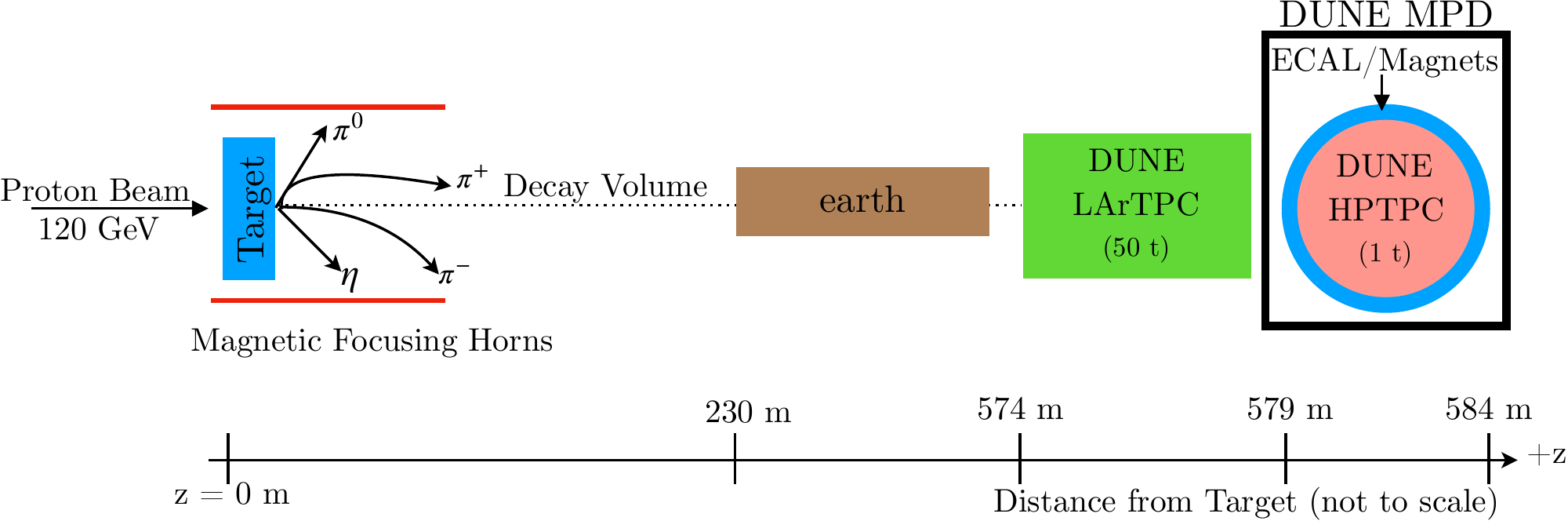}
    \caption{Schematic layout of the DUNE beam, target, and near detectors. The $z$ axis represents the beam direction; $z = 0$~m corresponds to the front of the target, $z = 574$~m corresponds to the front of the liquid argon near detector, and $z = 579$~m corresponds to the front of the multi-purpose detector. The masses listed with the two detector components are the expected fiducial masses.}
    \label{fig:Layout}
\end{figure}

Fig.~\ref{fig:Layout} provides a schematic of the features of the DUNE target and near detector hall of interest for this work. The figure is not drawn to scale. We consider that the proton beam strikes a target and that the particles emerging from the target enter a magnetic focusing horn system. The focused particles traverse a decay volume; any particles that reach the boundaries of the decay volume are assumed to come to rest and decay. The near detector hall is positioned downstream of the decay volume and the space between them is occupied by raw earth.

\subsection{Experimental Parameters and Assumptions}

Here we specify the exact experimental parameters that we assume in our simulations, including the beam setup, magnetic focusing horns, and the detectors.

\textbf{Beam:} We assume that the protons in the LBNF-DUNE beam have $120$ GeV of energy, and that $1.47 \times 10^{21}$ protons strike the target each year. Unless otherwise stated, we assume data collection of equal time in neutrino (``forward horn current'') and antineutrino (``reverse horn current'') modes. In general, we will assume ten years\footnote{Some projections of DUNE operation include the possibility of an upgraded beam (roughly twice the number of protons on target per year) that can operate in the second half of the experiment. We assume a constant number of protons on target per year, so our ten-year projection is equivalent to a shorter operation time with such an upgraded beam.} ($1.47 \times 10^{22}$ protons on target) of total data collection in our analyses. Different beam configurations, most notably the possibility of the beam protons having $80$ GeV of energy or focusing higher energy SM neutrinos, have also been considered by the collaboration. In Appendix~\ref{appendix:MesonProd}, we discuss the impact of different beam configurations on our results, which is marginal.

\textbf{Focusing Horns:} Consistent with simulations performed by the DUNE collaboration, we assume three aluminum focusing horns that produce toroidal magnetic fields. Depending on the polarity of the applied current, the horns focus either the positively or negatively charged particles that are produced in collisions of the protons with a graphite target embedded within the first horn.
As explained in detail in Section~\ref{sec:Simulation}, we include the magnetic focusing effects on the momentum distributions of long-lived, charged mesons -- $\pi^\pm$ and $K^\pm$, specifically. Short-lived ($D^\pm$, $D_s^\pm$) and neutral mesons ($\pi^0$, $\eta$, $K_L^0$, $K_S^0$) produced in our simulations are not deflected by the magnetic focusing horns.

\textbf{Decay Volume and Near Detector Distance:} We assume a total distance of $230$~m between the target and the end of the decay pipe, wherever relevant. The front of the LArTPC is $574$~m from the target, and the front of the MPD is  $579$~m from the target. 

\textbf{Liquid Argon Near Detector:} The liquid argon near detector has a width (transverse to the beam direction) of $7$ m, a height of $3$ m, and a length (in the beam direction) of $5$ m. The fiducial mass of the near detector is $50$ t of argon. See Ref.~\cite{LiquidSlides} for more detail.

\textbf{Multi-Purpose Detector:} The Multi-Purpose Detector (MPD) is a magnetic spectrometer with a cylindrical high-pressure gaseous argon time projection chamber (HPTPC) at its heart. The HPTPC has a diameter of $5$~m and a length of $5$~m. It is surrounded by an electromagnetic calorimeter (ECAL) and the HPTPC+ECAL system is situated inside a magnet with $0.5$~T central field. The axis of the cylinder is perpendicular to the beam direction. However, for simplicity, when simulating our signals (see Section~\ref{sec:Simulation}), we assume the symmetry axis of the detector is aligned with the direction of the beam, since this leads to a uniform apparent detector thickness for all incoming particles. The possibility of including a muon tagger interleaved with the ECAL is under exploration by the collaboration~\cite{NewGasSlides}. We discuss how such a muon tagger impacts our results wherever relevant.

The HPTPC operates at a pressure of 10 atmospheres, leading to a total mass of argon in the gas TPC of roughly $1.8$~tons ($1$~ton fiducial). See Refs.~\cite{GasSlides,DocDB,NewGasSlides} for more detail. Parameters associated to the detection of charged particles in the gaseous argon are given in the next subsection. 

\subsection{Particle Thresholds and Reconstruction Capabilities}

In this subsection, we discuss various thresholds required for particle identification in the HPTPC, as well as the capability of measuring charge and distinguishing pairs of particles. We make use of the information in Ref.~\cite{DocDB}.
\begin{table}[!htbp]
\begin{center}
\begin{tabular}{|c|c|}\hline
Particle & Kinetic Energy Threshold (MeV) \\ \hline\hline
$e^+/e^-$ & $0.14$ \\ \hline
$\mu^+/\mu^-$ & $1.8$ \\ \hline
$\pi^+/\pi^-$ & $1.8$ \\ \hline
$p$ & $3.7$ \\ \hline
\end{tabular}
\caption{Expected thresholds for track identification in gaseous argon, operated at 10 atm.\label{tab:Parameters:Thresholds}}
\end{center}
\end{table}

First, we focus on the energy thresholds for which a charged particle would leave a track long enough to be identified. We assume that $2$~cm is a reasonable length to identify a track. According to the continuous-slowing-down approximation~\cite{nist}, this corresponds to kinetic energies of $0.14$~MeV, $1.8$~MeV, and $3.7$~MeV for electrons, muons, and protons, respectively. These values assume a pressure of $10$ atm in the gaseous argon. Since the energy deposition $dE/dx$ is largely similar for muons and pions, we take their kinetic energy thresholds to be the same. These values are listed in Table~\ref{tab:Parameters:Thresholds} for clarity. We note that these are far lower than the thresholds for particle identification in the liquid argon near detector~\cite{Acciarri:2015uup}.

We assume that transverse momenta are measured at the $0.7\%$ level if they exceed $1$~GeV. Below $1$~GeV, we assume them to be measured at the $1\%$ level. 
Charge identification is expected to be efficient and accurate. Two particles with opposite charges emerging from a vertex will be identified as two separate tracks once they are separated by more than $1$~cm. For the purposes of this work, this corresponds to nearly perfect separation of positive-particle/negative-particle pairs, unless the vertex is very close to a detector boundary. Furthermore, there is no ambiguity in the charge assignment. Nearly all experimental signatures studied throughout this work will include at least two charged particles (with opposite signs) emerging from a common vertex. Events with only one charged particle, assuming that particle has a very high energy, could be more difficult to charge-identify.

\subsection{Particle Signatures in the Multi-Purpose Detector}
\label{subsec:ParticleSignals}

In the HPTPC, energy deposited in the form of ionization is sampled by a plane of charge-sensitive pads at each end of the cylindrical detector.  The characteristics of the charge (energy) deposited per unit length ($dE/dx$) for each track enable this detector to distinguish electrons from heavier particles for a large range of particle momenta, and some ability to distinguish among the tracks left by heavier particles (pions, muons, kaons, protons), especially if they come to a stop within the gas volume. The ECAL design is currently undergoing optimization, but in its current state it is only one interaction length thick. By itself, this ECAL should have some ability to separate interacting pions from (1) muons and (2) pions that punch through (all muons and roughly 30\% of pions will punch through and escape the ECAL). With the addition of a muon tagger (currently under consideration~\cite{NewGasSlides}) outside the ECAL, more separation power would be achieved.

Table~\ref{tab:Parameters:Signals} lists the types of signals that are expected for the new physics processes discussed in later sections, along with links to those sections. In general, the signals are oppositely charged pairs of leptons or hadrons whose reconstructed tracks should point back to the target. The reconstructed invariant mass will provide another handle for identifying some of the possible decay channels. Finally, the decaying particles are produced associated with the high-energy proton-beam--target interaction and will have $\mathcal{O}(10)$ GeV of energy, leading to decay products with several GeV each. In contrast, much of the background (produced in neutrino interactions) will have lower energy, giving another kinematical variable with which signal events can be distinguished.

\begin{table}[!htbp]
\begin{center}
\begin{tabular}{|c|c|c|}\hline
Signal & Features & New Physics Models \\ \hline\hline
$X \to e^+e^-$ & Invariant Mass, Direction & \hyperref[sec:DarkPhoton]{Dark Photon}, \hyperref[sec:Leptophilic]{Leptophilic Gauge Boson}, \hyperref[sec:DarkHiggs]{Dark Higgs} \\ \hline
$X \to \mu^+\mu^-$ & Invariant Mass, Direction & \hyperref[sec:DarkPhoton]{Dark Photon}, \hyperref[sec:Leptophilic]{Leptophilic Gauge Boson}, \hyperref[sec:DarkHiggs]{Dark Higgs} \\ \hline
$X \to \pi^+\pi^-/\pi^0\pi^0$ & Invariant Mass, Direction & \hyperref[sec:DarkPhoton]{Dark Photon}, \hyperref[sec:Leptophilic]{Leptophilic Gauge Boson}, \hyperref[sec:DarkHiggs]{Dark Higgs} \\ \hline
$X \to \nu e^+ e^-$ & Direction, Energy & \hyperref[sec:HNL]{Heavy Neutral Leptons} \\ \hline
$X \to \nu e^\pm \mu^\mp$ & Direction, Energy & \hyperref[sec:HNL]{Heavy Neutral Leptons} \\ \hline
$X \to \nu \mu^+ \mu^-$ & Direction, Energy & \hyperref[sec:HNL]{Heavy Neutral Leptons} \\ \hline
$X \to e^\pm \pi^\mp/\mu^\pm \pi^\mp$ & Invariant Mass, Direction & \hyperref[sec:HNL]{Heavy Neutral Leptons} \\ \hline
$X \to e^\pm \rho^\mp/\mu^\pm \rho^\mp$ & Invariant Mass, Direction & \hyperref[sec:HNL]{Heavy Neutral Leptons} \\ \hline
\end{tabular}
\caption{Expected decay signatures of new physics processes. In the left column, $X$ represents the new physics particle decaying.\label{tab:Parameters:Signals}}
\end{center}
\end{table}

In the HPTPC, possible background pairs of $e^+e^-$ arise from photon conversions. The conversion distance\footnote{This is in contrast with the LArTPC, where the conversion distance of photons is $\mathcal{O}(10~\mathrm{cm})$. Far more photons can fake the signal of $e^+e^-$ in the LArTPC than in the HPTPC.} of photons in the gas is $\mathcal{O}(10~$meters$)$; approximately 12\% of photons produced by standard beam neutrino interactions in the gas will convert before reaching the ECAL. However, these converted $e^+e^-$ pairs will typically be of much lower energy than the signal pairs considered in this paper and they will infrequently correlate with the beam direction. We discuss backgrounds of this nature in detail in Section~\ref{sec:DarkPhoton}.

For momenta above a few hundred MeV/$c$, it is not possible to distinguish muons from pions via $dE/dx$ alone in the HPTPC. With the addition of the ECAL (and potentially a muon tagger~\cite{NewGasSlides}), some fraction of the pions will be distinguishable by their hadronic interactions. We discuss the capability of distinguishing muons and pions in the relevant sections of this paper. Backgrounds could then arise from standard charged-current neutrino interactions in the gas, in which a muon and a pion of opposite charges are produced with no other visible particles, but these would still need to point back in the direction of the target to be considered background to signals discussed in this paper.

Neutral pions are only visible in the HPTPC if the photons from the pion decay convert. Most of the time, the photons will reach the ECAL, where they will convert and produce signals with a shower axis that points back to the $\pi^0$ decay point. Given the $\mathcal{O}($meters$)$-long photon conversion length, rarely (about 1\% of the time) will both photons convert within the gas. If one photon converts in the gas and the other reaches the ECAL before converting, then the energy depositions in the two subdetectors will still point back to a common vertex, allowing reconstruction of the $\pi^0$ invariant mass. Backgrounds to the new physics processes are expected to be small, since the background processes would need to produce multiple pions and have no other activity at the neutrino interaction vertex. Multiple pion production in standard neutrino interactions is generally accompanied by additional vertex activity, which is detectable down to very low energies in the HPTPC.

\subsection{Background Rates from the Neutrino Beam}
Table~\ref{tab:Backgrounds} lists the expected number of neutrino-related events for scattering on argon for a variety of different interaction types~\cite{DocDB,LauraFluxes}. For different search channels throughout this work, these events can contribute as background (assuming they have the right kinematics or could otherwise be misidentified as our desired signal). We will use these estimated rates to infer the total number of background events for each search.

\begin{table}[!htbp]
\begin{center}
\begin{tabular}{|c|c|}\hline
Event class & Number of events per ton-year \\ \hline \hline
$\nu_\mu$ CC Total & $1.64\times 10^{6}$ \\ \hline
$\nu_\mu$ NC Total & $5.17\times 10^5$ \\ \hline
$\nu_\mu$ CC $\pi^0$ inclusive & $4.47 \times 10^{5}$ \\ \hline
$\nu_\mu$ NC $\pi^0$ inclusive & $1.96\times 10^5$ \\ \hline\hline
$\nu_e$ CC Total & $1.89 \times 10^4$ \\ \hline
$\nu_e$ NC Total & $5.98 \times 10^3$ \\ \hline
\end{tabular}
\caption{Expected number of neutrino-related events scattering on argon (liquid or gas) assuming $1$ t of fiducial mass and $1$ year of data collection.\label{tab:Backgrounds}}
\end{center}
\end{table}

As an example, consider the total number of $\nu_\mu$ charged-current events in one year of operation -- $1.64 \times 10^6$ -- assuming one ton of fiducial mass. Charged-current $\nu_\mu$-scattering on argon will lead to the production of a single charged pion along with the muon (and no other detectable hadronic activity at the interaction vertex) $\mathcal{O}(10\%)$ of the time. 
Therefore, we naively expect $\mathcal{O}(10^5)$ of this type of background event  -- a $\mu^-\pi^+$ pair -- per year of data collection in the MPD. If, for example, we were searching for a signal that consists of $\mu^+ \mu^-$ pairs, then this background could be further reduced using the kinematic information of the charged particles, information regarding the hadronic system, particle identification capabilities, etc. We will discuss these techniques in detail in Sections~\ref{DarkPhoton:Backgrounds} and~\ref{sec:HNL:Backgrounds}.

%\afterpage{\clearpage} 
%--------------------------------
%Simulations Section
%--------------------------------

\section{Simulation Details}\label{sec:Simulation}\setcounter{footnote}{0}

All of the new physics scenarios studied in this work rely on simulating the flux of SM particles that then decay into beyond-the-Standard-Model (BSM) particles at or near the DUNE target. We will be interested in the flux of the BSM particles, both in terms of the direction and energy spectrum, in order to determine the probability for a given BSM particle to reach the DUNE MPD and decay inside it. In this section, we discuss the details of simulating the production of SM particles (predominantly mesons) that decay into BSM particles. 

\subsection{Production of Charged and Neutral Mesons}
The general strategy of DUNE, while operating as a neutrino beam experiment, is to produce a large flux of charged mesons (mostly $\pi^\pm$ and $K^\pm$) that decay leptonically, leading to a large flux of SM neutrinos (mostly $\nu_\mu$ or $\overline{\nu}_\mu$, depending on the mode of operation) at the near and far detectors. In order to produce a purer beam (i.e., one with a smaller contamination of $\overline{\nu}_\mu$ with respect to $\nu_\mu$ when operating in neutrino mode, or vice versa for antineutrino mode), focusing horns are utilized immediately downstream of the target, as depicted in Fig.~\ref{fig:Layout}. This allows for focusing of positive (forward horn current) or negative (reverse horn current) mesons. Forward (reverse) horn current corresponds to an enhancement of the (anti)neutrino flux, so we will refer to the modes as ``neutrino'' and ``antineutrino'' modes henceforth.

We will explore scenarios in which the flux of a new particle is produced via decays of charged or neutral mesons. To simulate the production rate of a given meson, we use the software {\sc Pythia8}~\cite{Sjostrand:2007gs}, simulating 120~GeV protons on target and using the \texttt{``SoftQCD:all''} option to generate mesons. In doing so, we obtain the overall rate of meson production from the primary proton interaction; this procedure underestimates the total meson flux as it neglects secondary production when mesons produced in the primary interaction scatter in or near the target. Our results may then be interpreted as moderately conservative.

The average numbers of different charged and neutral mesons produced per proton on target (POT), assuming a 120~GeV beam, are listed in Tables~\ref{tab:ChargedMesonProduction} and \ref{tab:NeutMesonProduction}, respectively.
{\sc Pythia8} also allows us to obtain the lab-frame four-momenta of the outgoing mesons, which will be of interest in simulating the flux of different new particles. Appendix~\ref{appendix:MesonProd} provides more details of this simulation as well as comparisons between 80~GeV and 120~GeV proton beam results.
\begin{table}[!htb]
    \centering
    \begin{tabular}{|c||c|c|c|c|c|c|c|c|}\hline
        Species & $\pi^+$ & $\pi^-$ & $K^+$ & $K^-$ & $D^+$ & $D^-$ & $D_s^+$ & $D_s^-$ \\ \hline
         Mesons/POT & $2.7$ & $2.4$ & $0.24$ & $0.16$ & $3.7\times 10^{-6}$ & $6.0\times 10^{-6}$ & $1.2\times 10^{-6}$ & $1.6\times 10^{-6}$ \\ \hline
    \end{tabular}
    \caption{Average number of charged mesons produced per proton-on-target assuming 120 GeV protons.}
    \label{tab:ChargedMesonProduction}
\end{table}

\begin{table}[!htb]
    \centering
    \begin{tabular}{|c||c|c|c|c|}\hline
        Species & $\pi^0$ & $\eta$ & $K_L^0$ & $K_S^0$ \\ \hline
         Mesons/POT & $2.9$ & $0.33$ & $0.19$ & $0.19$ \\ \hline
    \end{tabular}
    \caption{Average number of neutral mesons produced per proton-on-target assuming 120 GeV protons.}
    \label{tab:NeutMesonProduction}
\end{table}

\subsection{Focusing of Charged Mesons}

The quantities we explicitly require in order to simulate meson decays are not their four-momenta at production but instead their four-momenta at the time of decay. Neutral mesons ($\pi^0$, $K_L^0$, $K_S^0$) and charged mesons that decay promptly ($D^\pm$, $D_s^\pm$) are unaffected by the focusing horns, so,  in our simulations, we use the lab-frame four-momenta obtained with {\sc Pythia8} as the  four-momenta at decay, assuming they don't hit the boundaries of the decay volume.  Long-lived charged mesons (namely, $\pi^\pm$ and $K^\pm$), on the other hand, are subject to the effects of the focusing horns. For the distributions of the long-lived charged mesons, we use output from the DUNE Beam Interface Working Group (BIWG)~\cite{LauraFluxes}, which makes use of {\sc Geant4}~\cite{Agostinelli:2002hh,Allison:2016lfl} and {\sc Fluka}~\cite{Ferrari:2005zk,Bohlen:2014xx}. 

\begin{figure}
    \centering
    \includegraphics[width=\linewidth]{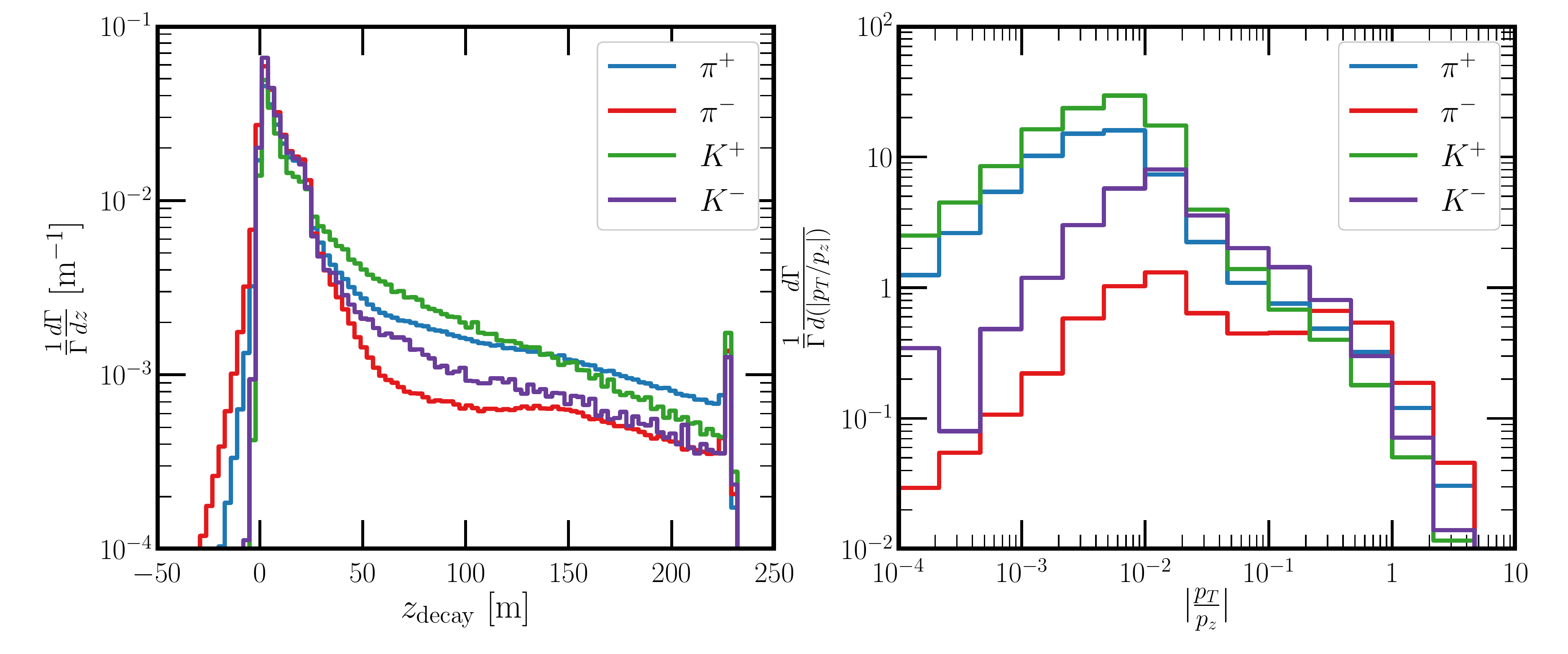}
    \caption{Decay distributions of long-lived charged mesons as determined by the DUNE collaboration~\cite{LauraFluxes} using {\sc Geant4} and {\sc Fluka} simulations of charged meson transport through the DUNE focusing horns. The left panel presents decay distributions in terms of $z_{\rm decay}$, the longitudinal position of the decay. The right panel presents decay distributions in terms of the ratio of transverse and longitudinal momentum, $|p_T/p_z|$, as a proxy for how well-focused the corresponding mesons are. In both panels, we show $\pi^+$ distributions (blue), $\pi^-$ distributions (red), $K^+$ distributions (green), and $K^-$ distributions (purple). This figure is for operation in neutrino mode, in which positively charged mesons are focused.}
    \label{fig:Focusing:DecayDists}
\end{figure}

Fig.~\ref{fig:Focusing:DecayDists} depicts the normalized differential decay rates of the long-lived charged mesons extracted from {\sc Geant4} and {\sc Fluka} (with operation in neutrino mode), as a function of the longitudinal distance $z_\mathrm{decay}$ in the left panel and 
\begin{equation}
\left\lvert \frac{p_T}{p_z}\right\rvert \equiv \frac{ \sqrt{p_x^2 + p_y^2}}{|p_z|}
\end{equation}
in the right panel. The left panel of Fig.~\ref{fig:Focusing:DecayDists} illustrates several effects of the focusing horns. First, we see that the positively charged mesons tend to decay at larger $z$ -- this is because they are the focused mesons, and travel through the horns and beam pipe to larger distances. Secondly, we also see a peak of decays for all four distributions located near $z_\mathrm{decay} \approx 230$ m, corresponding to the end of the decay pipe. Mesons that reach the end of the decay pipe will stop and decay at rest. In the right panel, $\left\lvert \frac{p_T}{p_z}\right\rvert$ characterizes how well-focused a given meson is at the time of its decay -- a perfectly focused meson would have $p_T = 0$ and point towards the near detector. We clearly see two effects here -- first, the distribution of the positively charged mesons peaks at smaller $\left\lvert \frac{p_T}{p_z}\right\rvert$ than that of the negatively charged ones, as the former are focused. Also, we observe that there is a greater separation, due to focusing, between the $\pi^+$ vs. $\pi^-$ distributions than for the $K^+$ vs $K^-$ distributions: pions are focused more efficiently than kaons.

\begin{figure}
    \centering
    \includegraphics[width=0.5\linewidth]{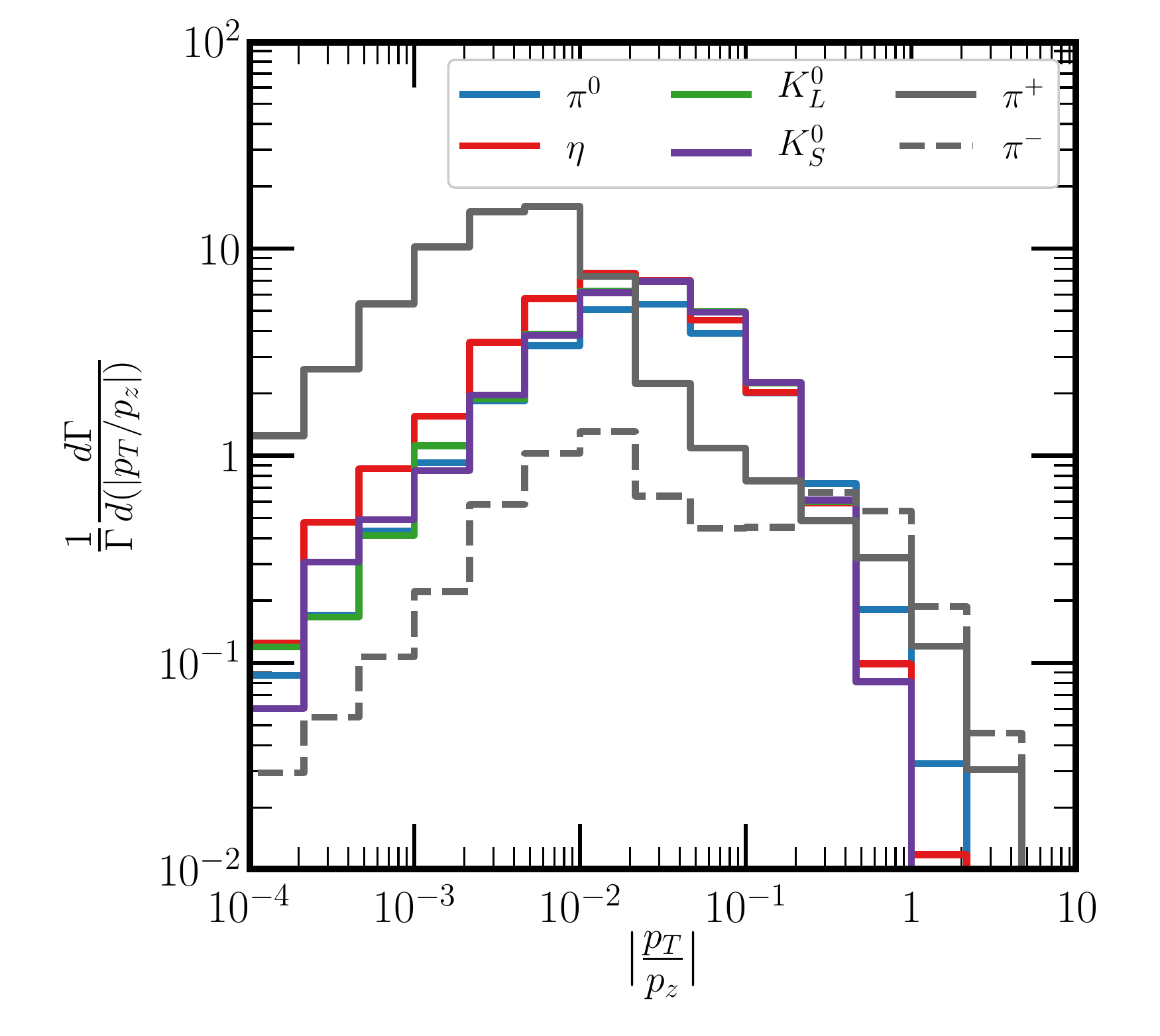}
    \caption{Identical to the right panel of Fig.~\ref{fig:Focusing:DecayDists}, except for unfocused neutral mesons: $\pi^0$ (blue), $\eta$ (red), $K^0_L$ (green), and $K^0_S$ (purple). We include the focused/unfocused $\pi^+$/$\pi^-$ distributions for comparison in grey solid/dashed lines.}
    \label{fig:Focusing:DecayDistsUnfocused}
\end{figure}

In contrast to the right panel of Fig.~\ref{fig:Focusing:DecayDists}, Fig.~\ref{fig:Focusing:DecayDistsUnfocused} depicts the normalized differential decay rates as a function of $\left\lvert \frac{p_T}{p_z}\right\rvert$ for a subset of the unfocused species -- namely, $\pi^0$, $\eta$, $K_L^0$, and $K_S^0$. The distributions for promptly decaying $D^\pm$ and $D_s^\pm$ are similar to those in Fig.~\ref{fig:Focusing:DecayDistsUnfocused} and we choose not to display them.
Comparing Fig.~\ref{fig:Focusing:DecayDistsUnfocused} with the right panel of Fig.~\ref{fig:Focusing:DecayDists}, the bulk of decays of the unfocused mesons lie near $\left\lvert \frac{p_T}{p_z}\right\rvert \sim 3\times 10^{-2}$, while those of the focused mesons lie near $\left\lvert \frac{p_T}{p_z}\right\rvert \sim 10^{-2}$. Instead, the bulk of decays of the anti-focused, wrong-sign mesons (specifically $\pi^-$ in this case) lie near $\left\lvert \frac{p_T}{p_z}\right\rvert \sim 10^{-1}$. As discussed, we simulate the production of neutral mesons using {\sc Pythia8} and determine the position of their decay according to the proper lifetime of the meson and its simulated energy. We insist that any meson that reaches the end of the decay pipe decays at rest at the rock surface, see Fig.~\ref{fig:Focusing:DecayDists} (left).

\subsection{Particle Flux at the DUNE Near Detector}
Once the parent distribution has been obtained, we must determine the flux of the BSM particle(s) of interest at the DUNE near detector. This is done on a particle-by-particle basis by Monte-Carlo simulating the rest-frame decays of the parent particle to obtain the four-momenta of the relevant daughter(s). These four-momenta are boosted from the parent's rest frame back to the lab frame. Combined with the position $(x, y, z)$ of the meson decay location, we can determine whether the BSM particle will pass through the MPD. The fraction of produced particles that pass through the near detector defines the geometrical acceptance of the detector. A careful treatment of this effect is critical for the DUNE near detector. Since it has a small solid angle [$\mathcal{O}(10^{-4}$ rad$^2$)], the flux of new particles is highly dependent on features such as the boost of the parent meson. Because of our assumption that the detector is coaxial with the beam direction (see Section~\ref{sec:ExperimentalParameters}), we have (effectively) reduced the solid angle by roughly $20\%$. This lower acceptance will be compensated in our signal event rates by the longer effective depth of the detector; these two effects will approximately cancel.

We are interested in decays of parent mesons into both two- and three-body final states. The two-body kinematics may be solved easily in the parent rest frame; however, the three-body kinematics is less straightforward. We discuss our recipe for handling three-body decays in the next subsection. 

\subsubsection{Three-Body Decays}
In some cases, we are interested in the production of a particle that comes from a three-body decay. Here we discuss the assumptions made in these situations and their validity. Consider the three-body decay of a particle $P$ into particles $S_1$, $S_2$, and $X$ (with masses $m_P$, $m_{S_1}$, $m_{S_2}$, and $m_X$, respectively); we will be interested in the lab-frame kinematics of $X$ (i.e., its energy and whether it is pointing in the direction of the detector). In all of the scenarios we consider in this work, $P$ is considered to be spin-0, so the angular distribution of the outgoing particles is isotropic in its rest-frame. In contrast to two-body decays, however, $X$ is not monoenergetic in that frame. Kinematics dictates that the rest-frame energy of $X$ must lie in the range
\begin{equation}
    m_X \leq E_X^\mathrm{(CM)} \leq \sqrt{\frac{ \left(m_P^2 - \left(m_{S_1}^2 + m_{S_2}^2 + m_X^2\right)\right) \left(m_P^2 - \left(m_{S_1}^2 + m_{S_2}^2 - m_X^2\right)\right)}{4m_P^2} + m_X^2}.
\end{equation}
Beyond this, different interaction structures governing the decay $P\to S_1 S_2 X$ determine the shape of the differential width $d\Gamma/dE_X$.

Our goal in this subsection is to investigate the dependence of our analysis on the choice of $d\Gamma/dE_X$, and to show that the dependence is fairly small. As an illustrative example, we focus on the decay $K^+ \to \mu^+ \pi^0 N$, where $N$ is a heavy neutral lepton (see Section~\ref{sec:HNL}) with a mass of 200 MeV.
\begin{figure}
    \centering
    \includegraphics[width=\linewidth]{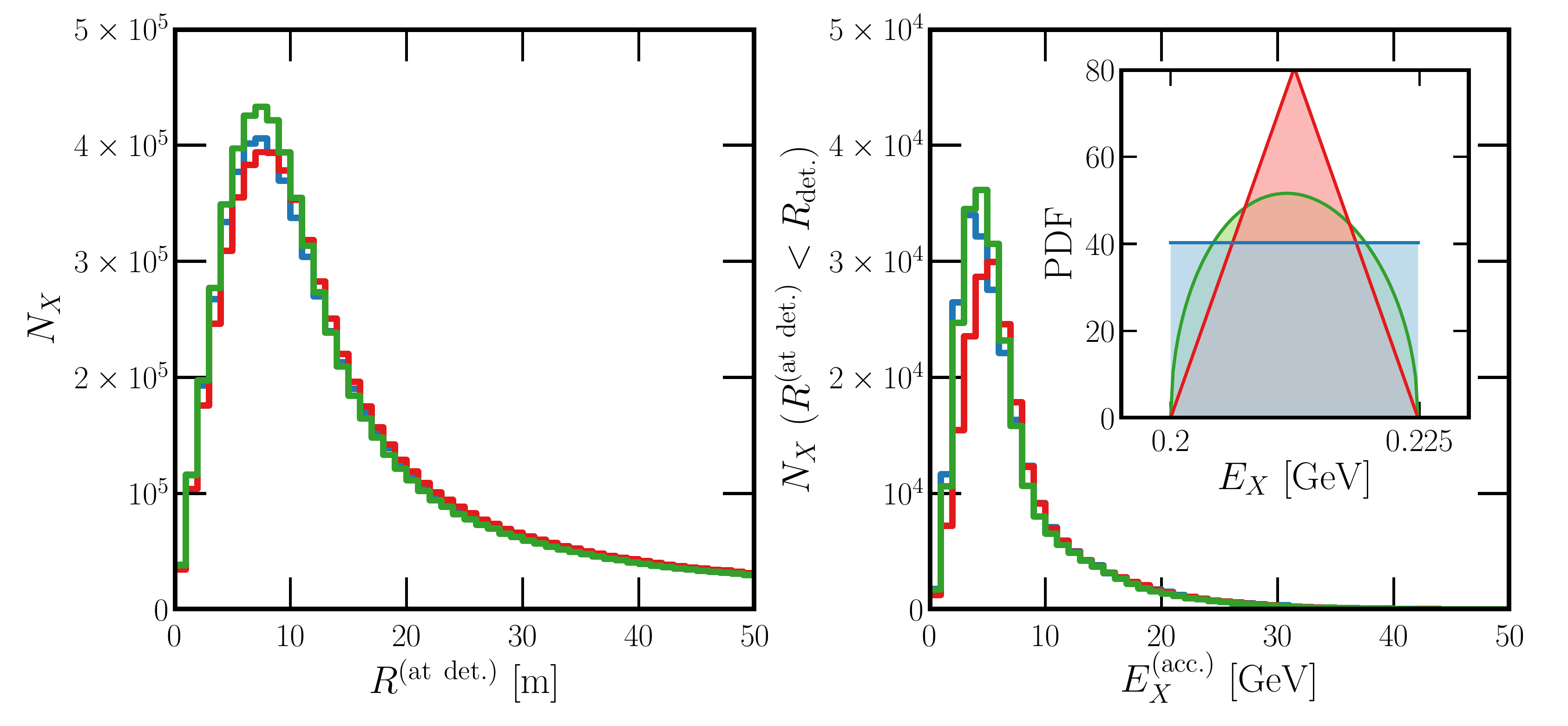}
    \caption{Distributions of $N$ produced in the decay $K^+ \to \mu^+ \pi^0 N$, assuming three different probability density functions for the energy of $N$ in the $K^+$ rest frame. For concreteness, we take $m_N = 200$ MeV. The three distributions are flat (blue), piecewise-linear (red), and the true distribution (green) if this is a three-body decay producing a heavy neutral lepton. The left panel displays the spatial distribution of $N$ at the front face of a detector located $579$ m from the original $K^+$ production point. Of those that have pass through the face of the MPD, we display their energy distribution in the right figure.}
    \label{fig:ThreeBodyFocusing}
\end{figure}

We simulate the three-body decay (isotropic in the $K^+$ rest frame) according to three different energy distributions: (a) flat between the energy endpoints (200 MeV to 225 MeV) (b) piecewise linear, symmetric about the middle of the energy range, and (c) the true energy distribution for HNL production, as described in Ref.~\cite{Gorbunov:2007ak}. These three energy distribution probability density functions are depicted in the inset of Fig.~\ref{fig:ThreeBodyFocusing} (right). The decays are simulated using the focused $K^+$ distributions (for simplicity, we concentrate on operation during neutrino mode only), and we determine (1) the spatial distribution of $N$ at the location of the MPD, and (2) the energy distribution of the $N$ particles that pass through the MPD. 

Fig.~\ref{fig:ThreeBodyFocusing} depicts the results of the simulations. The left panel depicts the radial distribution at $z = 579$ m (the front face of the MPD), as measured from the center of the beam axis.\footnote{As an aside, we note that the radial distribution peaks at $R\sim 10$ m. The standard model neutrino flux peaks at much lower radii. Using the DUNE-PRISM near detector concept~\cite{duneprism}, where the liquid and gaseous argon near detectors may move off-axis up to 36 m could allow for further optimization of these types of searches.} The right panel depicts the energy distribution of accepted $N$ (i.e., those with $R^{\mathrm{(at\ det.)}} < 2.5$ m). In all panels of Fig.~\ref{fig:ThreeBodyFocusing}, the blue curves correspond to the flat energy distribution, the red correspond to the piecewise-linear distribution, and the green correspond to the true energy distribution. In the left panel of Fig.~\ref{fig:ThreeBodyFocusing}, we see that the flat and true distributions both tend to agree in terms of radial distributions, so we expect that the acceptance fraction in simulating these two scenarios should be largely similar. The piecewise-linear distribution tends to prefer larger radii. Similarly, in the right panel, we see that the energy distributions of accepted events for the flat and true distributions match fairly well, where the piecewise-linear distribution prefers higher energies. Based on this evidence, when simulating three-body decays, we assume the distributions of the decay products are flat in the phase space of the final-state particle energy. For our ambitions here, this simplified analysis suffices, especially when it comes to estimating detector sensitivities.

\subsection{Signals of Decays in the Multi-Purpose Detector}
We are, in general, interested in the event rates of different decays of new particles in the DUNE MPD. After determining the flux of a given new particle at the MPD location and determining the associated probability for a decay of interest to occur, we also simulate the kinematics of the final-state particles. All of these are model dependent and discussed in detail in Sections~\ref{sec:DarkPhoton}-\ref{sec:HNL}.

Of interest are quantities such as the energy of visible particles and opening angles between different particles that leave a track in the MPD. When any threshold or efficiency effects (see Table~\ref{tab:Parameters:Thresholds}) are relevant, we apply them. We will discuss how the kinematic quantities will allow for an improved search for these new particles, where relevant. For instance, any new particle that decays completely visibly, such as a dark photon decaying to an electron-positron pair, will produce decay products with an invariant mass peak once the total four-momentum of the final-state particles is reconstructed. This allows for good separation of signal from background since the latter is not expected to have any such feature.

We will also be specifically focused on the decay products' kinematics when discussing disentangling different new physics scenarios. In particular, when discussing whether a newly-discovered HNL is a Dirac or Majorana fermion (see Section~\ref{sec:HNL:DiracMajorana}), we will focus on the lab-frame decay product energies as a proxy for the new particle's rest-frame decay distribution. Such kinematic information can provide an interesting handle into whether lepton number is violated, and we will exploit it when possible.

%\afterpage{\clearpage}  
%------------------------------
%Dark Photon Section
%------------------------------

\section{Dark Photon}\label{sec:DarkPhoton}\setcounter{footnote}{0}

In this section and Section~\ref{sec:Leptophilic}, we explore the possibility that a new vector boson exists in nature. Here, we will focus on the case in which this vector boson acquires small couplings to SM fermions only through kinetic mixing with the SM $U(1)$ hypercharge group -- the so-called dark photon -- interacting via the vector portal,
\begin{equation}\label{eq:DarkPhoton:Lagrangian}
\mathcal{L} \supset -\frac{1}{4}F^{\prime\mu\nu} F^\prime_{\mu \nu}- \frac{\varepsilon}{2} F^{\mu\nu} F^\prime_{\mu \nu} + \frac{M_{A^\prime}^2}{2} A^\prime_\mu A^{\prime\mu},
\end{equation}
where $M_{A^\prime}$ is the mass of the dark photon $A^{\prime}$ with field strength $F_{\mu\nu}'$ and $\varepsilon$ characterizes the strength of kinetic mixing with the SM hypercharge gauge-group, with field strength $F_{\mu\nu}$.  The kinetic mixing term is a renormalizable operator and the size of $\varepsilon$ depends upon how it is generated~\cite{Gherghetta:2019coi}; we will be interested in $\varepsilon < 10^{-4}$.  

As discussed in Section~\ref{sec:Introduction}, many new physics search strategies utilizing neutrino detectors have been proposed, specifically in searches for dark matter and associated dark sectors. One particular intensely studied new physics model, due to its simplicity, is one in which dark matter is charged under $U(1)^\prime$ and couples to the SM via the dark photon introduced in Eq.~(\ref{eq:DarkPhoton:Lagrangian}). Such dark matter can be fermionic or scalar, and current and future neutrino experiments are capable of probing well-motivated regions of parameter space in which the dark matter is a thermal relic, symmetric between dark matter particles and antiparticles in the early universe. The search strategy adopted in these proposals typically calls for the dark matter to be produced in meson decays either via an off-shell dark photon, or a dark photon that decays promptly into dark matter pairs. The dark matter then travels to the detector and scatters, depositing energy that can be measured at the neutrino near detector (typically scattering off nuclei or electrons, although more exotic signatures have been proposed~\cite{deGouvea:2018cfv,Kim:2019had}). 
In this section, we focus on searching only for the dark photon without requiring the existence of dark matter, $\chi$.  Should the dark photon be part of a larger sector containing dark matter, we require that decays of the dark photon, $A'\rightarrow 2\chi$, are forbidden, i.e., $m_\chi<M_{A'}<2m_\chi$.  
Thus, any dark photon that is produced on-shell will only decay into SM particles. Depending on the strength of the kinetic mixing $\varepsilon$, such decays may lead to long-lived dark photons. This is the region of parameter space in which we are interested for this study.

Previous experiments -- among them, E141~\cite{Riordan:1987aw,Bjorken:1988as,Bjorken:2009mm,Andreas:2012mt}, Orsay~\cite{Davier:1989wz}, NuCal~\cite{Blumlein:2011mv,Blumlein:2013cua,Tsai:2019mtm}, and E137~\cite{Bjorken:1988as} -- have probed a similar region of parameter space in this fashion.\footnote{Ref.~\cite{Bauer:2018onh} provides a thorough review of searches for dark photons and existing constraints.}  The main differences between the various experiments is the production mechanism and the distance between the target where the dark photons are produced and the instrumented detector region. 
The interplay between these means each experiment has a sweet spot where it will be best suited to search for a particular combination of dark photon mass and kinetic mixing parameter. 
For the DUNE MPD, we will be interested in $A^\prime$ produced via the decays of light mesons $\pi^0$ and $\eta$ into $\gamma A^\prime$, as well as production from the continuum process $p p \to p p A^\prime$. Dark photons produced in these ways have energies of several GeV. The DUNE MPD will be sensitive to regions of parameter space for which the lab-frame decay length of the $A^\prime$ is $\mathcal{O}(100~\rm m)$ or longer, where decays at the location of the MPD will be optimal. In this section, we discuss the production and decay mechanisms of dark photons in the parameter space of interest for the DUNE MPD. We discuss the associated backgrounds for searches of this type in Section~\ref{DarkPhoton:Backgrounds}, and provide our estimates for the DUNE MPD sensitivity to these dark photons in Section~\ref{DarkPhoton:Sensitivity}.

\subsection{Dark Photon Production}

For the DUNE MPD, the dominant production mechanism for dark photons is the decay of neutral mesons, specifically $\pi^0$ and $\eta$. We simulate the production of these as described in Section~\ref{sec:Simulation}. The branching ratio of a given neutral meson $\mathfrak{m}$ into $\gamma A^\prime$ is
\begin{equation}
\mathrm{Br}(\mathfrak{m} \to \gamma A^\prime) = 
\mathrm{Br}(\mathfrak{m} \to\gamma \gamma) \times 2\varepsilon^2 \left(1 - \frac{M_{A^\prime}^2}{m_\mathfrak{m}^2}\right)^3.
\end{equation}
From Ref.~\cite{Tanabashi:2018oca}, $\mathrm{Br}(\pi^0 \to \gamma \gamma) = 98.823\%$ and $\mathrm{Br}(\eta \to \gamma \gamma) = 39.41\%$. Given the spectra of $\pi^0$ and $\eta$ produced in the DUNE target and the kinematics of $\mathfrak{m} \to \gamma A^\prime$ decay, we estimate (a) the fraction of produced $A^\prime$ that reach the HPTPC detector and (b) the energy spectrum of the $A^\prime$ flux that reaches the detector, as described in Section~\ref{sec:Simulation}.

In addition to meson decays into $A^\prime$, the bremsstrahlung process $p p \to p p A^\prime$ allows us to probe larger values of $M_{A^\prime}$, here limited by $\sqrt{s} \simeq \sqrt{2m_p E_p} = 12$ GeV. The calculation of the production in this process is described in Refs.~\cite{Blumlein:2013cua,deNiverville:2016rqh}. 
The production rate is expressed in terms of properties of the outgoing $A^\prime$, with momentum $p_{A^\prime} \equiv (E_{A^\prime}, p_T \cos\phi, p_T \sin\phi, zP)$, where $p_T$ is the transverse momentum of the outgoing $A'$, $E_{A^\prime}$ is its energy, $P$ is the initial proton incoming momentum, $z$ is the fraction of the proton's initial momentum transferred to the longitudinal momentum of the $A^\prime$, and $\phi$ is an azimuthal angle. The double-differential production rate is
\begin{equation}\label{eq:DarkPhoton:BremProduction}
    \frac{d^2 N_{A^\prime}}{dz dp_T^2} =\frac{\sigma_{pN}(s^\prime)}{\sigma_{pN}(s)} \left\lvert F_{1,N}(M_{A^\prime}^2)\right\rvert^2 w_{ba}(z,p_T^2),
\end{equation}
where $\sigma_{pN}(s)$ is the cross section of a proton hitting a target nucleus $N$ at center-of-mass energy squared $s$, and $w_{ba}(z,p_T^2)$ is the photon splitting function,
\begin{align}\label{eq:DarkPhoton:Splitting}
    w_{ba}(z,p_T^2) = \frac{\varepsilon^2 \alpha_{EM}}{2\pi H} &\left[\frac{1 + (1-z)^2}{z} - 2z(1-z)\left(\frac{2m_p^2 + M_{A^\prime}^2}{H} - z^2 \frac{2m_p^4}{H^2}\right) \right.\nonumber \\
    &\left.+ 2z(1-z)(z+(1-z)^2) \frac{m_p^2 M_{A^\prime}^2}{H^2} + 2z(1-z)^2 \frac{M_{A^\prime}^4}{H^2}\right],
\end{align}
and $H = p_T^2 + (1-z)M_{A^\prime}^2 + z^2 m_p^2$, and $m_p$ is the proton mass.  In Eq.~(\ref{eq:DarkPhoton:BremProduction}), the ratio of total cross sections, evaluated at the two energies of interest, $s^\prime = 2m_p(E_p - E_{A^\prime})$ and $s = 2m_p E_p$, is close to one for the energies and masses of interest here.  The form factor, $F_{1,N}$, in Eq.~(\ref{eq:DarkPhoton:BremProduction}) encodes the mixing of the $A^\prime$ with SM vector mesons $\rho$ and $\omega$, leading to an enhancement in dark photon production when $M_{A^\prime} \approx$ 800 MeV~\cite{Faessler:2009tn}.

To determine the number of $A^\prime$ that are produced in the direction of the detector, one must integrate Eq.~(\ref{eq:DarkPhoton:Splitting}) over the appropriate range of $z$ and $p_T^2$ according to the detector geometry. The majority of the existing literature~\cite{Alekhin:2015byh,deNiverville:2016rqh,Berlin:2018pwi} focuses on detectors with a much larger solid angle (as viewed from the beam target) than we consider here.  The traditional approach is to integrate over a specific range of $z$ and $|p_T|$, e.g.,~$z \in [0.1, 0.9]$ and $|p_T| < 1$ GeV for SHiP~\cite{deNiverville:2016rqh}.  In contrast, here, we transform the variables of Eq.~(\ref{eq:DarkPhoton:BremProduction}) from $(z,p_T^2) \to (E_{A^\prime}, p_T^2)$.  
In these variables, the requirement that the $A^\prime$ enters the detector leads to the constraint  
\begin{equation}
    p_T^2 <  \left(E_{A^\prime}^2 - M_{A^\prime}^2\right)\sin^2\theta_\mathrm{det.}~,
    \label{eq:detectorrestriction}
\end{equation}
where $\theta_\mathrm{det.}$ is half the angular size of the detector, as viewed by the target. For the DUNE MPD, $\theta_\mathrm{det.} \approx 4 \times 10^{-3}$. Note that this relationship holds regardless of the detector's solid angle, and we advocate for its use whenever considering dark photon production in this process. As explained in Refs.~\cite{Blumlein:2013cua,Gorbunov:2014wqa}, the form of Eq.~(\ref{eq:DarkPhoton:Splitting}) is only valid in certain limits, requiring
\begin{equation}
    E_p, E_{A^\prime}, E_p - E_{A^\prime} \gg m_p, M_{A^\prime}, |p_T|.
    \label{eq:validitylimits}
\end{equation}
Some of these restrictions are automatic once Eq.~(\ref{eq:detectorrestriction}) is satisfied.  We include these restrictions in our calculation and only include contributions to the flux from regions in which Eqs.~(\ref{eq:detectorrestriction}) and (\ref{eq:validitylimits}) are satisfied and thus Eqs.~(\ref{eq:DarkPhoton:BremProduction}) and (\ref{eq:DarkPhoton:Splitting}) are valid.

\begin{figure}
\centering
\includegraphics[width=0.6\linewidth]{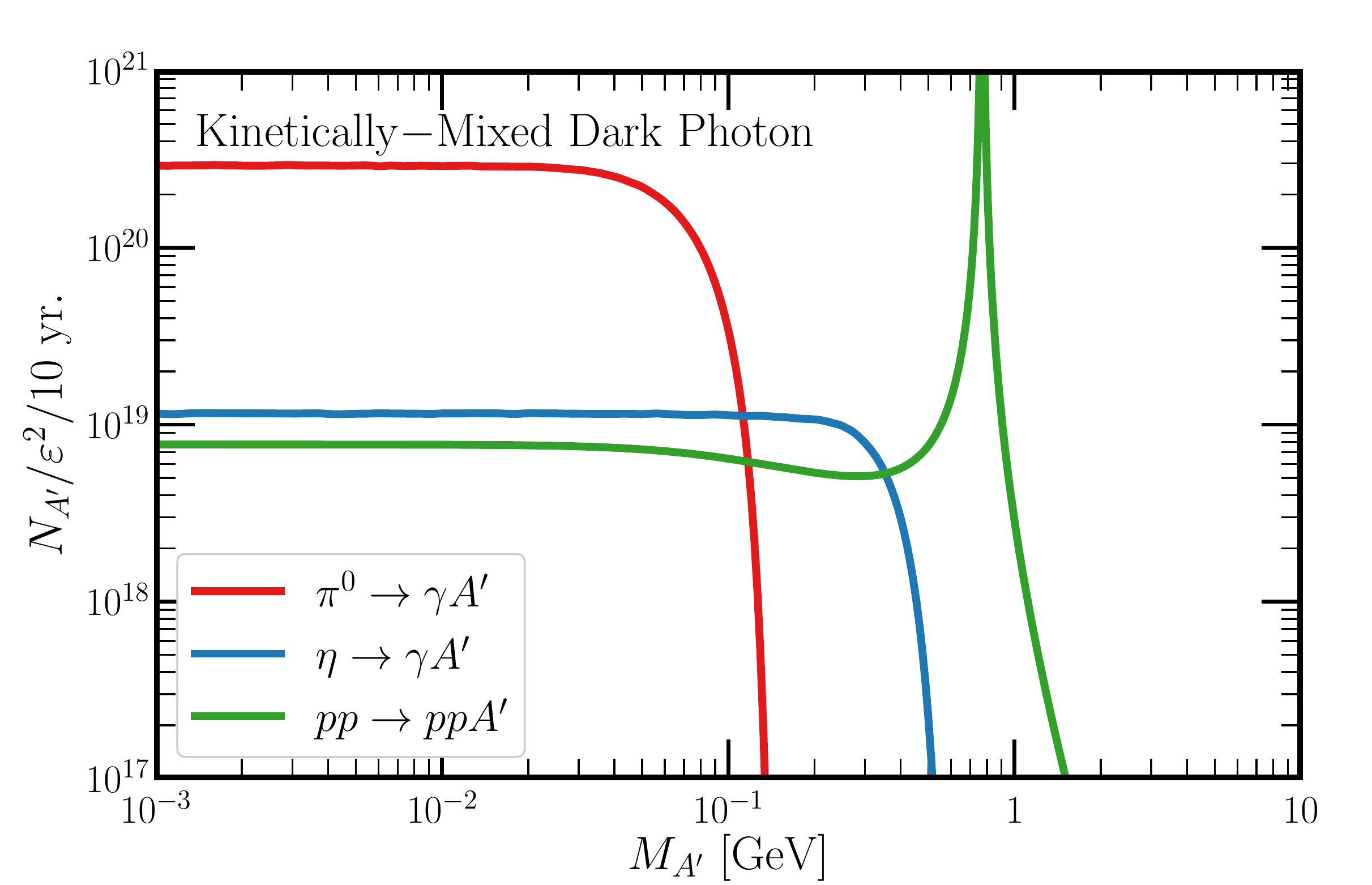}
\caption{Expected number of $A^\prime$ that travel towards the DUNE Multi-Purpose Near Detector, a distance of $579$ m from the production target, as a function of mass $M_{A^\prime}$. We assume ten years of beam operation, $\varepsilon = 1$, and a beam luminosity of $1.47 \times 10^{21}$ protons per year. The red curve displays $A^\prime$ produced via decays of $\pi^0$, the blue line displays those from $\eta$ decays, and the green displays those produced in the bremsstrahlung process $p p \to p p A^\prime$.}
\label{fig:AprimeSpectrum}
\end{figure}

Fig.~\ref{fig:AprimeSpectrum} displays the expected number of $A^\prime$ that traverse the HPTPC in 10 years of operation at DUNE, normalized to $\varepsilon = 1$; this flux scales with $\varepsilon^2$. As explained above, bremsstrahlung production allows us to reach larger $M_{A^\prime}$. The peak in the bremsstrahlung curve for $M_{A^\prime} \simeq 800$ MeV comes from enhanced mixing near the $\rho$ meson mass.
In Fig.~\ref{fig:AprimeSpectrum}, we see that the bremsstrahlung process (green) is comparable, even at $M_{A^\prime} \lesssim 400$ MeV, to production via $\eta$ meson decay. This is in contrast with projections for this process at, for instance, SHiP~\cite{Alekhin:2015byh} or SeaQuest~\cite{Berlin:2018pwi}, where the prediction is that the $A^\prime$ flux due to $\eta$ decays is an order of magnitude or so larger than that from bremsstrahlung. This difference is due to the fact that the DUNE near detector is much further from its target than SHiP or SeaQuest, and has a significantly smaller solid angle than either. This leads to a relatively larger geometrical acceptance for the bremsstrahlung process relative to meson decays for DUNE, resulting in a comparable $A^\prime$ flux, as seen in Fig.~\ref{fig:AprimeSpectrum}.

Direct production of dark photons via $q q^\prime \to A^\prime$ may contribute for high center-of-mass energy proton collisions. However, such a calculation requires robust knowledge of the proton parton density functions for $s = M_{A^\prime}^2$. This mechanism could increase sensitivity, but only for $M_{A^\prime} \gtrsim 2$ GeV -- we do not include this in our calculations, and expect that it would not affect the sensitivity we present here.

\subsection{Dark Photon Decay Channels and Lifetime}
\begin{figure}[!t]
\centering
\includegraphics[width=0.6\linewidth]{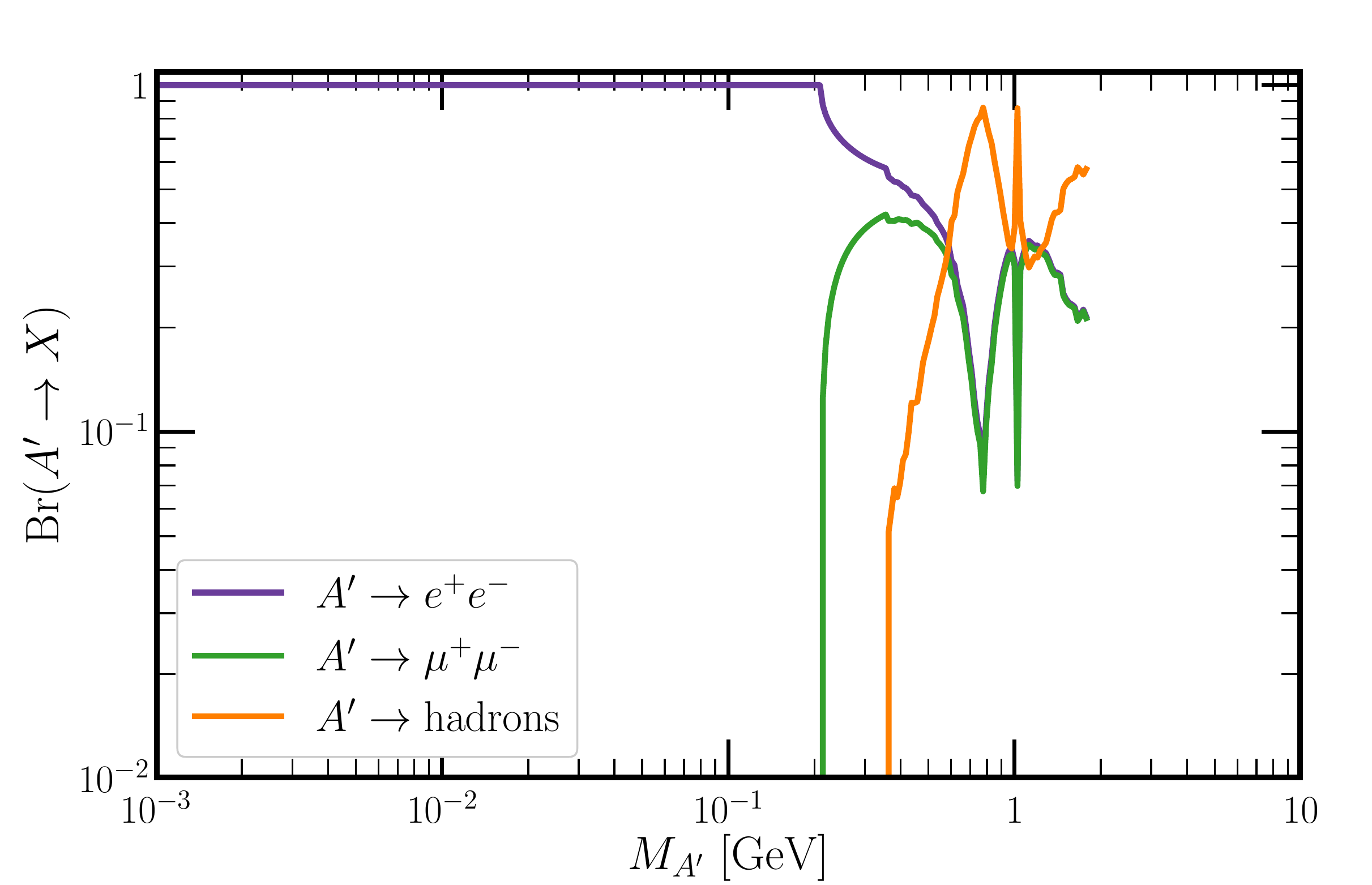}
\caption{The branching ratio of the dark photon $A^\prime$ into different final states of interest as a function of mass $M_{A^\prime}$: $A^\prime \to e^+e^-$ (purple), $\mu^+ \mu^-$ (green), and hadronic final states (orange). Adapted from Ref.~\cite{Buschmann:2015awa}.}
\label{fig:AprimeBR}
\end{figure}

We consider three decay modes of $A^\prime$: $A^\prime \to e^+ e^-$, $A^\prime \to \mu^+ \mu^-$, and $A^\prime \to$ hadrons. For the $A^\prime$ masses of interest, the decay channel $A^\prime \to$ hadrons consists largely of the final state $\pi^+ \pi^-$. However, depending on $M_{A^\prime}$, more complicated final states exist (see, e.g., Ref.~\cite{Buschmann:2015awa}, for a discussion of the different decay channels for $M_{A^\prime} \lesssim 1$ GeV). We reproduce the partial widths of $A^\prime$ here for clarity. For decays into lepton pairs $\ell^+ \ell^-$, the partial width is
\begin{equation}\label{eq:DarkPhoton:LeptonWidth}
    \Gamma(A^\prime \to \ell^+ \ell^-) = \frac{1}{3}\alpha%_\mathrm{EM}
    \,\varepsilon^2 M_{A^\prime} \sqrt{1 - \frac{4m_\ell^2}{M_{A^\prime}^2}}\left(1 + \frac{2m_\ell^2}{M_{A^\prime}^2}\right).
\end{equation}
In order to express the partial width of $A^\prime$ into hadron pairs, we rely on the $R$ ratio,
\begin{equation}
    R(s) \equiv \frac{\sigma(e^+ e^- \to \mathrm{hadrons})}{\sigma(e^+ e^- \to \mu^+ \mu^-)}~.
\end{equation} 
The partial width of $A'$ of mass $M_{A'}$ into hadrons is related to $R$ at $s=M_{A'}^2$ through
\begin{equation}
    \Gamma(A^\prime \to \mathrm{hadrons}) = \Gamma(A^\prime \to \mu^+ \mu^-) \times R(s=M_{A^\prime}^2).
\end{equation}
Combining these expressions, we display the branching fraction of $A^\prime$ into each final state as a function of $M_{A^\prime}$ in Fig.~\ref{fig:AprimeBR}.

Since we are interested in $c\tau \approx 100$ m, we can use these widths to determine the expected parameter range for which DUNE's sensitivity is maximized. We find, roughly, that the event rate peaks for $\varepsilon^2 \approx 10^{-15} (1\ \mathrm{GeV}/M_{A^\prime})$. Other factors, especially the boost of $A^\prime$, modify this relationship slightly.

\subsection{Backgrounds}
\label{DarkPhoton:Backgrounds}
In this section we discuss the expected background rates for the different search channels $A^\prime \to e^+ e^-$, $A^\prime \to \mu^+ \mu^-$, and $A^\prime \to$ hadrons. All of the backgrounds we discuss here come from neutrinos in the DUNE beam scattering off an argon nucleus in the HPTPC.\footnote{In addition to these backgrounds, there will be many events where a neutrino scatters off material in the ECAL/magnet (which, together, have a mass of roughly 400 tons), and particles from those interactions enter the gas TPC. These entering particles could contribute as background, which are, in principle, able to be vetoed using timing and fiducialization. Further study is required.} The signal comes from a decaying particle (that originated in or near the target), compared to a light neutrino scattering off a heavy nucleus, so we can take advantage of the fact that the outgoing charged particles point back to the target to greatly reduce background rates.

{\boldmath{$A^\prime \to e^+ e^-$}}: The dominant background for this channel is $\nu + \mathrm{Ar} \to \nu + \mathrm{Ar} + \pi^0$, i.e., neutral-current single-pion (NC$\pi^0$) events. This also includes $\pi^0$ produced in coherent production, where the recoiling target is too low-energy to be observable. Events of this class may be misinterpreted as signals with only $e^+e^-$ pairs if the following occur:
\begin{itemize}
    \item No other hadronic activity occurs at the vertex of the neutrino interaction.
    \item Of the two photons coming from $\pi^0$ decay, one is either missed by the ECAL, or is, for some reason, not attributed to a common vertex with the second photon.
    \item The other photon converts in the gas (which occurs for roughly 12\% of photons).
    \item The converted photon's electromagnetic shower is identified as an $e^+ e^-$ pair, with a direction consistent with the direction from the target/decay pipe.
\end{itemize}
In all, $\mathcal{O}(10^5)$ NC$\pi^0$ events are expected per ton-year of exposure. We expect that the neutrino-related background of $\mathcal{O}(10^6)$ events (for the exposure we consider) may be drastically reduced by the requirements listed above. Specifically, we expect that 80\% of NC$\pi^0$ events will have detectable hadronic activity (at least one charged particle with kinetic energy above 5 MeV) in the MPD. This means that only 20\%, or $2\times 10^5$ events in ten years, will be NC$\pi^0$ with no significant hadronic activity. As for the second requirement, we estimate (conservatively) that the DUNE MPD will miss 10\% of photons coming from $\pi^0 \to \gamma \gamma$ decay. This reduces our background sample to $2 \times 10^4$ events. The third requirement, that the other photon converts in the gas TPC (before reaching the ECAL) will occur in 12\% of events -- this brings our background down to 2400 events. Finally, we estimate the probability that a photon from $\pi^0$ decay will be pointing in the direction of the target. Because most of the $\pi^0$ are coming from nuclear emission, they will be emitted nearly isotropically. If we require that the angle of the photon is within $4$ mrad from the beam direction (the angular resolution of the gas TPC), then roughly $0.01\%$ of photons will pass this cut. A more conservative cut of $1^\circ$ ($5^\circ$) increases this to 0.1\% (2.6\%). Given all of these considerations, we expect an optimistic background rate of $< 1$ event in ten years. Even with a very conservative $5^\circ$ angular cut, this is a background rate of $\mathcal{O}(50)$ events.

If such a conservative cut is required (perhaps to retain a signal from decays downstream of the beam target, such as that discussed in Section~\ref{sec:HNL}), then we still have additional handles to separate signal from background. For the decay $A^\prime \to e^+ e^-$, the electron/positron pair will have an invariant mass consistent with $M_{A^\prime}$, where the electron/positron pair from $\pi^0 \to \gamma \gamma$, $\gamma$ conversion to $e^+ e^-$ will have zero invariant mass. Additionally, these background electrons/positrons will tend to have lower (around $500$ MeV) energy, where the signal electrons/positrons will have several GeV of energy. These kinematical separations may be exploited if a $1^\circ$ angular cut turns out to be too optimistic.

In addition to $\pi^0$ faking the $A^\prime \to e^+ e^-$ signal, there will be an irreducible background of events with $e^+ e^-$ being the only visible particles in the final state: neutrino trident events. Refs.~\cite{Ballett:2018uuc,Altmannshofer:2019zhy} estimate $\mathcal{O}(20)$ neutrino trident $e^+ e^-$ events in the MPD (we obtain this by scaling their LArTPC projections to the mass of the HPTPC) in ten years. Their kinematics should be different from our signature (in terms of invariant mass, direction, etc.) and should not contribute sizably to the search proposed here.

{\boldmath{$A^\prime \to \mu^+ \mu^-$}} \textbf{and} {\boldmath{$A^\prime \to$}} \textbf{hadrons:} We consider these two channels in the same category as they share the same background, predominantly $\nu_\mu$ charged-current-single-pion (CC$1\pi$) events, with an expected rate of $\mathcal{O}(10^{5})$ events per ton-year. The authors of Ref.~\cite{Ballett:2018uuc} explored a similar set of signals and backgrounds, focusing on the search for $\mu^+ \mu^-$ neutrino trident events in the DUNE LArTPC near detector (where the background rates are larger by a ratio of the fiducial mass in each detector). They discussed certain background rejection techniques, including the rejection of events with hadronic activity\footnote{The Multi-Purpose detector will be more sensitive than the LArTPC to smaller amounts of hadronic activity due to its lower thresholds, see Table~\ref{tab:Parameters:Thresholds}.} (around $10\%$ of background events survive this cut), and a variety of kinematical cuts that would apply to our search as well (roughly $1\%$ of background events survive). We also note that the ECAL surrounding the HPTPC will have moderate ability to identify charged pions vs. muons -- approximately $70\%$ of charged pions will interact hadronically in the ECAL, allowing for a positive identification of those that do interact. Additionally, the DUNE collaboration is considering adding a muon tagger~\cite{NewGasSlides} (likely consisting of alternating layers of scintillator and a high-density material like steel) that will further reduce pion/muon misidentification.

Finally, we note that the same angular cuts discussed regarding $A^\prime \to e^+ e^-$ can be applied here -- when a muon and a pion fake this signature, the direction of the pair will very rarely be in the direction of the beam. This allows for even further background reduction than when trying to reduce the backgrounds to neutrino trident events~\cite{Ballett:2018uuc}. Overall, we expect that ten signal events will constitute a significant excess over background, especially after considering that our signal events will share a common $A^\prime$ invariant mass, where the background events will be smoothly distributed in that variable.

\subsection{Dark Photon Sensitivity}\label{DarkPhoton:Sensitivity}

\begin{figure}[!t]
\centering
\includegraphics[width=0.75\linewidth]{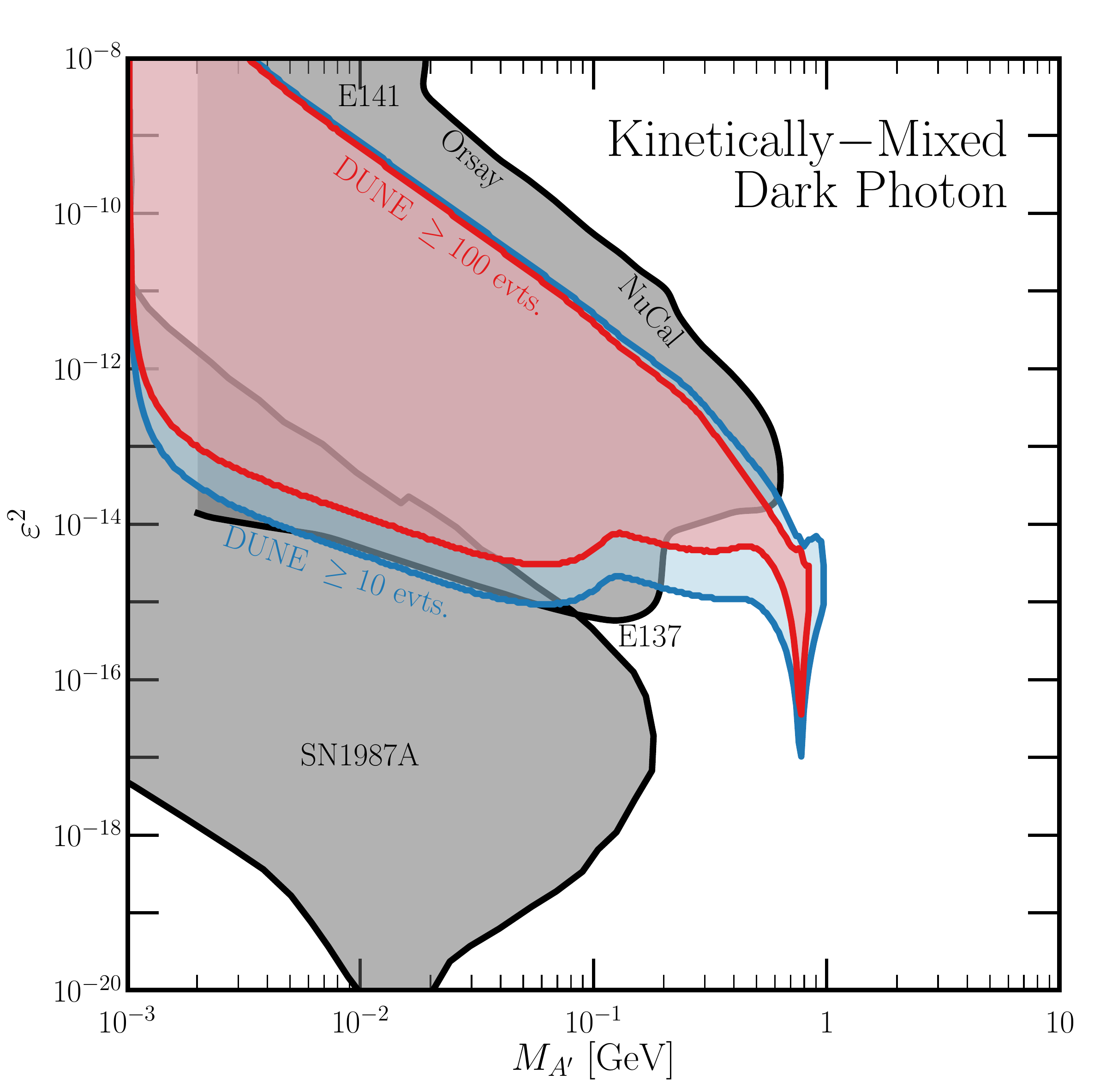}
\caption{Regions of parameter space of $A^\prime$ mass $M_{A^\prime}$ vs. kinetic mixing parameter squared $\varepsilon^2$ for which we expect $10$ ($100$) $A^\prime$ decays in the DUNE MPD after ten years of data collection, in blue (red). Currently excluded regions are in grey.}
\label{fig:AprimeSensitivity}
\end{figure}

Given the energy spectrum of the $A^\prime$ that reach the DUNE MPD, we calculate the expected number of $A^\prime$ decays as a function of  $M_{A^\prime}$ and $\varepsilon^2$ for a given amount of data collection. In Fig.~\ref{fig:AprimeSensitivity}, we display the regions where we expect at least $10$ (blue) and $100$ (red) such decays, assuming ten years of data collection.  In addition, we show the excluded region of this parameter space constrained by E141, Orsay, NuCal, and E137~\cite{Riordan:1987aw,Bjorken:1988as,Davier:1989wz,Bjorken:2009mm,Blumlein:2011mv,Andreas:2012mt,Blumlein:2013cua} and from emission during Supernova 1987A~\cite{Chang:2016ntp} in shaded gray.
These regions correspond to a total number of decays being $10$ or $100$ -- in order to determine the number of a certain type of decay, one must include the branching fraction into $e^+e^-$, $\mu^+\mu^-$, or hadrons shown in Fig.~\ref{fig:AprimeBR}. For instance, if $M_{A^\prime} \approx 300$ MeV and $\varepsilon^2 \approx 10^{-15}$, a total of $10$ decays are expected in ten years. According to the branching fraction of $A^\prime$, these events should be roughly $40\%$ $\mu^+\mu^-$ decays and $60\%$ $e^+e^-$. Searching for the correct combination of different final states can also improve the capabilities of a search for dark photons, and disentangle signals of dark photons from those of the other new physics scenarios we discuss in later sections.

In producing Fig.~\ref{fig:AprimeSensitivity}, we assume 100\% efficiency in identifying particle/antiparticle pairs -- this approximation likely breaks down for lighter $A^\prime$, but in those regimes, existing limits are more powerful than the DUNE MPD search for dark photons. 

Overall, we find that the DUNE MPD will be able to extend sensitivity to a dark photon in the sub-GeV mass regime for kinetic mixing in the $\varepsilon^2 \approx 10^{-16} - 10^{-14}$ range. This complements other future experiments, such as FASER~\cite{Feng:2017uoz,Ariga:2018uku}, SHiP~\cite{Alekhin:2015byh}, and SeaQuest~\cite{Berlin:2018pwi}, which will be sensitive to larger values of $\varepsilon^2$, as the detectors associated with these proposals are significantly closer to their production targets. If dark photons exist and are produced in this type of environment, then those produced at CERN (for detection at FASER/SHiP) would be much more boosted than those produced at DUNE. These detectors are then sensitive to dark photons with relatively shorter proper lifetimes, meaning that DUNE is sensitive to smaller $\varepsilon^2$ for similar mass dark photons.

%\afterpage{\clearpage} 
%------------------------------
%Leptophilic Section
%------------------------------

\section{Leptophilic Gauge Bosons}\label{sec:Leptophilic}\setcounter{footnote}{0}

While the dark photon couples to the Standard Model purely via kinetic mixing, as discussed in Section~\ref{sec:DarkPhoton}, new vector bosons may also interact with the SM if they are gauge bosons of a new group under which some SM particles are charged.  Such a gauge symmetry should be anomaly free, which often requires extending the field content of the SM.  However, if SM fields are charged in a vectorlike fashion, for instance differences of baryon/lepton number between two generations, then no additional fermions are necessary.
We expect that DUNE will have the strongest sensitivity to such new physics when the new symmetry is leptophilic, and therefore focus on the case of a $L_\alpha - L_\beta$ gauge boson with $\alpha$, $\beta = e,$ $\mu$, $\tau$ \cite{Harnik:2012ni,Heeck:2014zfa,Bilmis:2015lja,Jeong:2015bbi,Ilten:2018crw,Wise:2018rnb,Altmannshofer:2014cfa,Altmannshofer:2014pba,Kamada:2015era,Araki:2015mya,Araki:2017wyg,Kaneta:2016uyt,Babu:2017olk,Gninenko:2018tlp}.

The leptophilic gauge boson\footnote{We use the symbol $V$ to represent a leptophilic gauge boson in this section, to contrast with the purely kinetically-mixed dark photon $A^\prime$ in Section~\ref{sec:DarkPhoton}.} $V$ of mass $M_V$, associated with gauging the difference between lepton numbers $L_\alpha - L_\beta$, will couple directly to neutrinos or charged leptons in generations $\alpha$ and $\beta$.  This coupling means that the gauge boson will be produced in charged meson decays through final state bremsstrahlung and can decay to neutrinos and charged leptons, if kinematically accessible.  In addition to this direct coupling, kinetic mixing between $V$ and the SM photon is induced at one loop via SM particles that are charged under both SM hypercharge and the new $U(1)$. This loop-induced kinetic mixing, as defined in Eq.~(\ref{eq:DarkPhoton:Lagrangian}), is related to the $L_\alpha - L_\beta$ gauge coupling $g_{\alpha\beta}$ as~\cite{Kamada:2015era,Bauer:2018onh,Escudero:2019gzq}
\begin{equation}\label{eq:KineticMixing}
    \varepsilon_{\alpha\beta}(q^2) = -\frac{e g_{\alpha\beta}}{2\pi^2} \int_{0}^{1} x(1-x) \log{\left[ \frac{m_{\ell_\beta}^2 - x(1-x) q^2}{m_{\ell_\alpha}^2 - x(1-x) q^2}\right]} dx,    
\end{equation}
where $m_{\ell_\alpha}$ is the mass of charged lepton with flavor $\alpha$ and $q_\mu$ is the four-momentum of the boson $V$ that is mixed with the SM photon.  This contribution is in addition to any kinetic mixing that might be present in the UV Lagrangian.  From now on, we will assume that the bare kinetic mixing is zero and that the effective mixing is determined by Eq.~(\ref{eq:KineticMixing}).  This kinetic mixing allows for $V$ production through neutral meson decays and for $V$ to decay to a pair of charged leptons that are not in either generation $\alpha$ nor $\beta$.  Hence, even $L_\mu-L_\tau$ gauge bosons lighter than $2m_\mu$ will have a visible decay channel.

\begin{figure}
    \centering
    \includegraphics[width=0.6\linewidth]{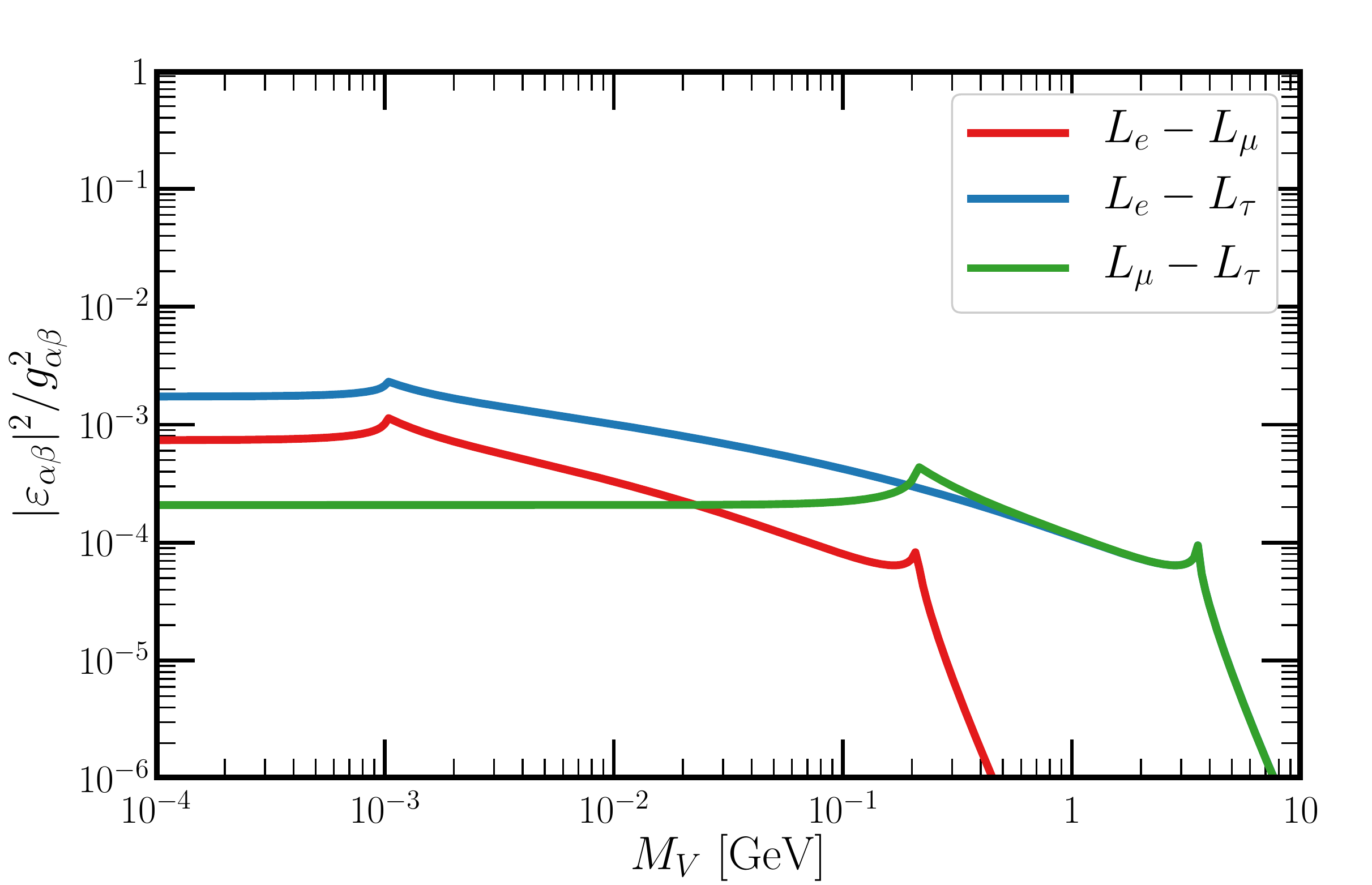}
    \caption{Kinetic mixing $|\varepsilon|^2$ (relative to the gauge coupling squared) given by Eq.~(\ref{eq:KineticMixing}) as a function of vector boson mass $M_V$ for the three different leptophilic gauge bosons: $L_{e} - L_{\mu}$ (red), $L_{e} - L_\tau$ (blue), and $L_\mu - L_\tau$ (green). Here, we assume the only contribution to the kinetic mixing is from the massive standard model fermions generating a loop process connecting the standard model photon and $V$.}
    \label{fig:LoopKinetic}
\end{figure}

We are interested in situations in which a $V$ is emitted on-shell due to this mixing, so $q^2 \equiv M_V^2$. 
In the limit where the momentum transfer (i.e., the $V$ mass) is well below the lighter of the two charged lepton masses, the kinetic mixing becomes a constant,
\begin{equation}
    \varepsilon_{\alpha\beta} \longrightarrow -\frac{eg_{\alpha\beta}}{12\pi^2} \log{\left(\frac{m_{\ell_\beta}^2}{m_{\ell_\alpha}^2}\right)}~,
\end{equation}
and $|\varepsilon_{\alpha\beta}|^2 \approx \mathcal{O}(10^{-4}-10^{-3}) g_{\alpha\beta}^2$.
In the opposite limit, where the momentum transfer (or $M_V$) is much larger than both of the charged lepton masses,
\begin{equation}
    \varepsilon_{\alpha\beta} \longrightarrow \frac{eg_{\alpha\beta}}{2\pi^2} \frac{m_{\ell_\beta}^2 - m_{\ell_\alpha}^2}{M_V^2}~.
\end{equation}
We show the kinetic mixing for each leptophilic $U(1)$ symmetry in Fig.~\ref{fig:LoopKinetic}. 
Assuming $V$ is produced on-shell, we show the kinetic mixing squared relative to the new gauge coupling squared for $L_e - L_\mu$ (red), $L_e - L_\tau$ (blue), and $L_\mu - L_\tau$ (green). The change of behavior of the loop integral at particle thresholds is apparent.

With any kinetic mixing, the production processes discussed in Section~\ref{sec:DarkPhoton} will apply here, suppressed according to Eq.~(\ref{eq:KineticMixing}) and Fig.~\ref{fig:LoopKinetic}. In addition, $V$ may be emitted in decays involving charged leptons and neutrinos by being radiated in the final state; we provide a derivation of the branching ratio for such decays in Appendix~\ref{appendix:derivation}. Which of these processes is most relevant depends on how strong the kinetic mixing is, as well as the lepton flavor(s) of interest. We discuss the production mechanisms in the following, and then the decay signatures in Section~\ref{subsec:LeptophilicDecays}. We provide our sensitivity estimates in Section~\ref{subsec:LeptophilicSensitivity}.

\subsection{Production of Leptophilic Gauge Bosons}\label{subsec:LeptophilicProduction}

In addition to the production via kinetic mixing, $V$ may be radiated in charged meson decays with final-state leptons. The greatest abundance of these will be from charged pion and kaon decays, $\pi^\pm/K^\pm \to \ell^\pm_\alpha \nu_\alpha V$. In Appendix~\ref{appendix:derivation}, we derive the branching ratios of these three-body decays.  The rate for these three-body decays is proportional to the square of the final-state lepton mass~\cite{Berman:1958gx,Unterdorfer:2008zz}.

For a $L_e - L_\mu$ gauge boson, the processes $\pi^\pm \to e^\pm \nu_e V$ and $\pi^\pm \to \mu^\pm \nu_\mu V$ (and the equivalent kaon decays) will contribute to $V$ production; the decays involving electrons, however, are helicity suppressed relative to those involving muons. Production via $\pi^\pm \to e^\pm \nu_e V$ allows for searches of heavier $V$, since the kinematic threshold will be $M_V = m_\pi - m_e$ rather than $M_V = m_\pi - m_\mu$, albeit at a suppressed rate. 
%%%%%%
\begin{figure}[!htbp]
    \centering
    \includegraphics[width=0.6\linewidth]{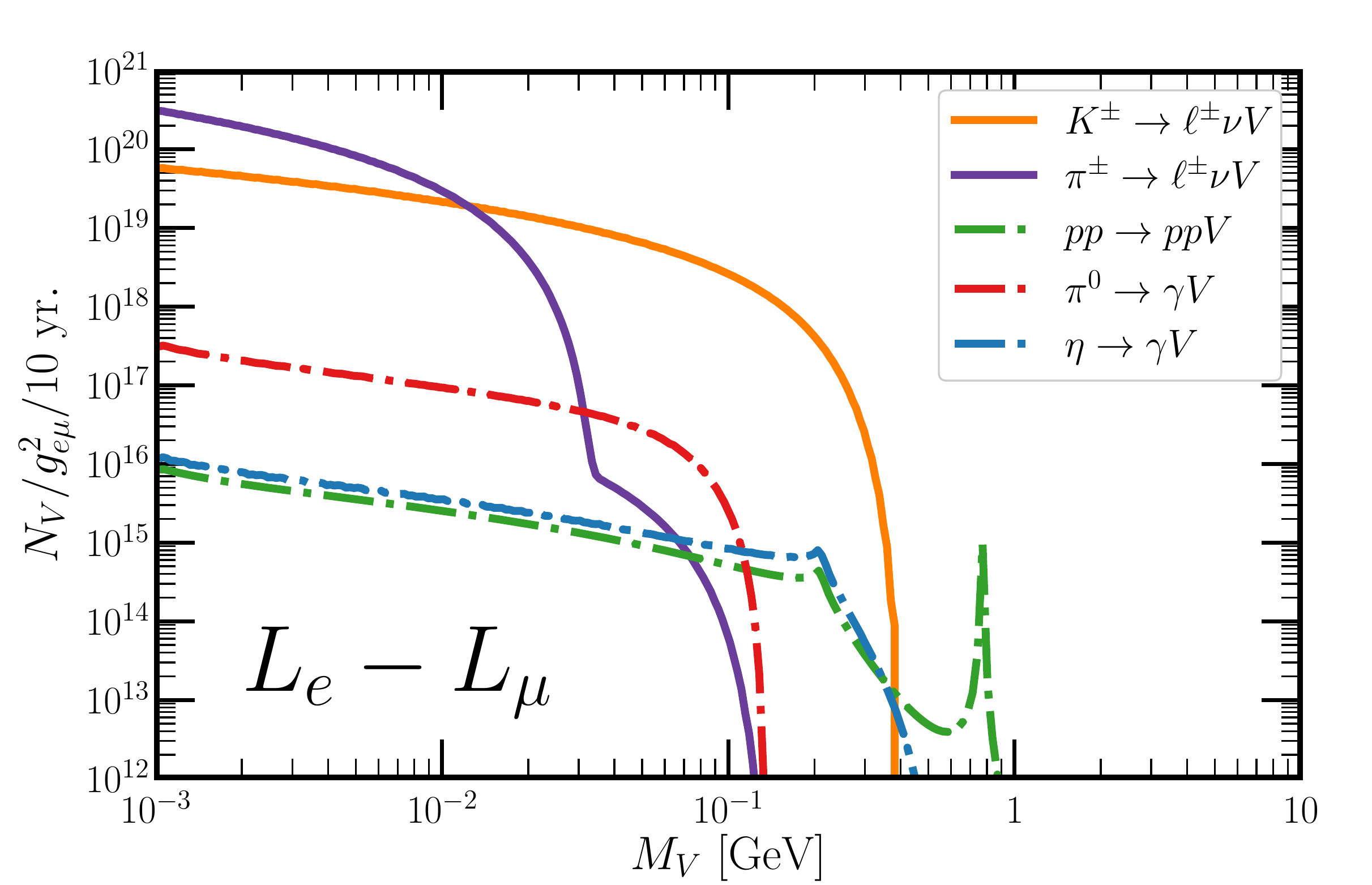}
    \caption{The number of $L_e - L_\mu$ gauge bosons $V$ traveling towards the DUNE MPD as a function of mass $M_V$ assuming ten years of data collection, and equal run time in neutrino and antineutrino modes. Solid lines display expected number due to charged meson decays -- purple ($\pi^\pm$) and orange ($K^\pm$). Dot-dashed lines display expected number due to processes relying on kinetic mixing -- red ($\pi^0$ decay), blue ($\eta$ decay), and green ($pp \to ppV$ bremsstrahlung).}
    \label{fig:Lem:Spectrum}
\end{figure}
%%%%%%
Fig.~\ref{fig:Lem:Spectrum} displays the expected number of $V$ passing through the DUNE MPD, normalized to $g_{e\mu}^2=1$, assuming ten years of data collection with equal run time in neutrino and antineutrino modes\footnote{We sum over the two different charges of meson decays, as well as the different final state leptons $e^\pm$ and $\mu^\pm$, in generating Fig.~\ref{fig:Lem:Spectrum}. The differences in production rate between neutrino and antineutrino modes, due to the focusing of charged mesons, is negligible.}.  We show contributions from charged mesons ($\pi^\pm$ in purple and $K^\pm$ in orange), as well as neutral-meson contributions ($\pi^0$ in red and $\eta$ in blue) via loop-induced kinetic mixing, as well as the bremsstrahlung process $pp\to pp V$ in green.  These rates assume that there is only loop-induced kinetic mixing, at the rate depicted in Fig.~\ref{fig:LoopKinetic}.  The features apparent in the $pp\to ppV$ and $\eta \to \gamma V$ production mechanisms (mostly near $M_V \approx 200$ MeV) are reflections of the variation of the kinetic mixing with $V$ mass, as depicted Fig.~\ref{fig:LoopKinetic}.

\begin{figure}[!htbp]
    \centering
    \includegraphics[width=0.6\linewidth]{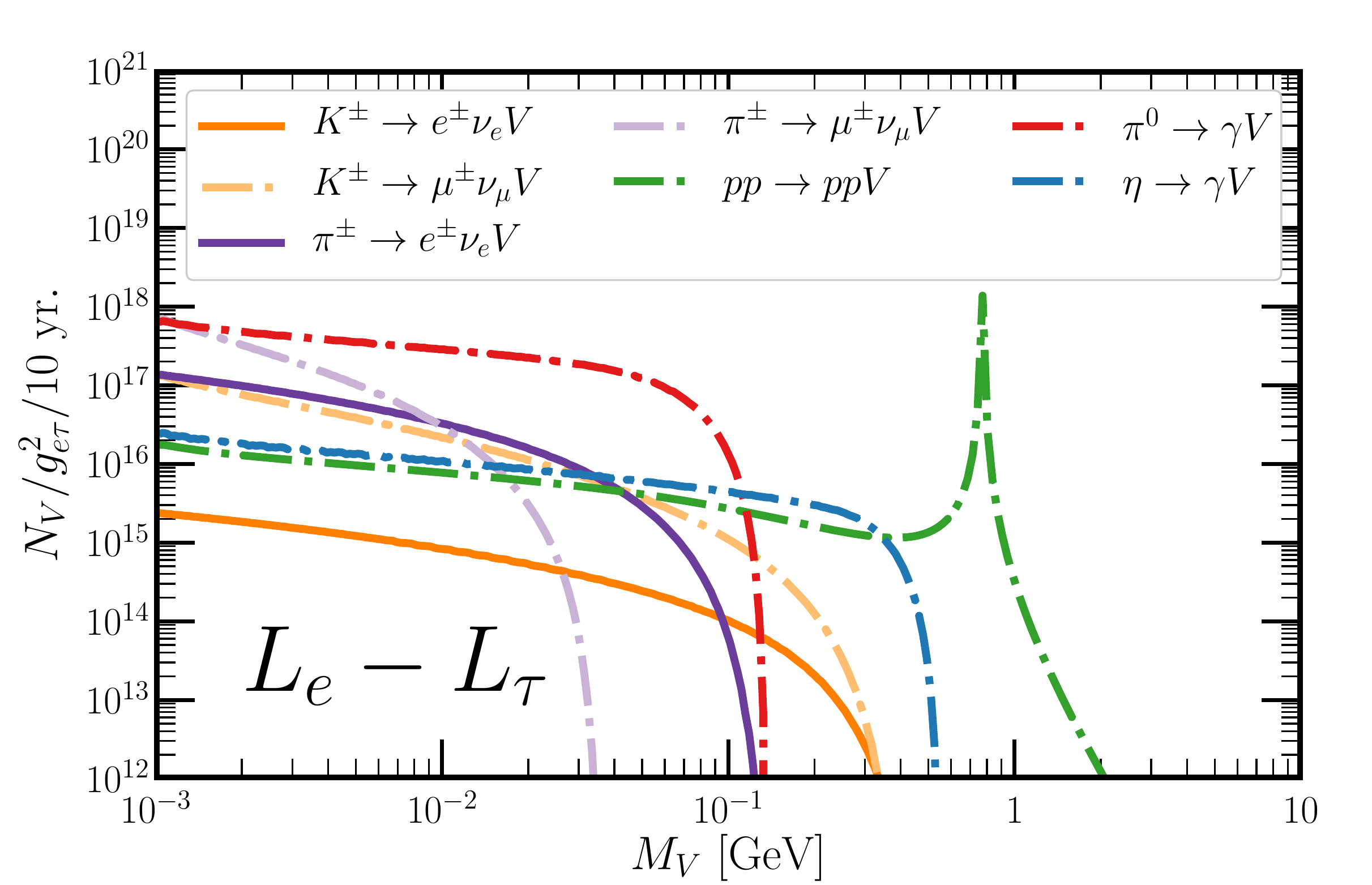}
    \caption{The number of $L_e - L_\tau$ gauge bosons $V$ traveling towards the DUNE MPD as a function of mass $M_V$ assuming ten years of data collection, and equal run time in neutrino and antineutrino modes. Solid lines display expected number due to charged meson decays -- purple ($\pi^\pm$) and orange ($K^\pm$). Dot-dashed lines display expected number due to processes relying on kinetic mixing -- red ($\pi^0$ decay), blue ($\eta$ decay), green ($pp \to ppV$ bremsstrahlung), purple ($\pi^\pm$ decays to muons), and orange ($K^\pm$ decays to muons).}
    \label{fig:Let:Spectrum}
\end{figure}

For a $L_e - L_\tau$ vector boson, charged meson decays to electrons are chirally suppressed while production via charged meson decays with final-state muons rely on kinetic mixing.  The loop-induced kinetic mixing is largest for $L_e - L_\tau$ over most of the $V$ mass range, see Fig.~\ref{fig:LoopKinetic}, and is larger than $(m_e/m_\mu)^2$ for $M_V\ltap 5$ GeV.  Thus, despite the small kinetic mixing, the processes depending on kinetic mixing dominate the production of $V$ across all masses. Specifically, we note that the loop-suppressed process $K^\pm \to \mu^\pm \nu_\mu V$ is larger than the helicity-suppressed $K^\pm \to e^\pm \nu_e V$. Fig.~\ref{fig:Let:Spectrum} depicts the expected number of $V$ passing through the DUNE MPD for $L_e - L_\tau$ gauge bosons as a function of mass $M_V$ from various channels.
%%%%%%
%%%%%%

\begin{figure}[!htbp]
    \centering
    \includegraphics[width=0.6\linewidth]{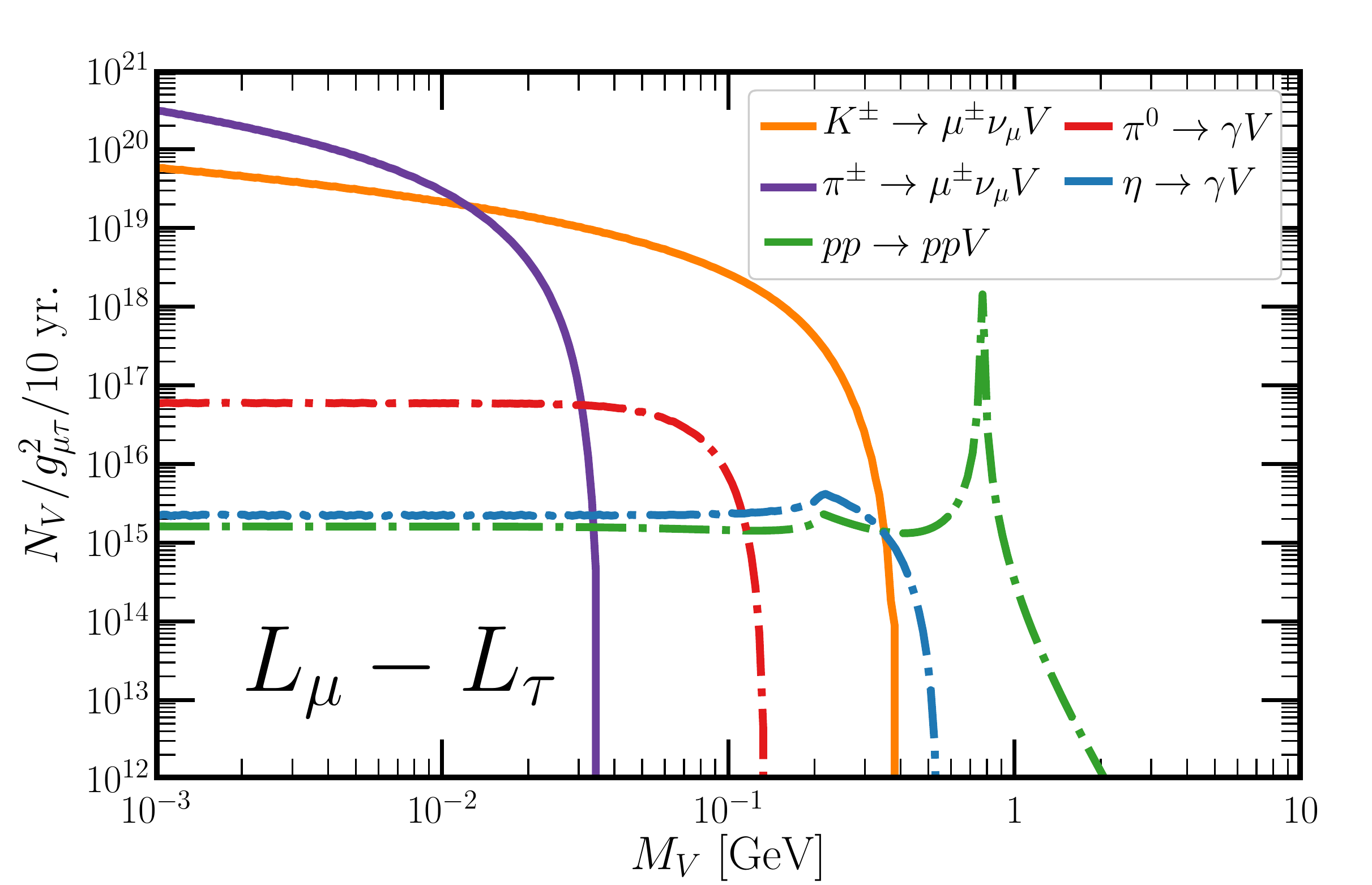}
    \caption{The number of $L_\mu - L_\tau$ gauge bosons $V$ traveling towards the DUNE MPD as a function of mass $M_V$ assuming ten years of data collection, and equal run time in neutrino and antineutrino modes. Solid lines display expected number due to charged meson decays -- purple ($\pi^\pm$) and orange ($K^\pm$). Dot-dashed lines display expected number due to processes relying on kinetic mixing -- red ($\pi^0$ decay), blue ($\eta$ decay), and green ($pp \to ppV$ bremsstrahlung).}
    \label{fig:Lmt:Spectrum}
\end{figure}

Finally, Fig.~\ref{fig:Lmt:Spectrum} depicts the number of $V$ produced for a $L_\mu - L_\tau$ gauge boson. Here, the small kinetic mixing (Fig.~\ref{fig:LoopKinetic}), causes the meson production mechanisms to final states containing muons to dominate as long as they are kinematically accessible. When the gauge boson mass is close to $770$ MeV and it can mix with the SM $\rho$, the proton-proton bremsstrahlung process causes significant production, even with the small loop-induced kinetic mixing.
%%%%%%
%%%%%%

\subsection{Decays of Leptophilic Gauge Bosons}\label{subsec:LeptophilicDecays}
At tree level, $L_\alpha - L_\beta$ gauge bosons will decay to pairs of neutrinos (flavor $\alpha$ or $\beta$), or pairs of charged leptons $V \to \ell_\alpha^+ \ell_\alpha^-$ and $V\to \ell_\beta^+ \ell_\beta^-$, assuming such a decay is kinematically accessible. 
%For completeness, we give the decay widths of these processes. 
Since we will not be able to observe neutrinos emerging from the decays of $V$, we sum over the flavors.  The decay width, assuming massless neutrinos, is
\begin{equation}
    \Gamma (V \to \nu \overline{\nu}) = \frac{ g_{\alpha\beta}^2 M_V}{12\pi}.
\end{equation}
The decay into charged leptons, similar to those of dark photons in Eq.~(\ref{eq:DarkPhoton:LeptonWidth}), is
\begin{equation}\label{eq:Lab:LeptonDecay}
    \Gamma (V \to \ell_\alpha^+ \ell_\alpha^-) = \frac{g_{\alpha \beta}^2 M_V}{12\pi} \left( 1 + \frac{2m_{\ell_\alpha}^2}{M_V^2}\right) \sqrt{ 1 - \frac{4m_{\ell_\alpha}^2}{M_V^2}}.
\end{equation}
When relevant, we also consider $V$ decays via loop-induced kinetic mixing. Specifically, we will be interested in $L_\mu - L_\tau$ gauge bosons that may decay into $e^+ e^-$ via the small kinetic mixing. This width is given by Eq.~(\ref{eq:Lab:LeptonDecay}) with the replacement $g_{\alpha\beta} \to e\varepsilon_{\alpha\beta}$. The resulting branching fraction for $L_\mu - L_\tau$ gauge bosons decaying this way is $\mathrm{Br}(V \to e^+ e^-) \approx 2\times 10^{-5}$.

\subsection{Sensitivity to Leptophilic Gauge Bosons}\label{subsec:LeptophilicSensitivity}

Here we present the expected sensitivity of DUNE to leptophilic gauge bosons decaying to $e^+ e^-$ or $\mu^+ \mu^-$ final states within the DUNE MPD. In all cases, we assume ten years of data collection and equal run time in neutrino and antineutrino modes (in agreement with the production rates depicted in Figs.~\ref{fig:Lem:Spectrum}-\ref{fig:Lmt:Spectrum}). 
For the $e^+ e^-$ final state, we choose to show regions of parameter space (the gauge boson mass and the new gauge coupling) for which greater than three or greater than ten signal events are expected in ten years. We discussed the backgrounds to a $e^+ e^-$ final state in Section~\ref{sec:DarkPhoton} and found that it should be background-free. As with the dark photon search, we assume here that the detector efficiency\footnote{We note that for invariant masses of ${\sim}1-100$ MeV, the invariant mass would be significantly harder to reconstruct than for larger invariant masses.} is 100\%.

We reproduce existing limits from experimental searches from Ref.~\cite{Bauer:2018onh}. Over the region of parameter space we are interested in, these are largely from electron beam dump experiments (E141, Orsay, and E137 providing the strongest constraints), as well as neutrino-electron scattering via the exchange of the new gauge boson (with the strongest constraints coming from the TEXONO experiment).
Ref.~\cite{Escudero:2019gzq} recently explored the impact of a light $L_\mu - L_\tau$ gauge boson on the cosmological history of the universe, and found that for certain combinations of masses and gauge couplings, large contributions to the number of effective relativistic species $\Delta N_\mathrm{eff.}$ at the time of big bang nucleosynthesis occur. Large values of $\Delta N_\mathrm{eff} \gtrsim 0.5$ are disfavored by measurements of big bang nucleosynthesis, however this is true only when data are analyzed under particular cosmological assumptions. No existing laboratory measurements constrain $L_\mu - L_\tau$ gauge bosons for $g_{\mu\tau} \lesssim 10^{-4}$, where we expect sensitivity.\footnote{One key theoretical motivation of $L_\mu - L_\tau$ gauge bosons is the possible explanation for the long-standing muon $(g-2)$ anomaly~\cite{Baek:2001kca,Pospelov:2008zw,Kamada:2015era,Araki:2015mya,Wise:2018rnb}. This solution requires $g_{\mu\tau} > 10^{-4}$ for $M_V > 1$ MeV, and will be probed by neutrino trident interactions in the DUNE LArTPC Near Detector~\cite{Altmannshofer:2019zhy,Ballett:2019xoj}.} Cosmological constraints should exist for $L_e - L_\mu$ and $L_e - L_\tau$ gauge bosons, but no such analysis exists in the literature. Likewise, no analysis for $M_V \lesssim 2$ MeV exists.

\begin{figure}
    \centering
    \includegraphics[width=0.6\linewidth]{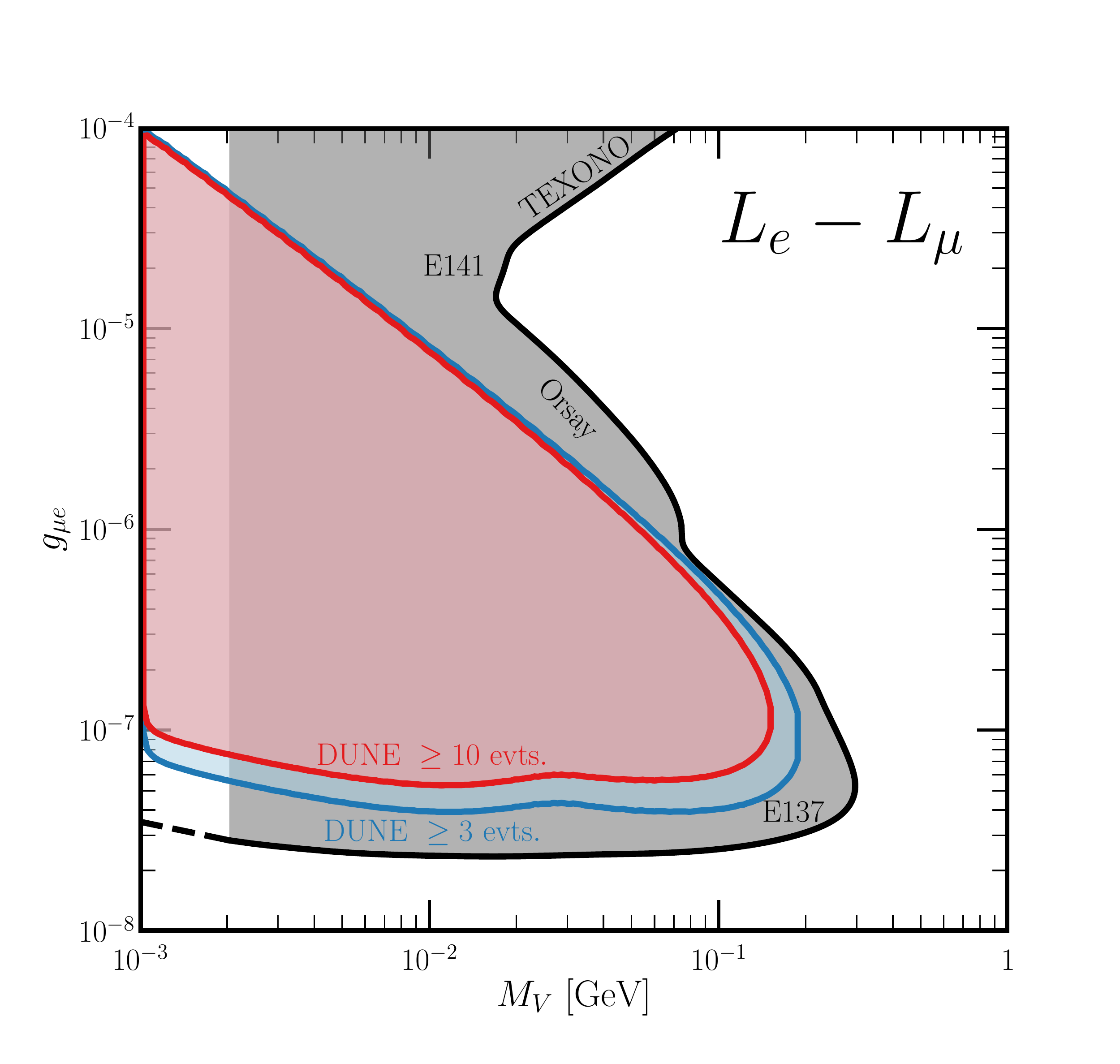}
    \caption{Expected sensitivity at the DUNE MPD to $L_e - L_\mu$ gauge bosons. The region in red (blue) indicates more than ten (three) signal events expected in ten years, a background-free search. Existing limits in this parameter space are in grey. The dashed line for small $M_V$ and $g_{\mu e}$ indicates our extrapolation of existing limits.}
    \label{fig:Lemu:Sensitivity}
\end{figure}

Fig.~\ref{fig:Lemu:Sensitivity} displays our expected DUNE MPD sensitivity to $L_e - L_\mu$ gauge bosons with masses between $1$ MeV and $1$ GeV. For points inside the blue (red) region, more than three (ten) signal events are expected. Given the discussion regarding backgrounds for $A^\prime \to e^+ e^-$ in Section~\ref{DarkPhoton:Backgrounds}, we expect the three event contour to represent the MPD sensitivity fairly well. For comparison, existing limits are depicted in grey. We see here that DUNE presents sensitivity that is nearly as powerful as electron beam dump experiments, but it is unlikely to improve on them. This is not terribly surprising since electron bremsstrahlung at existing experiments leads to a large flux of $V$. As stated above, the existing literature does not display limits below roughly $2$ MeV -- we expect that DUNE is no more powerful than the existing limits below this mass either, and include a dashed black line to extend the existing literature limits as we expect they would extend.

\begin{figure}
    \centering
    \includegraphics[width=0.6\linewidth]{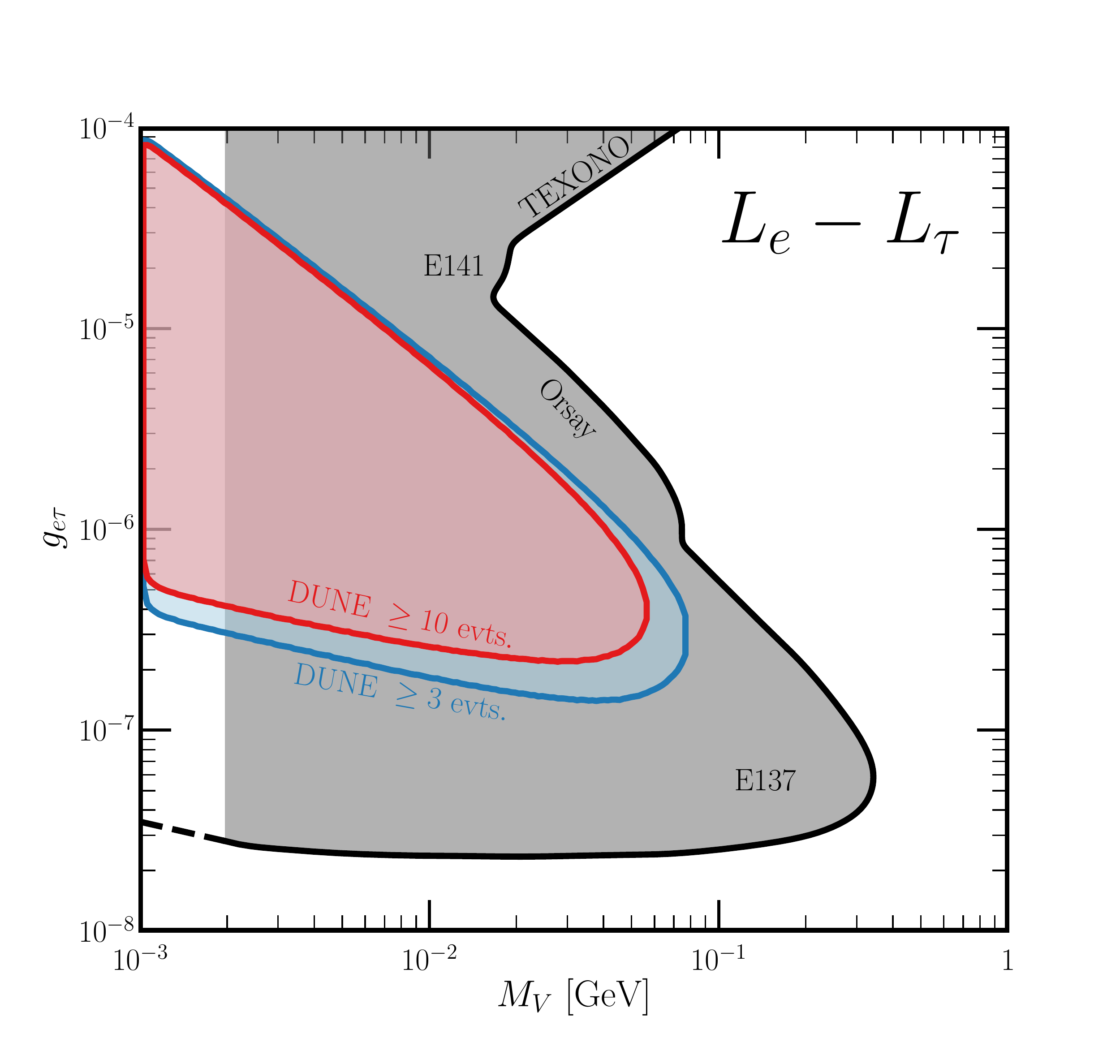}
    \caption{Expected sensitivity at the DUNE MPD to $L_e - L_\tau$ gauge bosons. The region in red (blue) indicates more than ten (three) signal events expected in ten years, a background-free search. Existing limits in this parameter space are in grey. The dashed line for small $M_V$ and $g_{e\tau}$ indicates our extrapolation of existing limits.}
    \label{fig:Letau:Sensitivity}
\end{figure}

The existing limits on $L_e - L_\tau$ are nearly identical to those from $L_e - L_\mu$, with the only exception being near $M_V \approx 200$ MeV, where decays into muons modify the lifetime of $L_e - L_\mu$ gauge bosons but not $L_e - L_\tau$ ones. Our expected DUNE MPD sensitivity, shown in Fig.~\ref{fig:Letau:Sensitivity}, is slightly weaker, due to the reduced production discussed cf. Fig.~\ref{fig:Let:Spectrum}. This results in the expected sensitivity being slightly weaker than, but comparable to, the existing limits.

\begin{figure}
    \centering
    \includegraphics[width=0.6\linewidth]{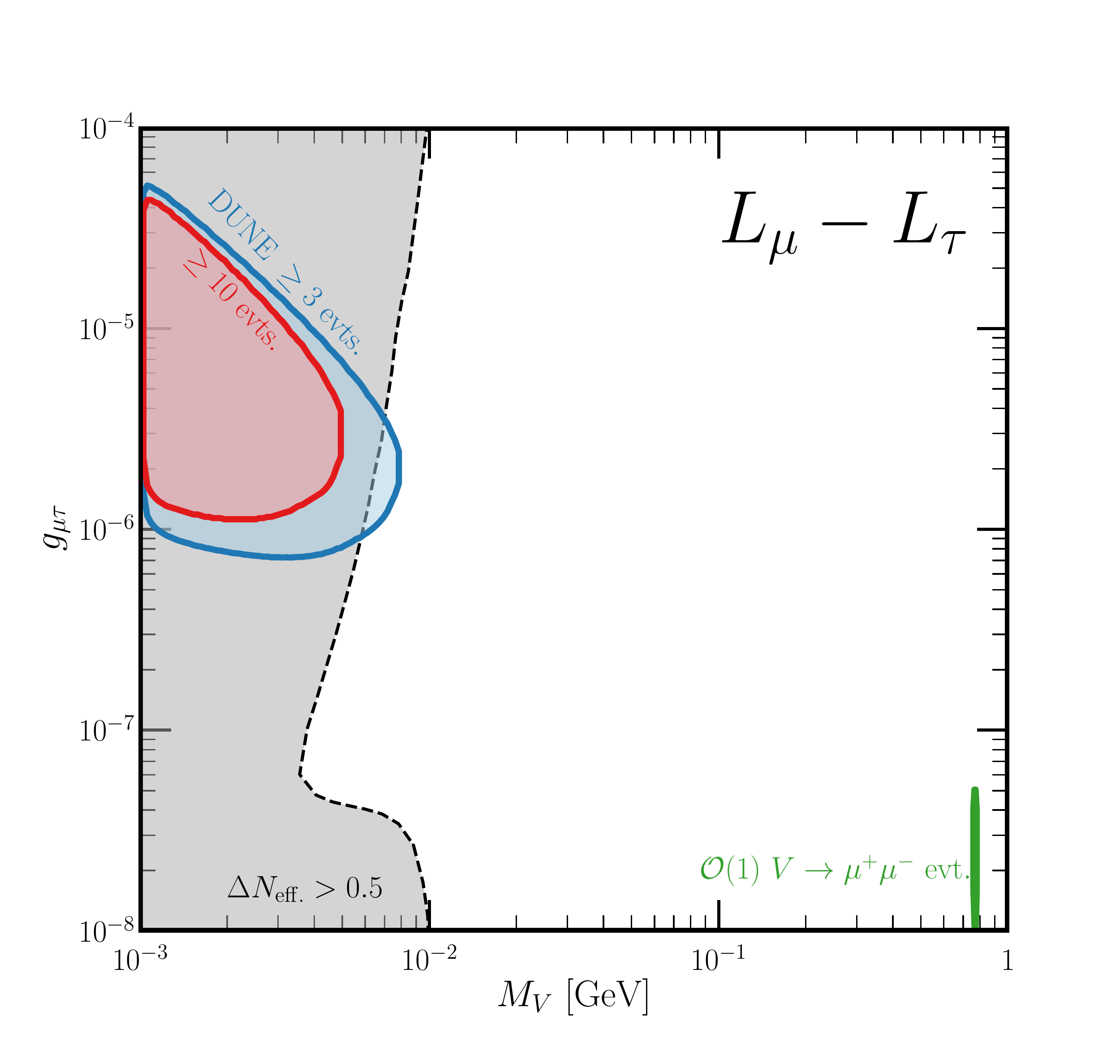}
    \caption{Expected sensitivity to $L_\mu - L_\tau$ gauge bosons at the DUNE MPD. The region in red (blue) indicates more than ten (three) signal events ($V \to e^+ e^-$) expected. In grey, the region for which $\Delta N_\mathrm{eff} > 0.5$ at the time of big bang nucleosynthesis~\cite{Escudero:2019gzq}. In green is the region of parameter space for which $V \to \mu^+ \mu^-$ may yield $\mathcal{O}(1)$ signal events, see text for further discussion.}
    \label{fig:Lmutau:Sensitivity}
\end{figure}

Lastly, we depict the expected sensitivity to $L_\mu - L_\tau$ gauge bosons in Fig.~\ref{fig:Lmutau:Sensitivity}. Here, we are hindered in such a search for $V \to e^+ e^-$ due to the kinetic-mixing-suppressed branching fraction to electrons. Regardless, such a search is complementary to the limit derived in Ref.~\cite{Escudero:2019gzq}, and we emphasize that the DUNE MPD approach would be the first terrestrial search for $L_\mu - L_\tau$ gauge bosons in this region of parameter space.
As discussed around Fig.~\ref{fig:Lmt:Spectrum}, for $M_V \approx m_\rho$, the bremsstrahlung process is enhanced due to $\rho-V$ mixing, leading to significant $V$ production. At this mass, $V$ may decay into $\mu^+ \mu^-$ unsuppressed. We find that for a very narrow range of parameter space, the DUNE MPD may expect to observe $\mathcal{O}(1)$ signal event with ten years of data collection. We indicate the region in which this occurs in green in Fig.~\ref{fig:Lmutau:Sensitivity} -- we encourage a more detailed study of this region of parameter space for future work, specifically in determining the neutrino-related backgrounds to such a search. However, combining searches for $V \to \mu^+ \mu^-$ both in the MPD and the LArTPC, as well as allowing the $V$ to decay upstream of both detectors, could enhance the overall DUNE sensitivity in this region, allowing for searches for $L_\mu - L_\tau$ gauge bosons in completely new areas of parameter space. 

%\afterpage{\clearpage}  

%--------------------------
%Dark Higgs Section
%--------------------------

\section{Dark Higgs Boson}\label{sec:DarkHiggs}\setcounter{footnote}{0}

In the previous sections, we discussed sensitivity to semi-long-lived vector bosons that couple to the Standard Model via kinetic mixing with the SM hypercharge group or as a result of a new gauge symmetry under which some SM fermions are charged. Those scenarios are a subset of the commonly-studied renormalizable dark sector portals, where new particles may be added to the SM and the interaction between the dark sector and the SM is via a renormalizable operator.

A second portal option is the Higgs portal, where a new scalar $\varphi$ interacts with the SM via one of two operators\footnote{The second option is more common in the literature -- the first is super-renormalizable, but forbidden if $\varphi$ carries any charge, SM or new-physics related. For this study, they are phenomenologically equivalent, as long as $\varphi$ and the SM Higgs boson may mix.}
\begin{equation}
    \mathcal{L} \supset \delta \left\lvert H\right\rvert^2 \varphi, \quad \mathrm{or}\quad \mathcal{L} \supset \lambda \left \lvert H\right\rvert^2 \left\lvert \varphi\right\rvert^2.
\end{equation}
In the case of the first operator, mixing between $H$ and $\varphi$ occurs after electroweak symmetry breaking, while for the second operator mixing requires both $H$ and $\varphi$ to acquire a vacuum expectation value.
This mixing may be expressed in terms of Lagrangian parameters.  Here, however, we choose to use the phenomenological mixing parameter, which we denote as $\sin\vartheta$, and the $\varphi$ mass, $M_\varphi$, as the free parameters.

When considering new vector bosons in Sections~\ref{sec:DarkPhoton} and~\ref{sec:Leptophilic}, the couplings to the SM were given by the electric charge of the SM particle, reweighted by the parameter that characterizes the kinetic mixing between the SM and new $U(1)$ gauge bosons. This led to roughly equivalent decays of $A^\prime$ to electrons and muons, up to kinematic effects. For a dark Higgs, the coupling between $\varphi$ and SM particles proceeds through the SM Higgs bosons, so couplings are related to SM particle masses, and are therefore hierarchical. This will drive production mechanisms to favor situations involving heavy quarks, which we will discuss in the following subsections.  In particular, the dominant production modes are $K\rightarrow \pi\, \varphi$ and the decay modes are $\varphi\rightarrow e^+e^-, \mu^+\mu^-, \pi\,\pi$ with the $\varphi$ decaying preferentially into the heaviest final state available.

\subsection{Dark Higgs Boson Production}

Refs.~\cite{Gunion:1989we,Leutwyler:1989xj,Bezrukov:2009yw,Bezrukov:2013fca,Krnjaic:2015mbs,Feng:2017vli}, among others, discuss the production of $\varphi$. Because $\varphi$ couples via mixing with the SM Higgs scalar, its production will be largest in processes that allow for couplings to heavy quarks.  At DUNE, the dominant contribution comes from the decays of $K^0_L$ and $K^\pm$: the one-loop penguin diagram involving a $W$ boson and top quark introduces $\varphi-t$ couplings, which are large.  While $D$-meson production is not negligible at DUNE, contributions of $D$ meson decay into $\varphi$ will be suppressed by elements of the CKM matrix and the bottom-quark Yukawa coupling. The meson with the largest branching fraction into $\varphi$ is the $B$ meson. However, $B$-meson production at DUNE is too small to be competitive with existing experiments, since $\sqrt{s} \approx 12$ GeV.

The matrix element for $K$ decay into a light meson and a dark Higgs $\varphi$ is~\cite{Gunion:1989we,Leutwyler:1989xj,Bezrukov:2009yw} 
\begin{equation}
\mathcal{M} = \sin\vartheta \frac{m_K^2}{v} \left(\gamma_1 \frac{7}{18} \frac{m_K^2 - M_\varphi^2 + m_\pi^2}{m_K^2} - \gamma_2 \frac{7}{9} + \frac{1}{2} \frac{3}{16\pi^2 v^2} \sum_{i=u,c,t} V_{id}^* m_i^2 V_{is}\right),
\label{eq:phiprod}
\end{equation}
where $\gamma_1$ and $\gamma_2$ are both negligibly small. Here, and throughout, $v\approx 246$ GeV is the vacuum expectation value of the neutral component of the Higgs field. The dominant contribution comes from the third term in Eq.~(\ref{eq:phiprod}), given by contributions with a $W$ boson and a loop involving up-type quarks. The largest contribution will come when $i = t$, the contribution from the top quark.

Following Refs.~\cite{Gunion:1989we,Leutwyler:1989xj,Bezrukov:2009yw,Bezrukov:2013fca,Krnjaic:2015mbs,Feng:2017vli}, we may write the branching fractions of interest as
\begin{eqnarray}
\mathrm{Br}(K^\pm \to \pi^\pm \varphi) &=& 2\times 10^{-3} \sin^2\vartheta \, \rho_\varphi\left(\frac{M_\varphi^2}{m_{K^\pm}^2}, \frac{m_{\pi^\pm}^2}{m_{K^\pm}^2}\right), \label{eq:DH:ChargedKaon}\\
\mathrm{Br}(K_L^0 \to \pi^0 \varphi) &=& 7\times 10^{-3} \sin^2\vartheta \, \rho_\varphi\left(\frac{M_\varphi^2}{m_{K_L^0}^2}, \frac{m_{\pi^0}^2}{m_{K_L^0}^2}\right),\label{eq:DH:KL}
\end{eqnarray}
where $\rho_\varphi(x, y)$ is a dimensionless function,
\begin{equation}
    \rho_\varphi(x, y) = \frac{1}{2} \sqrt{1 + x^2 + y^2 - 2(x + y + xy)}.
\end{equation}
The branching fraction for $K_L$ is larger than for $K^\pm$ due predominantly to its smaller total width. We disregard the contribution from $K_S \to \pi^0\varphi$, which has a small branching fraction due to the large total width of $K_S$.

One may also consider $\varphi$ produced via decays of other mesons. In comparison with Eqs.~(\ref{eq:DH:ChargedKaon})-(\ref{eq:DH:KL}), the branching fraction of a $B$ meson into a strange meson and a $\varphi$, for small $\vartheta$, is~\cite{Feng:2017vli}
\begin{equation}
    \mathrm{Br}(B\to X_s\varphi) \simeq 5.7 \left(1 - \frac{M_\varphi^2}{m_b^2}\right)^2 \sin^2\vartheta.
\end{equation}
While this branching fraction is relatively large, the production rate of $B$ mesons in the DUNE target is extremely small, see Appendix \ref{appendix:MesonProd}. On the other hand, mesons lighter than kaons are produced in abundance. However, branching fractions of interest, such as $\pi^\pm \to e^\pm \nu \varphi$, are as small as $10^{-9}\sin^2\vartheta$. The dominant sensitivity at the DUNE MPD then, will come from kaon-related production of dark Higgs bosons.

\begin{figure}
\centering
\includegraphics[width=0.45\linewidth]{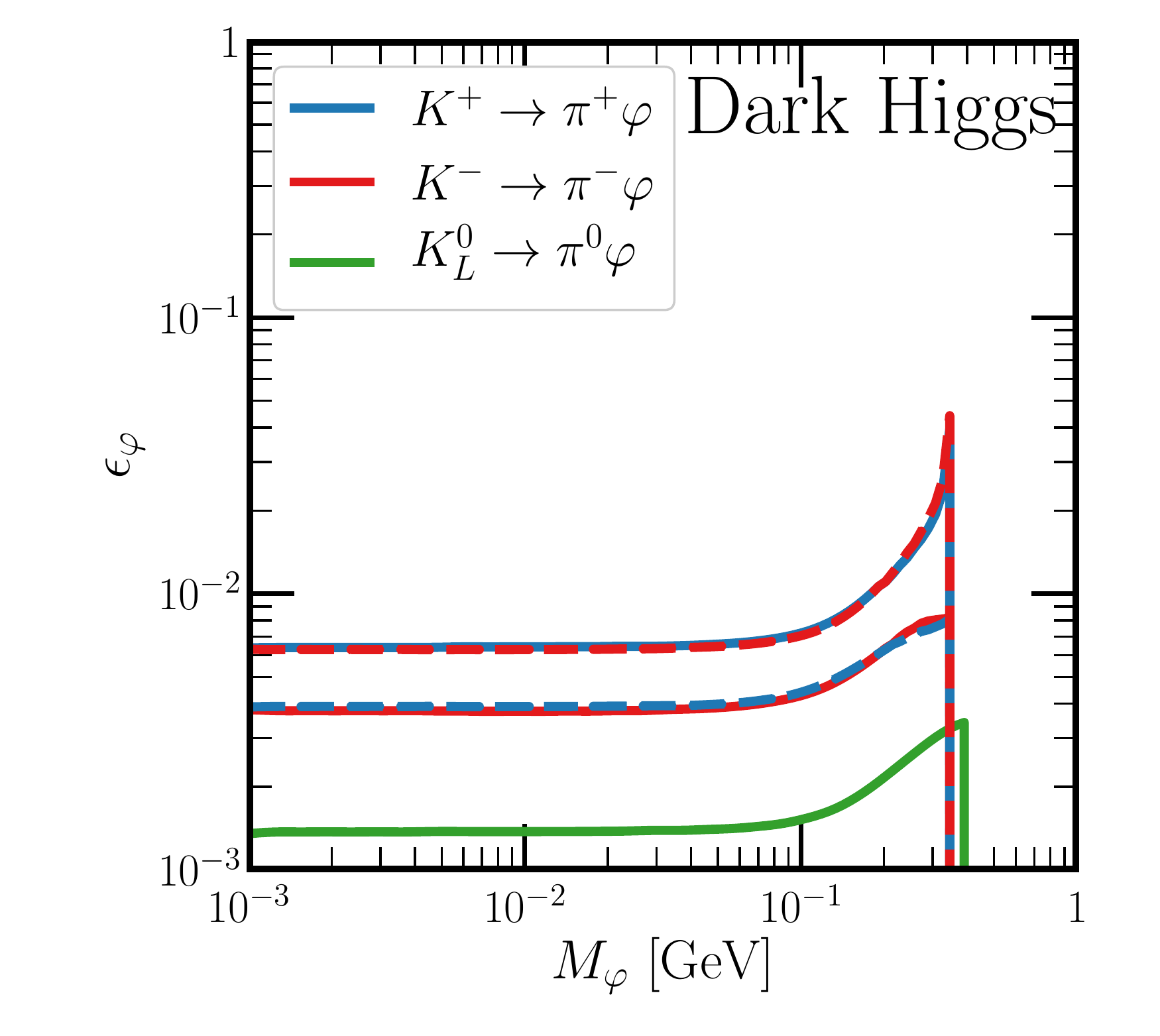}
\includegraphics[width=0.45\linewidth]{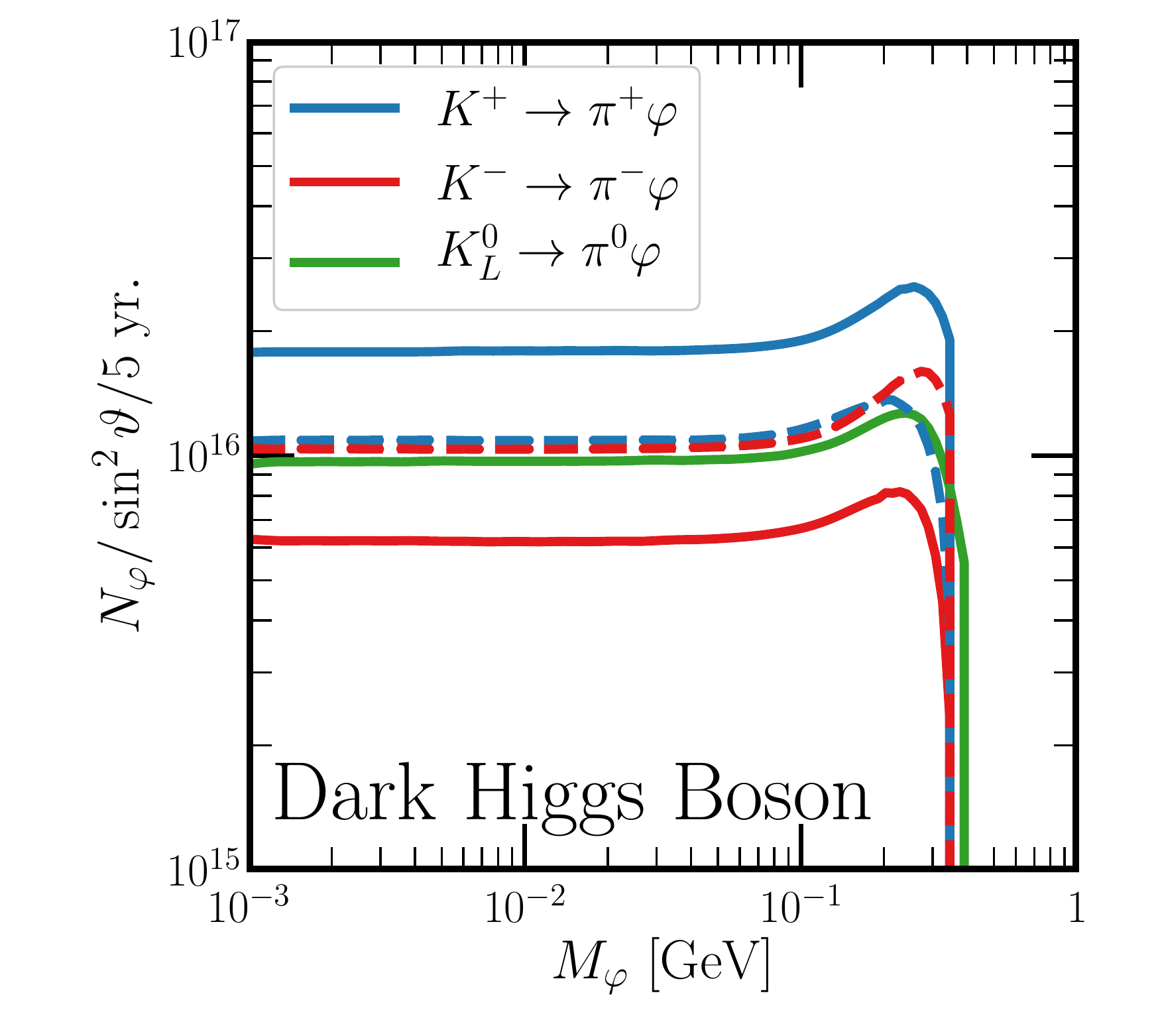}
\caption{Left panel: acceptance fraction $\epsilon_{\varphi}$ of dark Higgs particles as a function of their mass $M_\varphi$. The acceptance fraction is defined as the fraction of all produced $\varphi$ that have their momentum pointing in the direction of the DUNE MPD at the time of their production. Right panel: expected number of dark Higgs particles $\varphi$ directed towards the DUNE MPD as a function of their mass $M_\varphi$, assuming five years of data collection in both neutrino mode (solid lines) and antineutrino mode (dashed lines). We scale the number of particles to $\sin^2\vartheta = 1$. Solid (dashed) lines are for neutrino (antineutrino) mode, and blue, red, and green lines indicate $\varphi$ from $K^+$, $K^-$, and $K^0_L$ decays, respectively.  The $K_L^0$ distribution is unaffected by focusing and the distinction between neutrino/antineutrino modes, and so there is no corresponding dashed green line. $K_L^0$-produced $\varphi$ allow for reaching larger values of $M_\varphi$ as $m_{K_L^0} - m_{\pi^0} > m_{K^\pm} - m_{\pi^\pm}$.}
\label{fig:DH:Focusing}\label{fig:DH:Production}
\end{figure}

 In Fig.~\ref{fig:DH:Focusing}, we present the fraction of $\varphi$ that are produced in a given decay and are traveling in the direction of the MPD when they are produced.
Several features are worthy of note. Reversing the horn current and switching from neutrino to antineutrino mode nearly perfectly interchanges the acceptance fraction of $\varphi$ coming from $K^+$ and $K^-$ decays -- dashed lines are coincident with solid lines of the opposite color. In the limit $M_\varphi \to 0$, the acceptance fraction of the focused distribution ($K^+$ in neutrino mode, $K^-$ in antineutrino mode) is almost, but not quite, double that of the defocused distribution. If analyzing decays of charged pions instead of charged kaons, this ratio would be significantly larger; DUNE is designed to focus $\pi^\pm$ and defocus $\pi^\mp$, while kaon focusing is not the main priority of the experiment. Na{\"i}vely, we expect that the $K_L^0$-produced $\varphi$ would have an acceptance fraction somewhere between the focused and defocused distributions, as they are neither focused nor unfocused. This is not the case, as evident by the green line in Fig.~\ref{fig:DH:Focusing}, and is because a significant fraction (roughly $25\%$) of $K^0_L$ live long enough to reach the rock at the end of the DUNE beam decay pipe. Those $K_L^0$ decay at rest, and a very small fraction of the isotropic decay products are pointing in the direction of the MPD.

With this information about the direction of travel of the dark Higgs, we simulate the decays with the branching fractions from Eqs.~(\ref{eq:DH:ChargedKaon}) and~(\ref{eq:DH:KL}) and determine the number of $\varphi$ particles from these decays that pass through the DUNE MPD, assuming five years of data collection in both neutrino and antineutrino mode. This is depicted in the right panel of Fig.~\ref{fig:DH:Production}.
Here, solid (dashed) lines indicate (anti)neutrino mode, and blue, red, and green lines indicate $K^+$, $K^-$, and $K_L^0$ decays, respectively. Because $K_L^0$ are unaffected by the focusing horns, there is no distinction between this production mechanism for neutrino and antineutrino modes. The $K_L^0$ distribution also extends to slightly larger masses, as $m_{K^\pm} - m_{\pi^\pm} \approx 354$ MeV whereas $m_{K^0_L} - m_{\pi^0} \approx 394$ MeV.

%%%%%%%%%%%%%%%%%%%%%%%%%%%%%

\subsection{Dark Higgs Boson Decays}

For very light $\varphi$, $M_\varphi<2m_\pi$, the decay widths of $\varphi$ are given by%~\cite{Feng:2017vli}
\begin{equation}
\Gamma(\varphi \to \ell^+ \ell^-) = \frac{M_\varphi \sin^2\vartheta}{8\pi v^2} m_\ell^2 \left( 1 - \frac{4 m_\ell^2}{M_\varphi^2}\right)^{3/2}~.
\end{equation}
For masses above $2m_\pi$, but below a few GeV, the width to mesons is hard to determine due to strong QCD effects and the effects of meson resonances~\cite{Clarke:2013aya}.
For the mass range of interest here, $M_\varphi\lesssim 0.5$ GeV, the only meson decay open is to a pair of pions.  This decay width can be expressed in terms of a hadronic matrix elements as,
\begin{equation}
\Gamma(\varphi \to \pi^+ \pi^-) = 2\Gamma(\varphi\to \pi^0 \pi^0) = \frac{M_\varphi^3 \sin^2\vartheta}{16\pi v^2} \sqrt{ 1 - \frac{4m_\pi^2}{M_\varphi^2}}|G(M_\varphi^2)|^2~.
\end{equation}
Here $|G(M_\varphi^2)|$ is the (dimensionless, in contrast with some results in the literature) transition amplitude for the process $\varphi \to \pi\pi$.  To estimate $G(M^2_{\varphi})$, we follow the discussion of  Ref.~\cite{Donoghue:1990xh}, which we now briefly outline.  This transition amplitude may be expressed, in the limit of isospin conservation, in terms of three independent form factors, which are functions of $s$, the invariant mass-squared of the outgoing pion pair: 
\begin{equation}
    G(s) = \frac{2}{9}\theta_\pi(s) + \frac{7}{9}\left[\Gamma_\pi(s) + \Delta_\pi(s)\right]~.
\end{equation}
At next-to-leading order in chiral perturbation theory,\footnote{Note that these quantities are, like $G$, dimensionless, and differ in definition from those in Ref.~\cite{Donoghue:1990xh}.}
\begin{eqnarray}
\theta_\pi(s) &=& \left(1 + \frac{2m_\pi^2}{s}\right)\left(1 + \psi(s)\right) + b_\theta s, \label{eq:theta} \\
\Gamma_\pi(s) &=& \frac{m_\pi^2}{s} \left(1 + \psi(s) + b_\Gamma s\right), \label{eq:gamma} \\
\Delta_\pi(s) &=& d_F \left(1 + \psi(s) + b_\Delta s\right). \label{eq:delta}
\end{eqnarray}
In these formulae, the dimensionful quantities $b_\theta$, $b_\Gamma$, and $b_\Delta$ are extracted from derivatives of the form factors at $s = 0$. The dimensionless parameter $d_F$ is related to the strangeness content of the pion. Finally, the function $\psi(s)$ is defined as 
\begin{equation}
    \psi(s) = \frac{2s-m_\pi^2}{16\pi^2 F_\pi^2} \left[\kappa \log{\left(\frac{1-\kappa}{1+\kappa}\right)} + 2 + i\pi\kappa\right] + \frac{s}{96\pi^2 F_\pi^2},
\end{equation}
where $F_\pi = 130$ MeV is the pion decay constant and $\kappa \equiv (1 - 4m_\pi^2/s)^{1/2}$ is a kinematic factor. Ref.~\cite{Donoghue:1990xh} extracts the relevant parameters from measurements of pion-pion scattering. In our calculations, we use $d_F = 0.09$, $b_\theta = 2.7$ GeV$^{-2}$, $b_\Gamma = 2.6$ GeV$^{-2}$, and $b_\Delta = 3.3$ GeV$^{-2}$. As discussed in Ref.~\cite{Donoghue:1990xh}, this leads to an enhancement of the partial width into pions by a factor of approximately 4.4 at $M_\varphi = 500$ MeV, relative to the leading-order result from Ref.~\cite{Voloshin:1985tc}.
This leading order result is obtained by setting $b_{\theta,\Gamma,\Delta} = 0$, $d_F = 0$, and $\varphi(s) = 0$ in Eqs.~(\ref{eq:theta})-(\ref{eq:delta}).

\begin{figure}[!hbtp]
    \centering
    \includegraphics[width=0.6\linewidth]{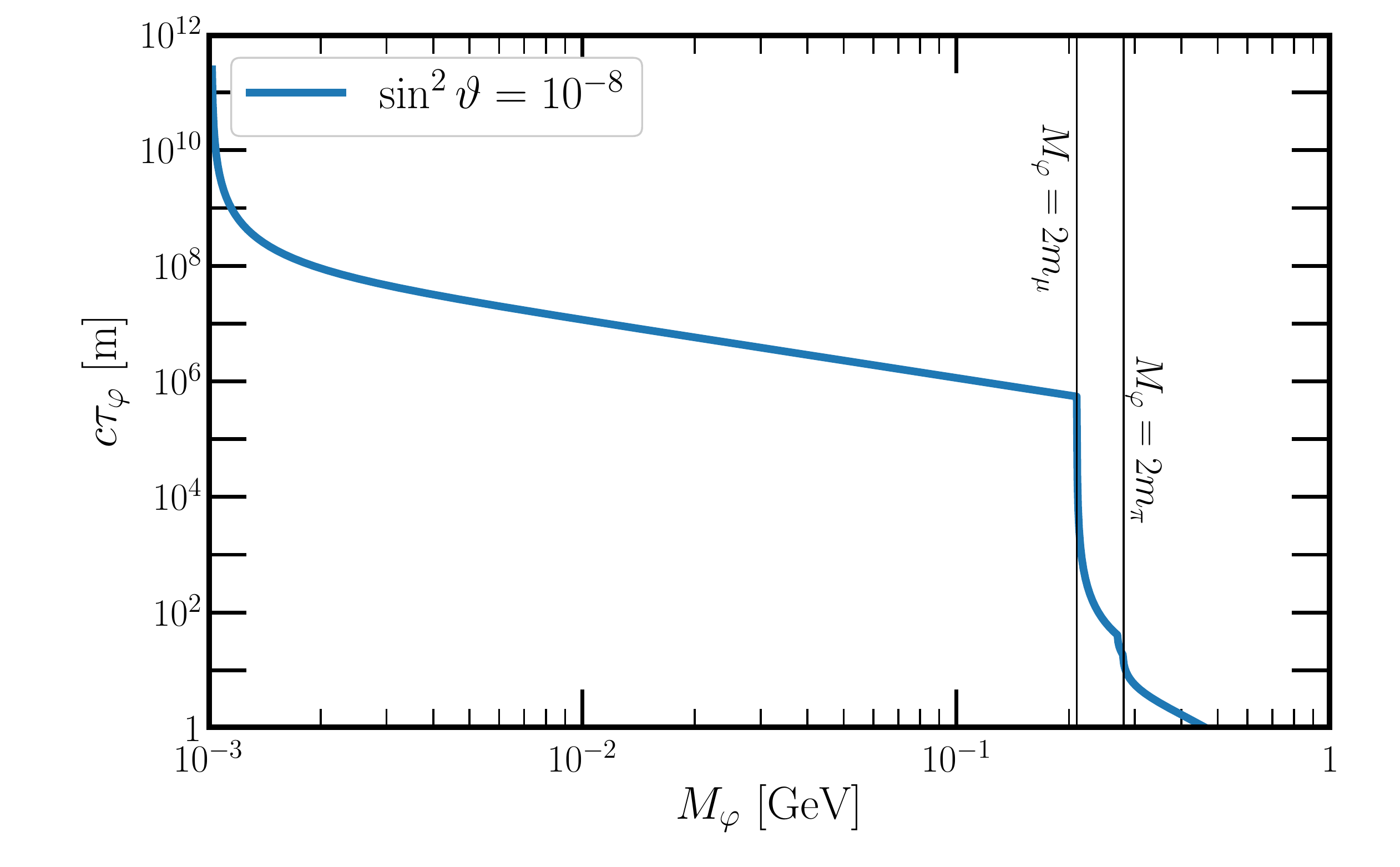}
    \caption{Rest-frame lifetime of a dark Higgs with mixing $\sin^2\vartheta = 10^{-8}$ as a function of mass $M_\varphi$. We label the points at which the decays $\varphi \to \mu^+\mu^-$ and $\varphi \to \pi\pi$ become kinematically accessible for clarity.}
    \label{fig:DH:Lifetime}
\end{figure}

Instead of depicting branching fractions of different final states as a function of mass, we choose to present instead the total lifetime of $\varphi$ as a function of its mass in Fig.~\ref{fig:DH:Lifetime}, keeping in mind that, given its hierarchical couplings to SM fermions, $\varphi$ decay is dominated by the heaviest allowed final state. 
For light masses, the lifetime (assuming $\sin^2\vartheta = 10^{-8}$) is significantly longer than the distance to the DUNE near detector. For masses between roughly $200-400$ MeV, the lifetime is of order of the distance between the DUNE target and the MPD, meaning a $\varphi$ decay in the detector becomes significantly more likely.
\begin{comment}
If we compare our result, depicted in Fig.~\ref{fig:DH:Lifetime}, with the discussion in Ref.~\cite{Clarke:2013aya}, we find that our result agrees with the longest-lifetime option from the literature. This implies that the results we obtain will be conservative in the small-$\sin^2\vartheta$ region of interest, and mildly optimistic in the large-$\sin^2\vartheta$ region\footnote{The large-$\sin^2\vartheta$ region will correspond to where $\varphi$ decays are so prompt that the majority of produced $\varphi$ decay before reaching the detector. Taking the longest lifetime consistent with the literature will give sensitivity to larger values of $\sin^2\vartheta$ in this region than if assuming a smaller lifetime, however such effects are tiny when presented in log-log space.}.
\end{comment}

%%%%%%%%%%%%%%%%%%%%%%%%%%%%

\subsection{Backgrounds and Sensitivity}

In Section~\ref{sec:DarkPhoton}, we discussed the possible backgrounds for searches of final states involving $e^+ e^-$, $\mu^+ \mu^-$, or pion pairs and argued that, especially when utilizing the invariant mass peak of $\mu^+ \mu^-$, we may safely consider the expected backgrounds in the DUNE MPD to be small. The signal for a dark Higgs scalar decaying into charged leptons will appear practically identical to that for a dark photon decaying into charged leptons. We discuss here two possible ways of disentangling the two hypotheses.

In contrast to the dark photon scenario studied in Section~\ref{sec:DarkPhoton}, where, for a particular mass $M_{A^\prime}$, one expects a certain ratio of different final states, such as $60\%$ of events being $A^\prime \to e^+ e^-$ and $40\%$ being $A^\prime \to \mu^+\mu^-$, here we expect nearly every event to be in a single search channel because the individual partial widths for $\varphi$ decay are so hierarchical. For lighter dark Higgs/vector bosons where decays into $\mu^+ \mu^-$ are kinematically forbidden, this strategy will not allow us to disentangle the two hypotheses. Secondly, because the $\varphi$ flux is coming from decays of both charged and neutral kaons, where $A^\prime$ comes from decays of neutral mesons (or continuum processes), the $A^\prime$ spectrum should be less focused than the $\varphi$ one. One proposal for the DUNE Near Detector Complex is that the liquid and gas TPCs move off axis (the DUNE-PRISM proposal~\cite{duneprism}) to measure different portions of the neutrino spectrum. In principle, one can calculate the distribution of $\varphi$ and $A^\prime$ as a function of how off-axis the near detector is, and search for such a shape when operating on- and off-axis, making it possible to distinguish the different new physics hypotheses.
\begin{figure}[!hbtp]
\centering
\includegraphics[width=0.6\linewidth]{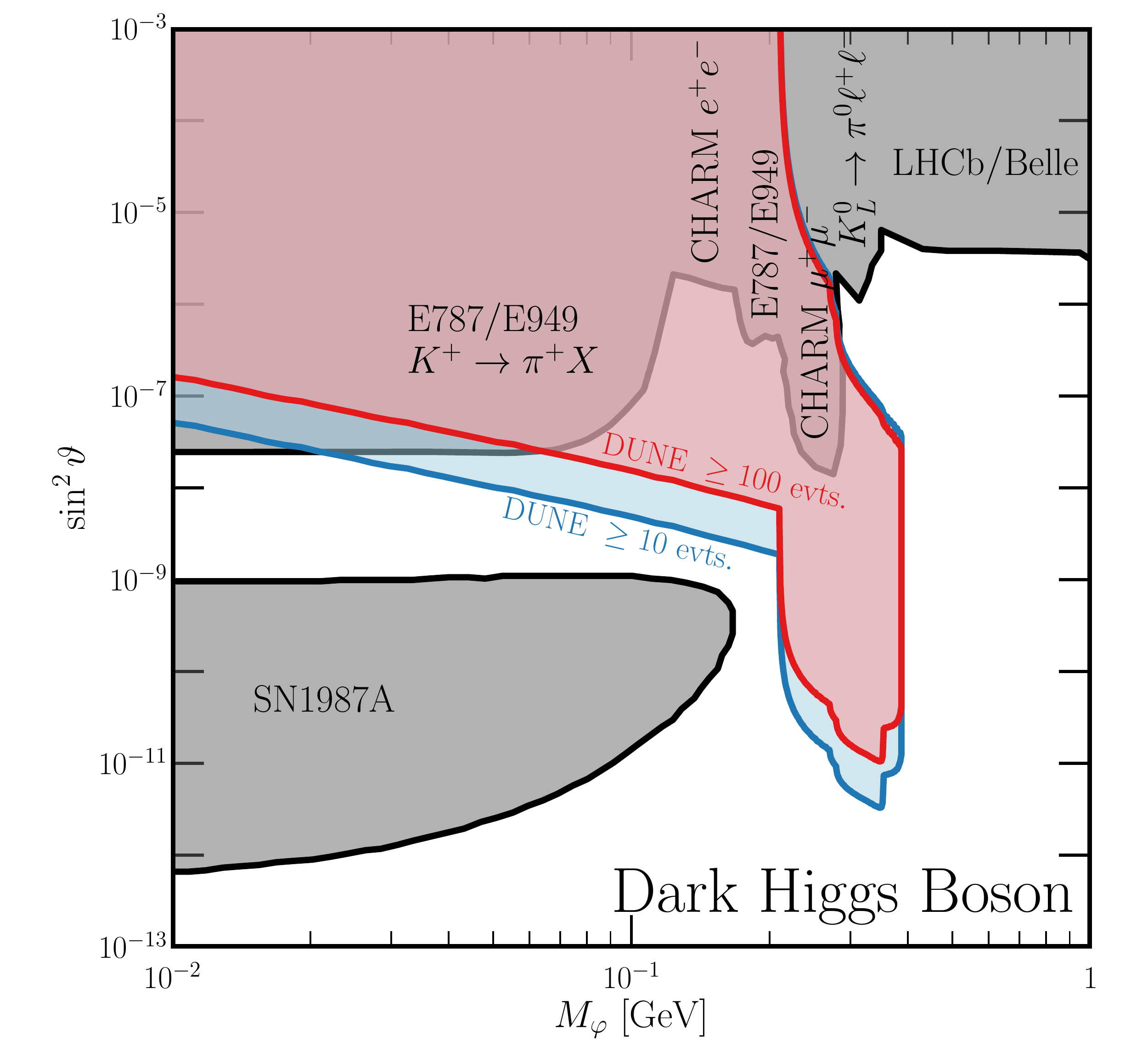} 
\caption{Region of the dark Higgs mass $M_{\varphi}$ versus mixing between the dark and SM Higgs bosons $\sin^2\vartheta$ parameter space where one expects more than 10 (100) dark Higgs scalar decays in the DUNE MPD, in blue (red). In grey is the region of parameter space currently experimentally excluded.}
\label{fig:DH:Sensitivity}
\end{figure}

Combining the fluxes and the decay width of $\varphi$, Fig.~\ref{fig:DH:Sensitivity} depicts in blue (red) regions of $M_\varphi$ vs. $\sin^2\vartheta$ parameter space for which we expect at least ten (one hundred) total decays (into $e^+ e^-$, $\mu^+ \mu^-$, or $\pi\pi$) in ten years of data collection at DUNE. 
For comparison, existing limits are depicted in grey~\cite{Bezrukov:2009yw,Krnjaic:2015mbs,Winkler:2018qyg}. We also include a limit from searches for charged kaons decaying to a charged pion and an invisible particle, reported in Ref.~\cite{Artamonov:2009sz}, translated into $M_\varphi$ vs. $\sin^2\vartheta$ parameter space. 

We see here that the DUNE MPD will allow for an improvement on current limits between roughly $20-350$ MeV, improving on the BNL search and CHARM for this wide range of masses. A DUNE MPD search may also allow us to comprehensively search in the window still open between searches of this type (displaced decays) and limits from considerations of supernova luminosity from Supernova 1987A~\cite{Krnjaic:2015mbs}.  A recent study \cite{Batell:2019nwo} of the sensitivity of Fermilab's short baseline neutrino detectors (e.g., ICARUS and SBND) to dark Higgs models projects bounds on $\sin^2\vartheta$ a factor of a few weaker than those anticipated at DUNE.

%-----------------------------------------
%Heavy Neutral Leptons Section
%-----------------------------------------

\section{ Heavy Neutral Leptons}\label{sec:HNL}\setcounter{footnote}{0}

In this section, we consider the existence of new fermions that are not charged under the SM gauge interactions. Assuming these are not charged under any other gauge symmetries, they couple, at the renormalizable level, to lepton doublets and the Higgs doublet. After electroweak symmetry breaking, for one new fermion $N$, the neutrino weak-interaction eigenstates $\nu_e,\nu_{\mu},\nu_{\tau}$ can be written as linear combinations of the neutrino mass eigenstates $\nu_1,\nu_2,\nu_3,N$:
\begin{equation}
\nu_{\alpha} = \sum_{i=1,2,3}U_{\alpha i}\nu_i + U_{\alpha N}N,
\end{equation}  
$\alpha=e,\mu,\tau$. The fourth mass eigenstate $N$ has mass $M_N$ while the lighter three states have masses $m_1,m_2,m_3$. The latter, along with the $U_{\alpha i}$ matrix elements, are constrained by neutrino oscillation experiments. This setup allows for the possibility that $\nu_1,\nu_2,\nu_3$, and $N$ are Majorana or Dirac fermions. 

Here, we refer to $N$ as a heavy neutrino or a heavy neutral lepton (HNL).\footnote{Such gauge-singlet fermions which mix with light neutrinos are often referred to as ``sterile neutrinos.'' We avoid this terminology, as it often refers to mixing that coherently modifies the active neutrino oscillations.} We are interested in $1~{\rm MeV}\lesssim M_N\lesssim 1$~GeV, when the HNLs can be produced by the same mechanisms -- meson and lepton decays -- as the light neutrinos in the DUNE target and decay pipe and are still heavy enough to decay into charged SM particles. The production and decay rates of such HNLs are governed by the weak interactions and the mixing parameters $U_{\alpha N}$. Furthermore, the same parameters allow for a variety of decay modes for the HNLs, depending on the HNL mass. These production and decay mechanisms allow us to search for HNLs in the DUNE MPD. For simplicity, we will consider that $N$ couples to only one charged lepton flavor at a time, i.e., we will assume that at most one of $U_{eN},U_{\mu N}, U_{\tau N}$ is nonzero for a given analysis.

Section~\ref{sec:HNL:Production} discusses the production channels for $N$ considered here and Section~\ref{sec:HNL:Decays} discusses $N$ decays that could be detected in the DUNE MPD. In Section~\ref{sec:HNL:Sensitivity}, we discuss backgrounds associated with each signal and present the expected sensitivity for HNLs coupling to the SM via nonzero $|U_{eN}|^2$, $|U_{\mu N}|^2$, or $|U_{\tau N}|^2$. Finally, Section~\ref{sec:HNL:DiracMajorana} discusses how, if such an HNL were detected in DUNE, measurements of the decay products could determine whether $N$ is a Dirac or Majorana particle, hence whether lepton-number symmetry is violated in nature.

Ref.~\cite{Ballett:2019bgd} recently explored the DUNE near detector's sensitivity to HNLs in great detail. Their analysis utilized the combined decay volume of the LArTPC and the MPD, and did not separate them for the sake of background reductions, as we do. Nevertheless, our results and theirs, when asking similar questions, are qualitatively equivalent (specifically, for example, Figs.~\ref{fig:HNL:ELimits}, \ref{fig:HNL:MuLimits}, and \ref{fig:HNL:TauLimits} in our work and Fig.~5 of Ref.~\cite{Ballett:2019bgd}).\footnote{We also emphasize that the simulation of HNL production in Ref.~\cite{Ballett:2019bgd} and our work is different for the bulk of results -- Ref.~\cite{Ballett:2019bgd} simulated HNL fluxes by reweighting the well-studied light neutrino fluxes at the DUNE near detector hall, where we simulate HNL fluxes from the parent mesons. The fact that the results agree validates the two different approaches.} Whereas Ref.~\cite{Ballett:2019bgd} focused on connecting such HNLs to the origin of the light neutrino masses, we choose a different lens and focus on diagnosing whether such HNLs are Dirac or Majorana particles.

\subsection{Production of Heavy Neutral Leptons}\label{sec:HNL:Production}

Refs.~\cite{Gorbunov:2007ak,Atre:2009rg,Alekhin:2015byh,Ballett:2016opr,Drewes:2018gkc,Abe:2019kgx,Ballett:2019bgd,Krasnov:2019kdc} have explored the possibility of HNL production in fixed-target experiments such as DUNE in some detail. In such a scenario, the relevant production mechanisms are leptonic decays of mesons, either via two-body decays (such as $\pi^+ \to \mu^+ N$) or three-body decays (such as $K^+ \to \pi^0 e^+ N$). Section~\ref{sec:Simulation} discusses how we simulate the generation of relevant mesons (in this case, $\pi^\pm$, $K^\pm$, $D^\pm$, and $D_s^\pm$) and the two- and three-body decays. We also include secondary production from the decays of $\tau^\pm$ produced in two-body $D^\pm$ and $D_s^\pm$ decays, discussed in some detail below.

For large enough beam energy, the direct production of $N$ via $q q^\prime \to W^* \to \ell N$ (where $q$ and $q^\prime$ are partons) may be considered. Additionally, $B$ mesons can also be produced in the target, contributing further to HNL production. Given that DUNE uses 120 GeV protons on target, however, HNL production from both of these channels is negligibly small.

The possible decay channels resulting in a flux of $N$ will depend on which coupling ($U_{eN}$, $U_{\mu N}$, or $U_{\tau N}$) is considered -- we discuss each case in turn below. In general, we express the two-body meson decay branching fractions into $N$ via
\begin{equation}\label{eq:HNL:TwoBodyDecay}
\mathrm{Br}(\mathfrak{m}^+ \to \ell_\alpha^+ N) = \mathrm{Br}(\mathfrak{m}^+ \to \ell_\alpha^+ \nu ) \left(\frac{|U_{\alpha N}|^2}{1-|U_{\alpha N}|^2}\right) \rho_N \left( \frac{m_{\ell_\alpha}^2}{m_\mathfrak{m}^2}, \frac{M_N^2}{m_\mathfrak{m}^2} \right),
\end{equation}
where $\mathrm{Br}(\mathfrak{m}^+ \to \ell_\alpha^+ \nu)$ is the two-body branching fraction into SM neutrinos, and $\rho_N(x,y)$ is a function that accounts for the available phase space in the decay, including helicity suppression.\footnote{Note that $\rho_N(x,y)$ can be greater than one, indicating helicity \textit{enhancement} relative to light neutrino production, where the $N$ prefers to emerge from the two-body $\mathfrak{m^+}$ decay right-handed.} This function is
\begin{equation}
\rho_N(x,y) = \frac{\left( x+y - (x-y)^2 \right) \sqrt{1 +x^2 + y^2 - 2(x + y + xy)}}{x(1-x)^2}.
\end{equation}
These decays are isotropic in the rest frame of the parent meson. We also calculate the polarization asymmetry of $N$ stemming from this decay; we will discuss this and how it affects the determination of whether $N$ is a Dirac or Majorana fermion in Section~\ref{sec:HNL:DiracMajorana}. The expressions for three-body decays are more complicated than Eq.~(\ref{eq:HNL:TwoBodyDecay}) --- we direct the reader to Refs.~\cite{Bondarenko:2018ptm,Krasnov:2019kdc} for details. See Section~\ref{sec:Simulation} for the approximations made in our three-body decay simulations and their validity.

\begin{figure}
    \centering
    \includegraphics[width=1.0\linewidth]{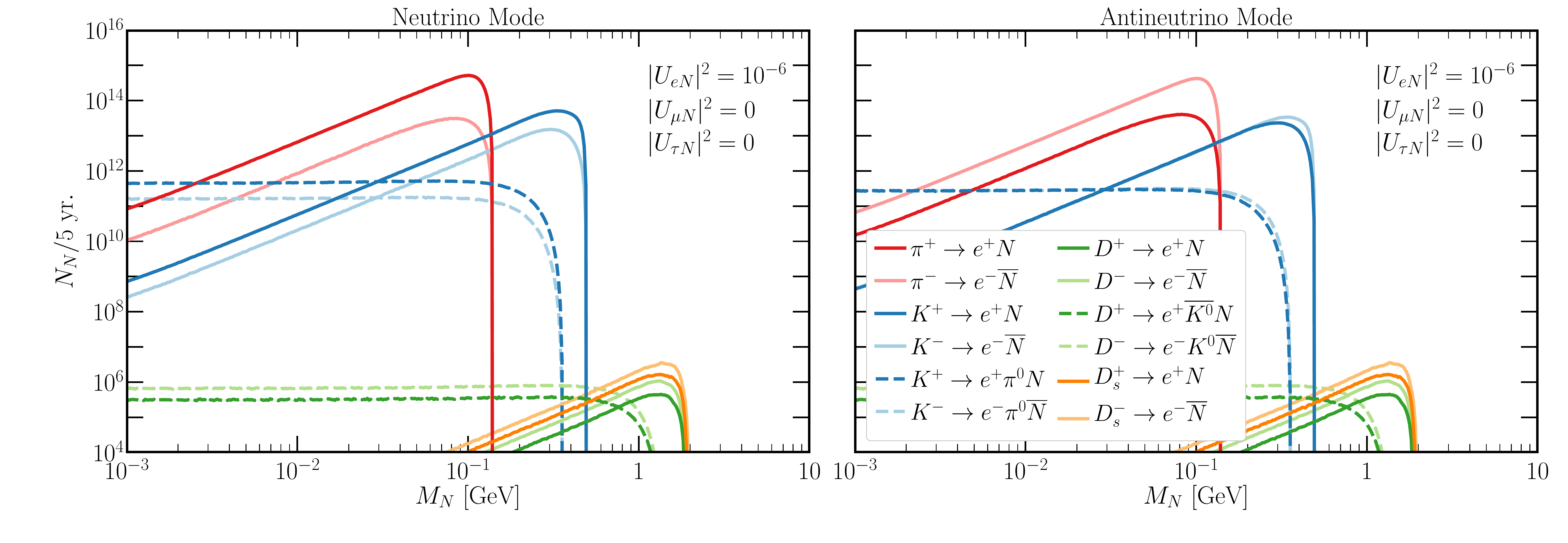}
    \caption{Number of electron-coupled heavy neutral leptons $N$ traveling toward the DUNE MPD as a function of mass $M_N$, assuming five years of data collection in neutrino mode (left) or antineutrino mode (right). We have fixed $|U_{eN}|^2 = 10^{-6}$. Different meson-decay contributions are indicated by different colors: pions (red), kaons (blue), $D$ mesons (green), and $D_s$ mesons (orange). For each meson, darker colors indicate positively-charged meson decays, where fainter colors indicate negatively-charged meson decays. Dashed curves indicate three-body meson decays.}
    \label{fig:HNL:ElectronProduction}
\end{figure}

\textbf{Production with electron coupling:} For HNLs that couple only to electrons via $|U_{eN}|^2$, the $N$ production processes of interest are as follows. For simplicity, we focus our discussion on positively charged mesons decaying, which would be the dominant production channels in neutrino mode at DUNE. The charge-conjugated processes dominate in antineutrino mode.

The meson decays considered are $\pi^+ \to e^+ N$, $K^+\to e^+ N$, $K^+ \to \pi^0 e^+ N$, $D^+ \to e^+ N$, $D^+ \to \pi^0 e^+ N$, $D^+ \to \overline{K^0} e^+ N$, $D_s^+ \to e^+ N$,  the most dominant channels for every possible $M_N$ considered. Fig.~\ref{fig:HNL:ElectronProduction} displays the expected number of $N$ that, when produced, are traveling in the direction of the MPD as a function of $M_N$, assuming five years of data collection in (anti)neutrino mode for the left (right) panel.
We take $|U_{eN}|^2 = 10^{-6}$ here. For all production mechanisms, the darker color corresponds to the decay of a positively charged meson into an HNL with $L = +1$ (assuming lepton number is a good quantum number), where the lighter color corresponds to a negatively charged meson decay into an HNL with $L = -1$. Dashed curves indicate three-body decays.

\textbf{Production with muon coupling:} As in the electron-coupled channel, we consider decays when only $|U_{\mu N}|^2$ is nonzero. The list is as follows: 
\begin{figure}
    \centering
    \includegraphics[width=1.0\linewidth]{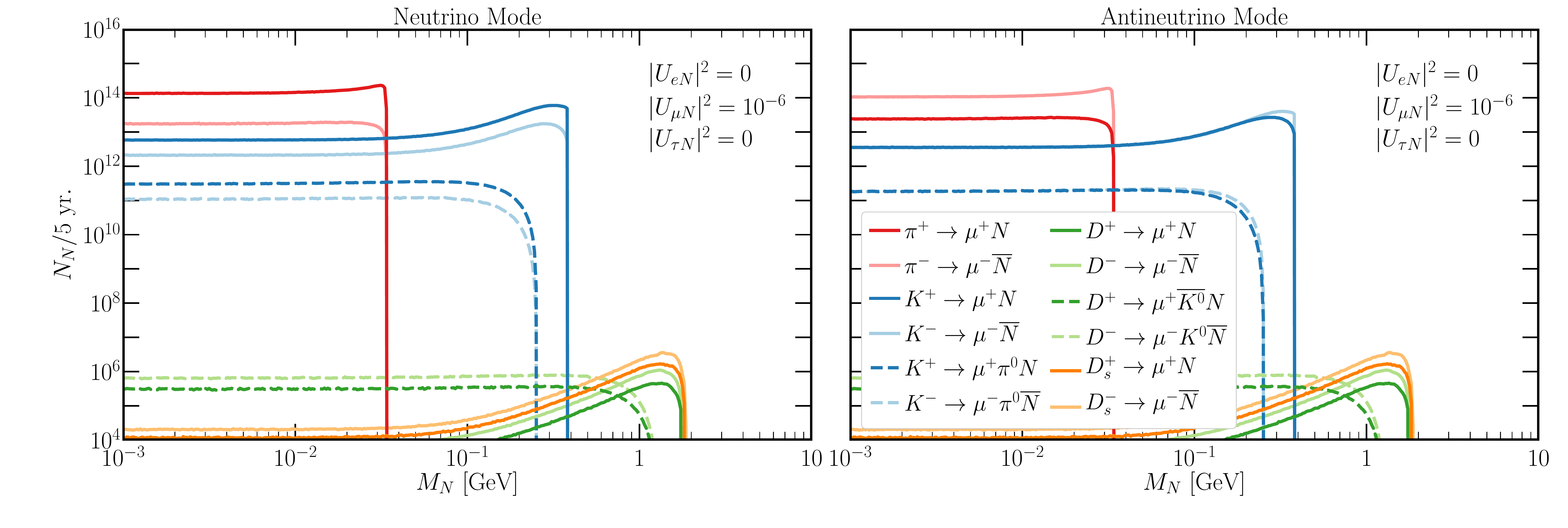}
    \caption{Number of muon-coupled heavy neutral leptons $N$ traveling toward the DUNE MPD as a function of mass $M_N$, assuming five years of data collection in neutrino mode (left) or antineutrino mode (right). We have fixed $|U_{\mu N}|^2 = 10^{-6}$. Different meson-decay contributions are indicated by different colors: pions (red), kaons (blue), $D$ mesons (green), and $D_s$ mesons (orange). For each meson, darker colors indicate positively-charged meson decays, where fainter colors indicate negatively-charged meson decays. Dashed curves indicate three-body meson decays.}
    \label{fig:HNL:MuonProduction}
\end{figure}
$\pi^+ \to \mu^+ N$, $K^+\to \mu^+ N$, $K^+ \to \pi^0 \mu^+ N$, $D^+ \to \mu^+ N$, $D^+ \to \pi^0 \mu^+ N$, $D^+ \to \overline{K^0} \mu^+ N$, $D_s^+ \to \mu^+ N$. The $N$ flux from each of these contributions is depicted in Fig.~\ref{fig:HNL:MuonProduction}. Again, we separate HNL production in neutrino mode (left panel) from antineutrino mode (right panel).

\begin{figure}
    \centering
    \includegraphics[width=0.6\linewidth]{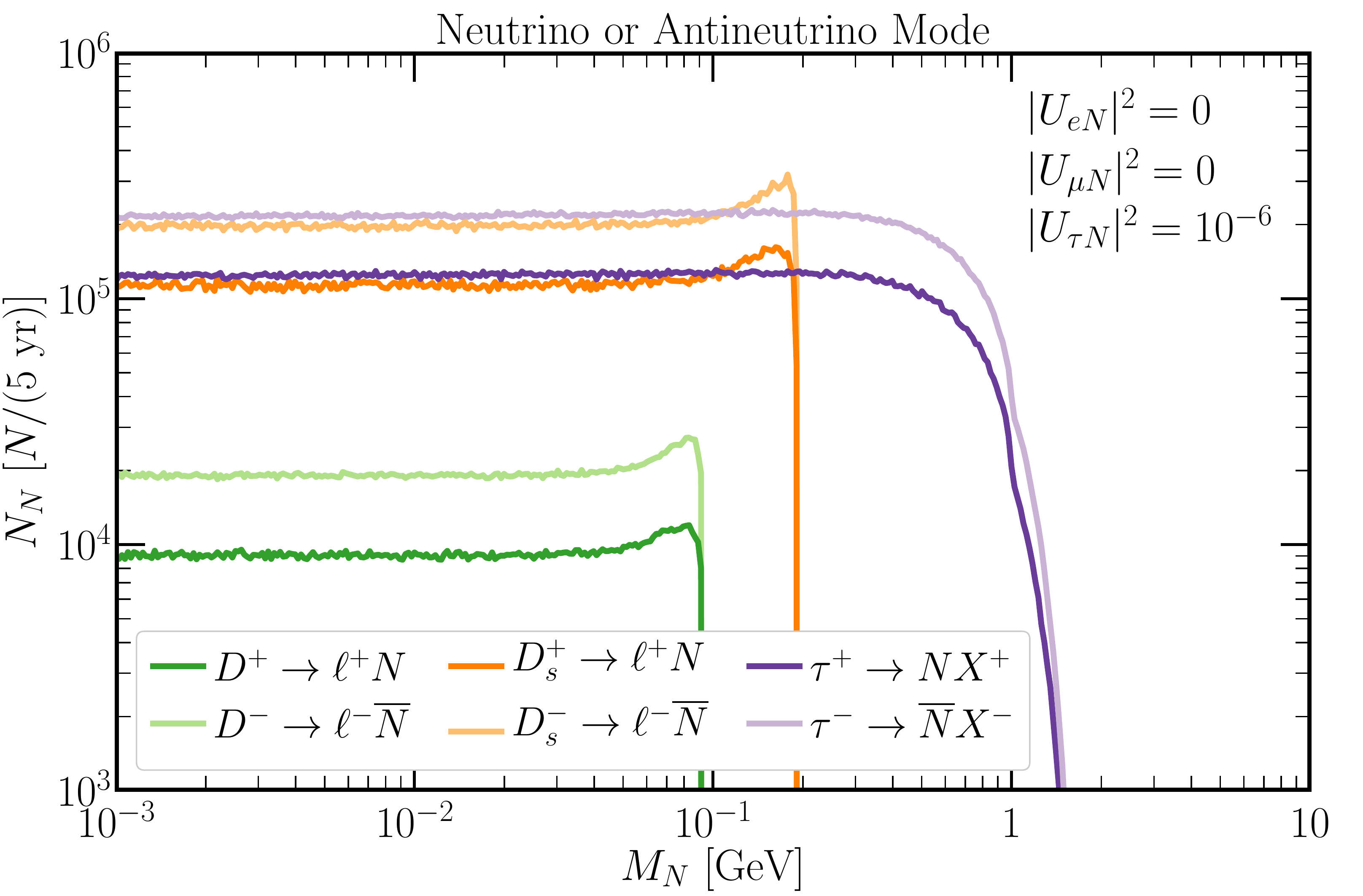}
    \caption{Number of tau-coupled heavy neutral leptons $N$ traveling toward the DUNE MPD as a function of mass $M_N$, assuming five years of data collection. We have fixed $|U_{\tau N}|^2 = 10^{-6}$.  The expected rate is independent of which mode the beam is operating in; all parent particles are unaffected by the focusing magnets. The different meson-decay contributions are indicated by different colors: $D$ mesons (green), $D_s$ mesons (orange), and secondary $\tau$ decays (purple). Dark colors indicate positively-charged particle decays, where faint ones indicate negatively-charged particle decays.}
    \label{fig:HNL:TauProduction}
\end{figure}

\textbf{Production with tau coupling:} Lastly, we list the meson decays in which HNLs are produced when the only nonzero coupling is $|U_{\tau N}|^2$. Because these decays involve the $\tau$ lepton, we need only consider decays of mesons with mass greater than $m_\tau$: $D^+ \to \tau^+ N$, $D_s^+ \to \tau^+ N$. Because $D_s$ is heavier than $D$ and its branching fraction into $\tau^+ \nu_\tau$ is larger, $D_s$ decays contribute more than $D$ decays; see the orange and green curves in Fig.~\ref{fig:HNL:TauProduction}. We also consider decays of secondary $\tau^\pm$ produced in two-body $D/D_s$ decays. Here, we assume that both the $D/D_s$ and subsequent $\tau$ decays are sufficiently prompt that the magnetic horns focus neither. We consider the following decays of $\tau$ leptons: 
$\tau \to N \pi$, $\tau \to N K$, $\tau \to N \rho$, $\tau \to N \nu_e e$, $\tau \to N \nu_\mu \mu$.
The number of $\tau$-produced HNLs corresponds to the purple curves in Fig.~\ref{fig:HNL:TauProduction}. These decays provide sensitivity to significantly heavier $\tau$-coupled HNL than $D/D_s$ decays on their own --- nearly up to the $\tau$ mass. Because the focusing has no effect on the $D$ mesons or $\tau$ leptons, Fig.~\ref{fig:HNL:TauProduction} represents the $N$ flux for five years of data collection in either neutrino or antineutrino mode.

\subsection{Decays of Heavy Neutral Leptons}
\label{sec:HNL:Decays}

Depending on the flavor to which an HNL couples and $M_N$, it will be able to decay into a variety of final states. At the DUNE MPD, several of these final states will either be fully invisible or swamped by backgrounds. We classify these decay channels as irrelevant and will combine their branching fractions/decay widths for simplicity. Our relevant final states are those with \textit{at least one charged lepton}.

Refs.~\cite{Gorbunov:2007ak,Berryman:2017twh,Bondarenko:2018ptm} give the partial widths of HNL decays of interest. For a given $|U_{\alpha N}|^2$ and $M_N$, we determine the available decay channels, the total rest-frame lifetime of $N$ and its branching fractions into channels of interest. We list the decay channels of interest for each coupling below.

If HNLs are Dirac particles so that lepton number is conserved, then the $N$ produced from positively charged meson decay (along with $\ell_\alpha^+$) have lepton number $+1$ and their decays must conserve lepton number. For instance, the decay $N \to \mu^- \pi^+$ would exist, but $N \to \mu^+ \pi^-$ would be forbidden. If $N$ is a Majorana fermion so that lepton number is violated, then both decays would occur, each with the same rate as that for $N \to \mu^- \pi^+$ when $N$ is a Dirac particle. When presenting sensitivities,\footnote{In the literature, assuming Majorana $N$ is more common, rendering comparisons between our results and existing limits a little more indirect.} we assume $N$ is a Dirac fermion, which corresponds to a more conservative reach to lower $|U_{\alpha N}|^2$ than if it were a Majorana fermion. We will discuss the nuances of the differences between Dirac and Majorana HNLs in Section~\ref{sec:HNL:DiracMajorana}.

\subsubsection{Relevant and Irrelevant Decay Channels}\label{HNL:subsec:Decays}

\begin{figure}[!htbp]
\centering
\includegraphics[width=0.6\linewidth]{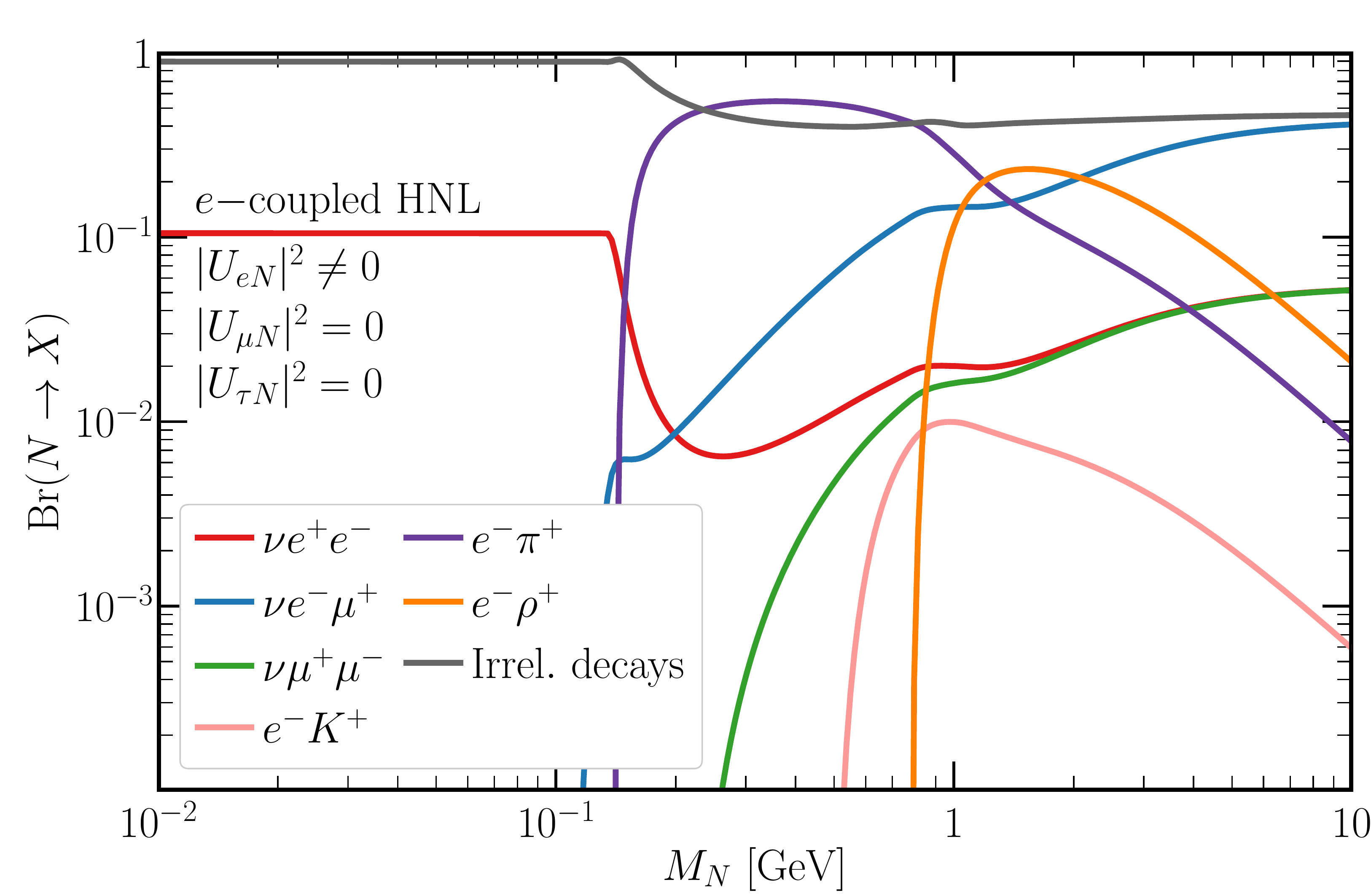}
\caption{The branching ratio of an electron-coupled HNL $N$ into different final states as a function of its mass $M_N$. The grey curve is a collection of decays considered to be irrelevant, as described in the text. \label{fig:HNL:BranchingFractionsE}}
\end{figure}

In creating the following list of possible decays, we have continued to assume for simplicity that only one of the mixing matrix elements $|U_{eN}|$, $|U_{\mu N}|$, and $|U_{\tau N}|$ is nonzero at a time. In the list, the daughter $\nu$ is one of the three light neutrino mass eigenstates, $\nu_1$, $\nu_2$, or $\nu_3$. Of course, in practice, this daughter neutrino will not be observed. This list also assumes that $N$ is a Dirac particle, and only gives the decays for the $L = +1$ particle $N$. We include the decays of $\overline{N}$ in our simulations.

\begin{figure}[!htbp]
\centering
\includegraphics[width=0.6\linewidth]{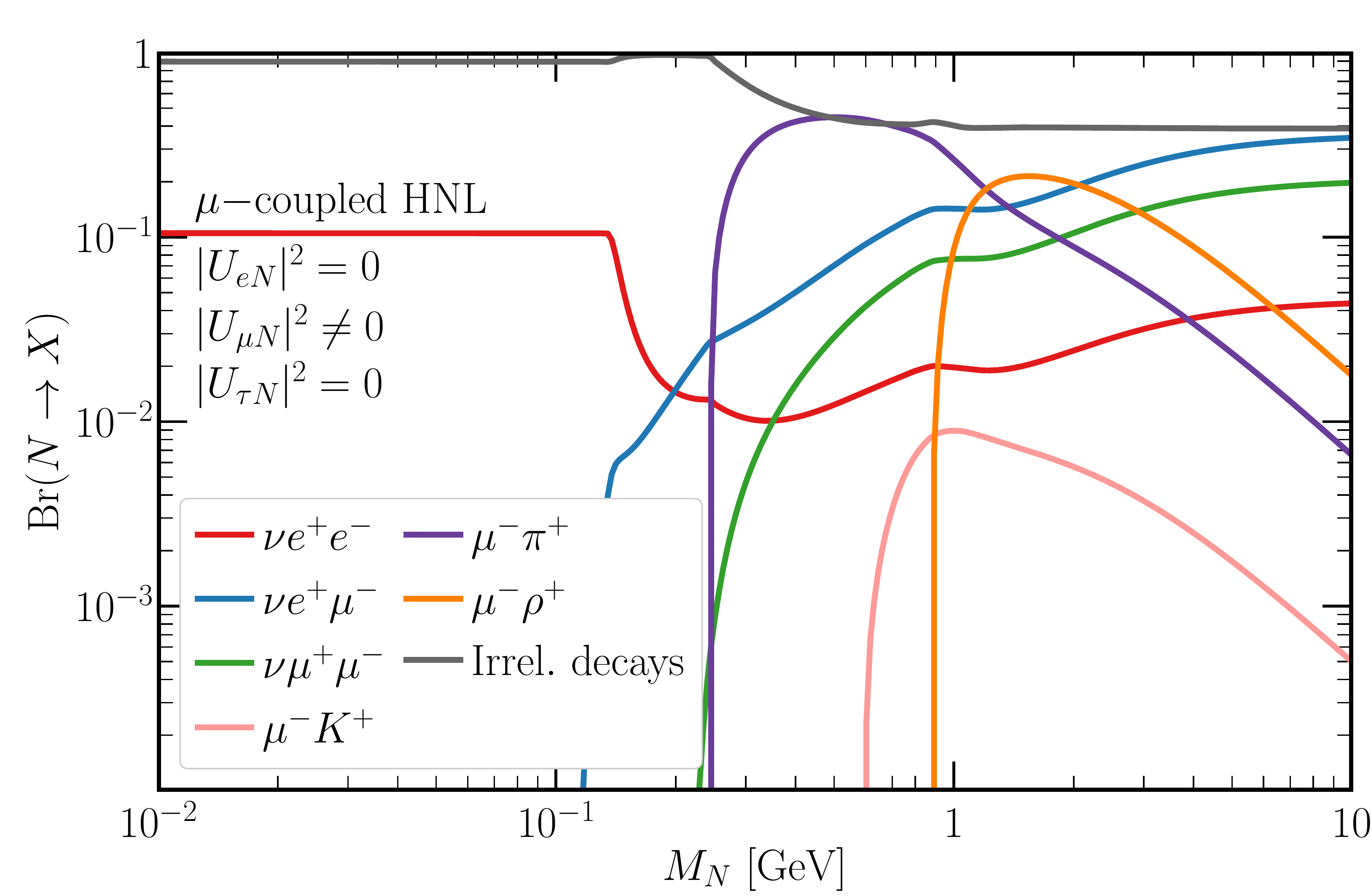}
\caption{The branching ratio of a muon-coupled HNL $N$ into different final states as a function of its mass $M_N$. The grey curve is a collection of decays considered to be irrelevant, as described in the text. \label{fig:HNL:BranchingFractionsMu}}
\end{figure}

\begin{figure}[!htbp]
\centering
\includegraphics[width=0.6\linewidth]{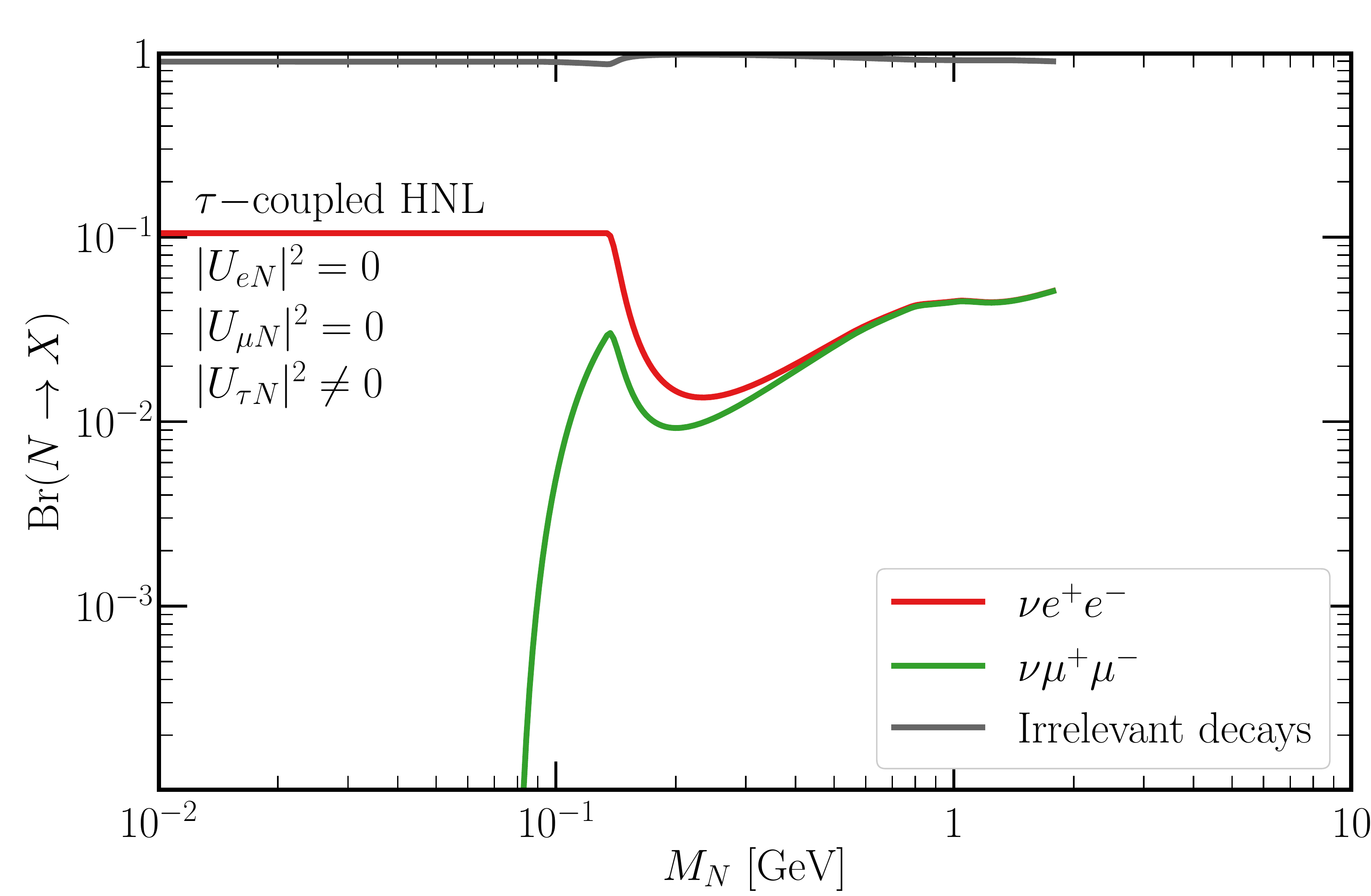}
\caption{The branching ratio of a tau-coupled HNL $N$ into different final states as a function of its mass $M_N$. The grey curve is a collection of decays considered to be irrelevant, as described in the text. Because we do not consider any $M_N > m_\tau$, we do not present any branching fractions in this region.\label{fig:HNL:BranchingFractionsTau}}
\end{figure}
\noindent \textbf{If \boldmath{$|U_{eN}|$} is nonzero:}

\textit{Relevant}: $N\to \nu e^+ e^-$, $N \to e^- \nu \mu^+$, $N \to \nu \mu^+ \mu^-$, $N \to e^- \pi^+$, $N \to e^- K^+$, $N \to e^- \rho^+$

\textit{Irrelevant}: $N \to 3\nu$, $N \to e^- \nu \tau^+$, $N \to \nu \pi^0$, $N \to \nu \eta$, $N \to \nu \eta^\prime$, $N\to \nu \rho^0$, $N\to \nu \omega$, $N \to \nu \phi$

\noindent \textbf{If \boldmath{$|U_{\mu N}|$} is nonzero:}

\textit{Relevant}: $N \to \nu e^+ e^-$, $N\to \mu^- \nu e^+$, $N\to \nu \mu^+ \mu^-$, $N \to \mu^- \pi^+$, $N \to \mu^- K^+$, $N \to \mu^- \rho^+$

\textit{Irrelevant}: $N\to 3\nu$, $N\to \nu \pi^0$, $N\to \nu \eta$, $N\to \nu \eta^\prime$, $N\to \nu \rho^0$, $N\to \nu \omega$, $N\to \nu \phi$

\noindent \textbf{If \boldmath{$|U_{\tau N}|$} is nonzero:}

\textit{Relevant}: $N \to \nu e^+ e^-$, $N\to \nu \mu^+\mu^-$

\textit{Irrelevant}: $N \to 3\nu$, $N \to \nu \pi^0$, $N\to \nu \eta$, $N\to \nu \eta^\prime$, $N\to\nu \rho^0$, $N\to\nu \omega$, $N\to\nu \phi$.

We expect all of the decays we deemed irrelevant to be background-dominated; for instance, the $N \to \nu \pi^0$ channel will have a large NC$\pi^0$ background from incoming beam neutrinos. Additionally, since all irrelevant decays have at least one outgoing neutrino, in general the invariant mass of $N$ cannot be accurately reconstructed, meaning that this analysis handle is not available to distinguish signal from background.

Figs.~\ref{fig:HNL:BranchingFractionsE}-\ref{fig:HNL:BranchingFractionsTau} display the branching fractions into the relevant and irrelevant decay channels under the assumptions that $N$ couples to the SM via coupling only to electrons (Fig.~\ref{fig:HNL:BranchingFractionsE}), muons (Fig.~\ref{fig:HNL:BranchingFractionsMu}), and taus (Fig.~\ref{fig:HNL:BranchingFractionsTau}). The different relevant channels are in color with corresponding labels, where the irrelevant curves, which enforce that the branching fractions sum to $1$ for all masses, are in grey.

\subsection{Sensitivity to Heavy Neutral Leptons}
\label{sec:HNL:Sensitivity}

Here we discuss the sensitivity as a function of the HNL mass $M_N$ and mixing $|U_{\alpha N}|^2$ for the different possible $N$ decay channels. We also discuss existing limits in this parameter space and how DUNE will be able to improve on them.

\subsubsection{Backgrounds for Relevant HNL Decay Channels}\label{sec:HNL:Backgrounds}

We will discuss each of our relevant decay channels in turn, and their associated backgrounds.

\textbf{Searches for \boldmath{$N \to \nu e^+ e^-$}:} This final state is relevant regardless of which $|U_{\alpha N}|^2$ is nonzero and will contribute nontrivially to the resulting sensitivity of each coupling. Because of the final-state neutrino, the reconstructed $e^+ e^-$ pair will neither point perfectly in the beam direction nor reconstruct the invariant mass of $N$. However, as discussed in Section~\ref{sec:DarkPhoton}, the expected number of neutrino-related $e^+ e^-$ background events is very small for ten years of data collection.\footnote{This is true even if the signal events are produced along the decay pipe (in contrast to in the target, as was relevant in Section~\ref{sec:DarkPhoton}). Additionally, even though the neutrino carries away momentum, the $e^+ e^-$ pair should still be within a degree or so of the direction of the beam, allowing us to reduce neutrino-related backgrounds significantly.} In the DUNE MPD, roughly 20 $e^+ e^-$-trident events are also expected \cite{Ballett:2018uuc,Altmannshofer:2019zhy}. This rate, and the kinematics of such events, may be predicted in a data-driven way by using the sample of $\sim 1000$ events expected in the liquid argon TPC. Therefore, we expect that 10 signal events of $N \to \nu e^+ e^-$ are  a statistically significant sample for ten years of data collection.

\textbf{Searches for \boldmath{$N \to \nu e^- \mu^+$} or \boldmath{$\nu e^+ \mu^-$}:} As with $N \to \nu e^+ e^-$, these decays are expected to have very low backgrounds, mostly associated with the misidentification of a charged pion as an electron or muon. Particle identification would help reduce this background, as would the background reduction strategies discussed in Section~\ref{sec:DarkPhoton}. We expect that 10 signal events will correspond to a statistically significant sample in these channels.

\textbf{Searches for \boldmath{$N \to \nu \mu^+ \mu^-$}:} Like the dark photon search with $\mu^+ \mu^-$, this signal channel will be competing against $\nu_\mu$ charged-current backgrounds with single pion emission where the pion is misidentified to be a muon. In the case of the dark photon, the $\mu^+ \mu^-$ pair reconstructed the dark photon mass; here, they will not reconstruct the $N$ mass. Therefore, we would not expect to find a peak on top of a smooth background in the $\mu^+\mu^-$ invariant mass, as we would for a dark photon. We expect $\mathcal{O}(100)$ background events in ten years of data collection with the DUNE MPD. For this reason, we expect that at least 20 signal events, instead of 10 events, will correspond to a statistically significant sample in this channel.

\textbf{Searches for two-body final states:} The remainder of our relevant channels consist of decays of $N$ with two particles in the final state, one being the charged lepton and the other being a charged meson. While certain channels will have relatively large neutrino-related scattering backgrounds (for instance, $\nu_e$ CC1$\pi$ scattering as a background for $N \to e^- \pi^+$), the invariant mass reconstruction of $N$ and the direction of the total final-state particle momentum will allow for significant background suppression. Channels with a final-state meson that decays within the detector volume may allow for more distinguishing signatures. For instance, the channel $N \to \mu^- \rho^+$ will result in the $\rho^+$ decaying to $\pi^+ \pi^0$. This final state would be relatively easy to reconstruct. In the DUNE MPD,  $\pi^0$s decay to photons which deposit their energy in the ECAL, allowing the $\pi^+ \pi^0$ system to reconstruct the $\rho^+$ mass; the $\mu^- \pi^0 \pi^+$ system would have an invariant mass peak corresponding to the $N$ mass of interest. As with most other relevant channels, we expect 10 signal events to be statistically significant here.

\subsubsection{Existing Limits}
For each nonzero $|U_{\alpha N}|^2$, we compare the sensitivity reach of the DUNE MPD against existing limits. We assume that $U_{\alpha N}$ is the only coupling between the HNL and the SM; HNL production and decay are then completely determined by $M_N$ and $|U_{\alpha N}|^2$. Additional interactions between the HNL and SM particles or interactions between $N$ and other new particles modify the $N$ lifetime, and therefore its probability of decaying in the detector, in a nontrivial way.

Existing limits for electron-coupled HNL come from a variety of probes. For $M_N \lesssim 30$ MeV, the strongest constraints come from searches for deviations from predicted nuclear $\beta$ decay~\cite{Galeazzi:2001py,Hiddemann:1995ce,Belesev:2013cba,Holzschuh:1999vy,Holzschuh:2000nj,Derbin:1997ut,Schreckenbach:1983cg,Deutsch:1990ut} and $\pi\to e\nu$~\cite{Britton:1992xv,Britton:1992pg} rates and spectra; Ref.~\cite{deGouvea:2015euy} provides a thorough review of constraints on heavy neutrinos under minimal theoretical assumptions, see also \cite{Bryman:2019ssi,Bryman:2019bjg}. For larger $M_N$, the strongest constraints come from the PIENU~\cite{Aguilar-Arevalo:2017vlf}, PS191~\cite{Bernardi:1985ny,Bernardi:1987ek}, T2K~\cite{Abe:2019kgx}, JINR~\cite{Baranov:1992vq}, CHARM~\cite{Bergsma:1985is}, and BEBC~\cite{CooperSarkar:1985nh} experiments. Above $M_N \gtrsim 2$ GeV, the strongest constraints come from DELPHI~\cite{Abreu:1996pa} and lepton universality measurements~\cite{deGouvea:2015euy}. Recently, precise measurements of the ratio Br($\pi^+ \to e^+ \nu_e$)/Br($\pi^+\to \mu^+ \nu_\mu$) have allowed for improved limits for $M_N \lesssim 60$ MeV~\cite{Bryman:2019ssi,Bryman:2019bjg}. For $M_N$ between roughly $150$ MeV and $450$ MeV, we include a preliminary limit from the NA62 experiment~\cite{NA62Talk} that is stronger than any existing ones in this range.

For muon-coupled HNL, many of the same probes apply. In addition to the probes discussed for $e$-coupled HNL, the NuTeV experiment~\cite{Vaitaitis:1999wq} contributes limits for 400 MeV $ \lesssim M_N \lesssim$ 2 GeV. The Brookhaven E949 experiment provides constraints in the ${\sim}175$ MeV - ${\sim}300$ MeV region with precision measurements of the decay $K\to\mu\nu$~\cite{Artamonov:2009sz}. Measurements of the $\mu$ Michel decay spectrum~\cite{deGouvea:2015euy} and searches for $\pi \to \mu N$ at PSI~\cite{Daum:1987bg} provide constraints at low $M_N$, $K\to\mu\nu$ decays provide constraints for masses between roughly $70$ MeV and $175$ MeV~\cite{Hayano:1982wu}. Recently, Ref.~\cite{Coloma:2019htx} proposed a search using atmospheric-produced HNL that can travel to and decay inside Super-Kamiokande and Ref.~\cite{Arguelles:2019ziu} proposed a similar search using IceCube. Moreover, the MicroBooNE collaboration has recently reported a constraint in Ref.~\cite{Abratenko:2019kez}. These last three limits are slightly weaker than those discussed above.

HNLs that couple purely to the tau are relatively less constrained. Existing searches from CHARM~\cite{Orloff:2002de} and DELPHI~\cite{Abreu:1996pa} provide the most stringent constraints for the mass ranges of interest. IceCube could have sensitivity to $\tau$-coupled HNL via double-bang signatures~\cite{Coloma:2017ppo}; however, no such analysis has yet been performed by the collaboration. Current $B$ factories could also reanalyze data to search for such HNLs using kinematical arguments~\cite{Kobach:2014hea}. The existence of such HNLs can also be investigated in the NA62 experiment~\cite{Drewes:2018gkc}, the forthcoming FASER experiment~\cite{Kling:2018wct,Ariga:2018uku}, or the proposed SHiP experiment~\cite{SHiP:2018xqw,Tastet:2019nqj}.

\subsubsection{Expected DUNE MPD Sensitivity}
We present the expected sensitivity of the DUNE MPD assuming only one nonzero $|U_{\alpha N}|^2$ at a time. The search channels are as listed in Section~\ref{HNL:subsec:Decays} and the number of signal events deemed statistically significant is as discussed in Section~\ref{sec:HNL:Backgrounds}. For simplicity, we show only the search channels which drive the overall sensitivity at DUNE -- the collection of channels that dominate the others for some range of $M_N$. For a given $M_N$ of interest, a combined search for all kinematically-accessible channels would lead to slightly greater sensitivity than what we present.
\begin{figure}[!h]
    \centering
    \includegraphics[width=0.75\linewidth]{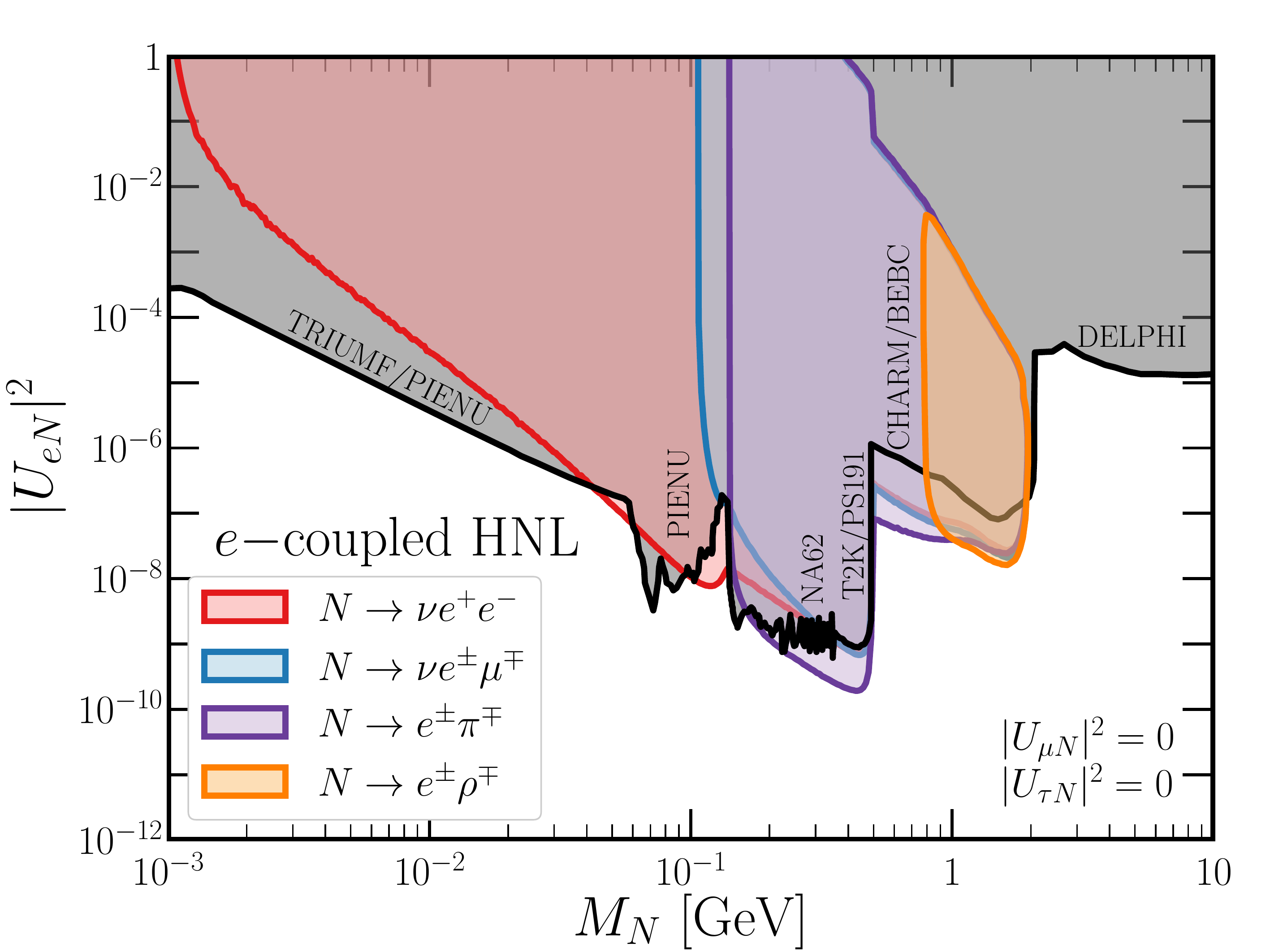}
    \caption{Expected sensitivity of the DUNE MPD to an electron-coupled HNL, assuming ten years of data collection (equal time in neutrino and antineutrino modes), assuming that the HNL is a Dirac fermion. Specific decay channels are considered: $N/\overline{N}\to \nu e^+ e^-$ (red), $N/\overline{N} \to \nu e^\pm \mu^\mp$ (blue), $N/\overline{N} \to e^\pm \pi^\mp$ (purple), and $N/\overline{N}\to e^\pm \rho^\mp$ (orange). Each channel corresponds to ten signal events, as discussed in Section~\ref{sec:HNL:Backgrounds}. See text for more detail.}
    \label{fig:HNL:ELimits}
\end{figure}

As discussed above, in determining the number of signal events for a given point in parameter space, and therefore the DUNE MPD sensitivity, we assume that the HNL is a Dirac fermion. If it were a Majorana fermion, then its total decay width would be a factor of two larger, leading to twice as many signal events for a given point in parameter space.\footnote{This assumes the HNL is longer-lived than the distance to the detector.} The difference in sensitivity under the Dirac and Majorana assumptions was explored in Ref.~\cite{Ballett:2019bgd}. In Section~\ref{sec:HNL:DiracMajorana}, we will explore how a discovered HNL could be diagnosed to be Dirac or Majorana.

Fig.~\ref{fig:HNL:ELimits} depicts the expected DUNE MPD sensitivity to an electron-coupled HNL assuming ten years of data collection with equal time in neutrino and antineutrino modes. The channels included here are $N, \overline{N} \to \nu e^+ e^-$ (red); $N \to \nu e^- \mu^+$ or $\overline{N} \to \overline{\nu} e^+ \mu^-$ (blue); $N \to e^- \pi^+$ or $\overline{N} \to e^+ \pi^-$ (purple); and $N \to e^- \rho^+$ or $\overline{N} \to e^+ \rho^-$ (orange). Three of these search channels are sensitive to regions of the parameter space that lie beyond what is constrained by existing limits (shown in grey) for some range of $M_N$. Most notably, in the absence of a discovery, the $N \to e^\pm \pi^\mp$ channel (which should produce a relatively clean signal in the DUNE MPD) will improve limits on $|U_{eN}|^2$, compared to those from CHARM and BEBC, by roughly an order of magnitude.

\begin{figure}
    \centering
    \includegraphics[width=0.75\linewidth]{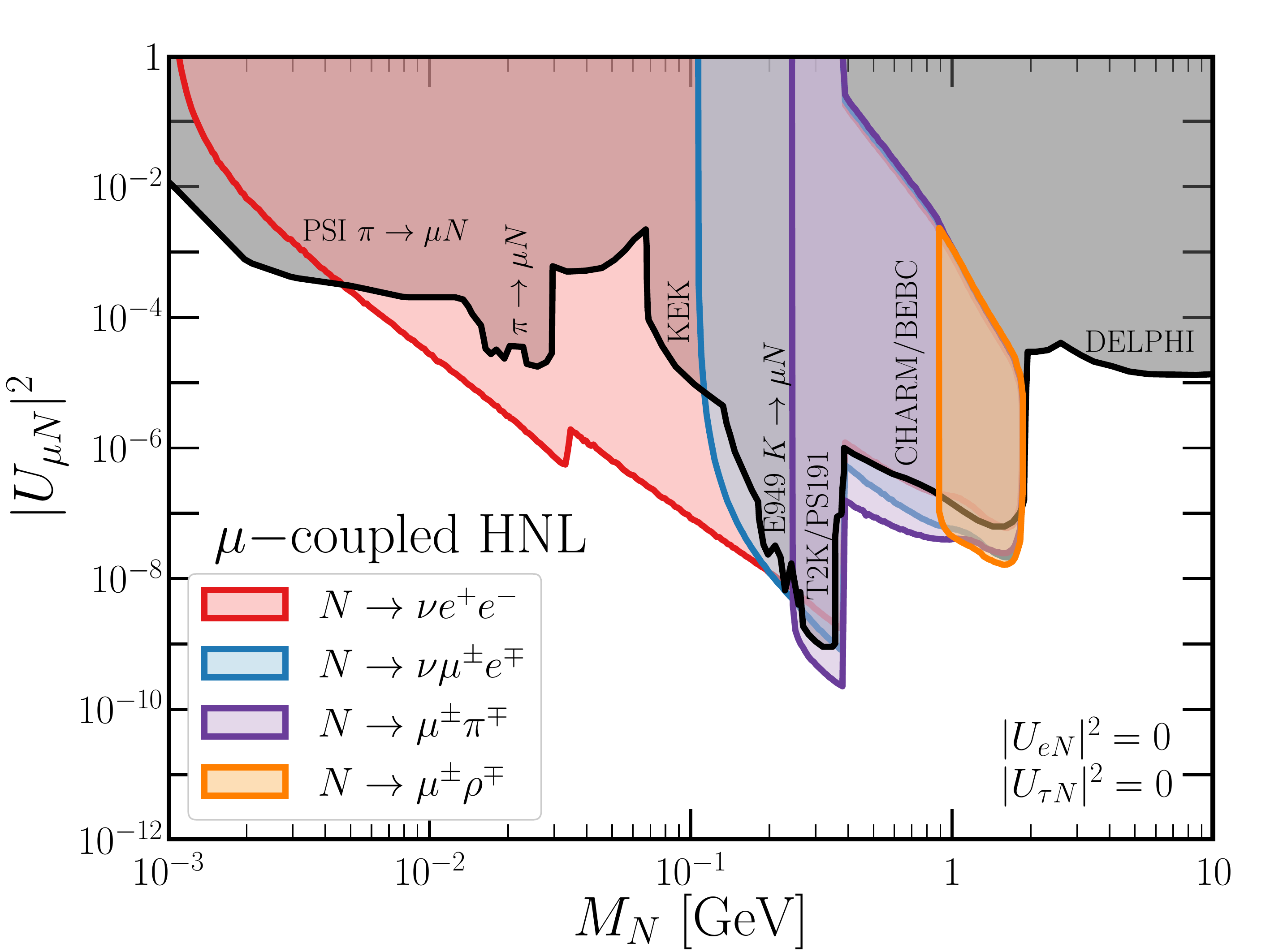}
    \caption{Expected sensitivity of the DUNE Multi-Purpose Near Detector to a muon-coupled heavy neutral lepton, assuming ten years of data collection (equal time in neutrino and antineutrino modes). We assume that the HNL is a Dirac fermion. Specific decay channels are considered: $N/\overline{N}\to \nu e^+ e^-$ (red), $N/\overline{N} \to \nu \mu^\pm e^\mp$ (blue), $N/\overline{N} \to \mu^\pm \pi^\mp$ (purple), and $N/\overline{N}\to \mu^\pm \rho^\mp$ (orange). Each channel corresponds to ten signal events, as discussed in Section~\ref{sec:HNL:Backgrounds}. See text for more detail.}
    \label{fig:HNL:MuLimits}
\end{figure}

Fig.~\ref{fig:HNL:MuLimits}, depicts the expected DUNE MPD sensitivity if the HNL couples to the muon.
The channels of interest are similar to those in Fig.~\ref{fig:HNL:ELimits} -- $N\to \nu e^+ e^-$ (red), $N \to \nu \mu^\pm e^\mp$ (blue), $N \to \mu^\pm \pi^\mp$ (purple), and $N \to \mu^\pm \rho^\mp$ (orange). As with the electron-coupled HNL, in the absence of a discovery, searches in the DUNE MPD for a muon-coupled HNL will improve on existing limits for a wide range of $M_N$. We highlight the sensitivity to low-mass $M_N$ driven by the search for $N \to \nu e^+ e^-$. We discuss this search channel and how an HNL decaying in this way could be diagnosed to be a Dirac or Majorana fermion in Section~\ref{sec:HNL:DiracMajorana}. 

\begin{figure}
    \centering
    \includegraphics[width=0.75\linewidth]{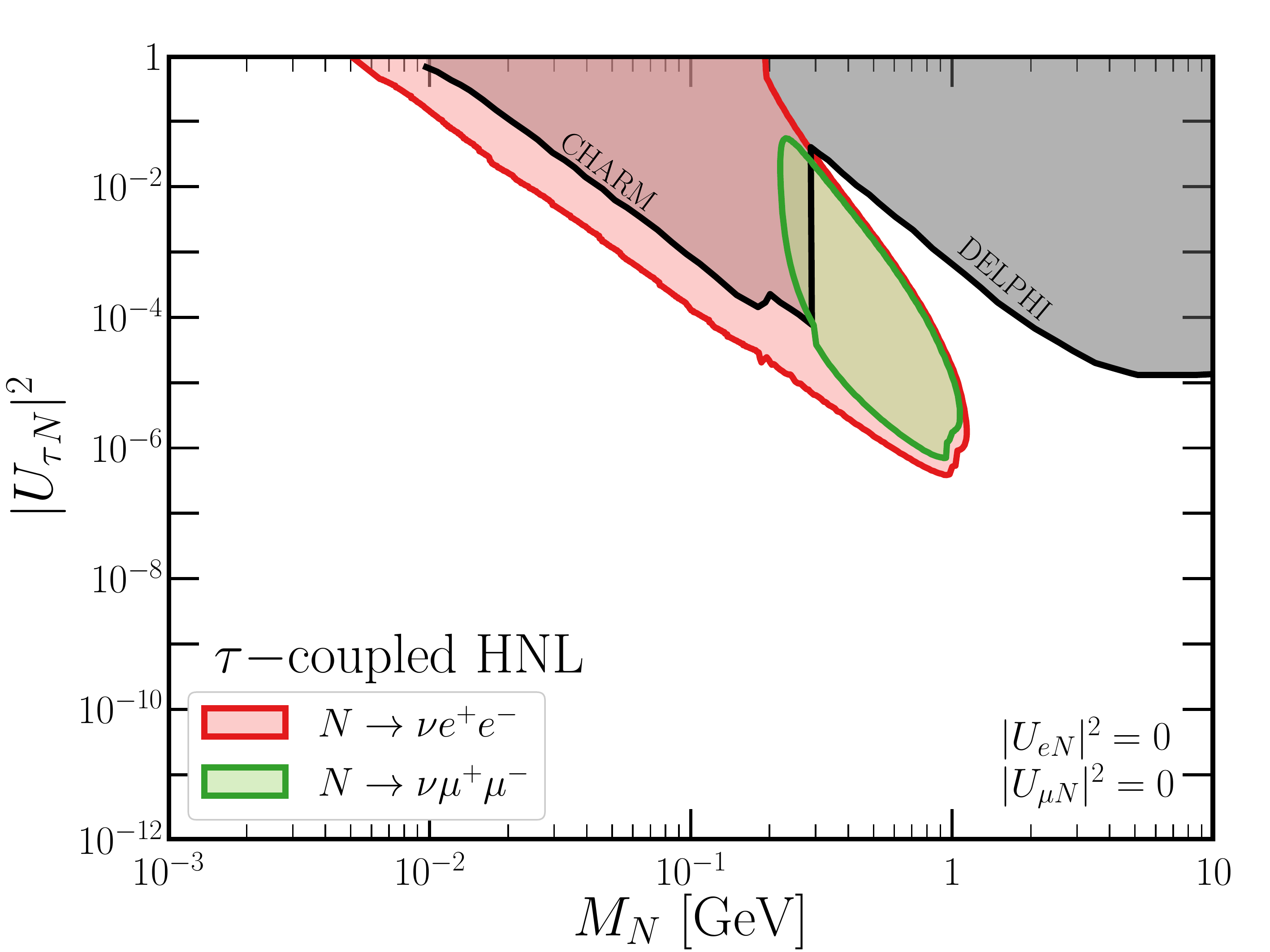}
    \caption{Expected sensitivity of the DUNE MPD to a $\tau$-coupled heavy neutral lepton, assuming ten years of data collection (how this time is apportioned between neutrino and antineutrino modes is irrelevant). We assume that the HNL is a Dirac fermion. We show sensitivity to the channels $N/\overline{N}\to \nu e^+ e^-$ (red) and $N/\overline{N}\to \nu \mu^+ \mu^-$ (green), compared against existing limits from CHARM~\cite{Orloff:2002de} and DELPHI~\cite{Abreu:1996pa}. For the final state $\nu e^+ e^-$, the red region encompasses parameter space for which ten signal events are expected, whereas for the final state $\nu \mu^+ \mu^-$, the green region indicates twenty signal events, due to background considerations. See text for more detail.}
    \label{fig:HNL:TauLimits}
\end{figure}

Finally, Fig.~\ref{fig:HNL:TauLimits} displays the expected sensitivity to $\tau$-coupled HNL at the DUNE MPD. Because no production mechanism exists for HNLs with $M_N > m_\tau$, we have no access to any final states with $\tau^\pm$ and a charged meson. The DUNE MPD sensitivity to a tau-coupled HNL is independent of how the data collection time is divided between neutrino and antineutrino modes since the production mechanisms are not impacted by the magnetic focusing horns. The two channels we display are $N \to \nu e^+ e^-$ (red) and $N \to \nu \mu^+ \mu^-$ (green). Due to background considerations discussed above, the $N \to \nu \mu^+ \mu^-$ curve corresponds to 20 signal events.
In the absence of a discovery, the $N\to \nu e^+ e^-$ channel will provide a more powerful constraint than existing CHARM constraints~\cite{Orloff:2002de}, and a combination of both channels will allow us to search for tau-coupled HNL in a window where the CHARM and DELPHI~\cite{Abreu:1996pa} searches do not have much sensitivity. This same window can be probed, as discussed above, by IceCube~\cite{Coloma:2017ppo}, $B$ factories~\cite{Kobach:2014hea}, NA62~\cite{Drewes:2018gkc}, and FASER~\cite{Kling:2018wct}. Such searches would be complementary in probing this gap with the DUNE MPD.

\subsection{Deducing the Nature of the HNL -- Dirac vs.~Majorana}\label{sec:HNL:DiracMajorana}\setcounter{footnote}{0}

Above, we assumed that HNLs are Dirac fermions and that lepton number is conserved. In that case, when $N$ is assumed to be the HNL produced in association with a positively charged lepton and having its own lepton number $L=1$, its decays always contain negatively charged leptons or neutrinos. On the other hand, if HNLs are Majorana fermions, then they will also violate lepton number in their decays, i.e., the $N$ produced via positively charged meson decay may then decay, for instance, into either $N \to \mu^- \pi^+$ or $N \to \mu^+ \pi^-$, with equal probability. Since the magnetized DUNE MPD has charge- and particle-identification capabilities, it will be able to search for differences between the rates of decay into these final states. We will show in Section~\ref{sec:HNL:DM:Charged} that perfect charge/particle identification are not required to perform this analysis.

As explored in Refs.~\cite{Balantekin:2018azf,Balantekin:2018ukw, Tastet:2019nqj}, Majorana HNLs also differ from Dirac ones in their decay kinematics. When the HNLs are produced in weak-interaction meson decays, they will in general have a net polarization, potentially leading to asymmetry in their rest-frame decay angular distributions. Different rest-frame angular distributions will be reflected in different decay-product energy spectra in the lab frame. Even with a charge-blind detector, these different spectra could be distinguished.

We explore the potential of determining whether HNLs that are discovered by DUNE are Dirac or Majorana fermions. Two classes of HNL final states are considered: those with unobservable lepton number due to the presence of a neutrino in the final state (e.g., $N \to \nu e^+ e^-$ or $N \to \nu \pi^0$) and those with only charged particles, where the total lepton number may be measured (e.g., $N\to \mu^\pm \pi^\mp$).

Because of leptonic mixing, if we can establish experimentally that an HNL $N$ is a Majorana particle, then it will follow that \emph{all} neutrinos, including the light ones, are Majorana particles. For, if $N$ is a Majorana particle, its exchange will lead, via mixing, to diagrams that give Majorana masses to all the neutrino mass eigenstates, and once these mass eigenstates have Majorana masses, it follows that they will be Majorana particles. Thus, establishing that an HNL $N$ is a Majorana particle would have far-reaching implications.

\subsubsection{Final States with Unobservable Lepton Number}\label{sec:HNL:DM:Uncharged}
If there are SM neutrinos in the final state of an HNL decay, then it is not possible to determine the state's total lepton number, and counting these events will not readily indicate whether the HNL is a Dirac or Majorana fermion. Instead of identifying the final state's lepton number, one must rely exclusively on kinematic information. 

\textbf{Searches for \boldmath{$N\to \nu \pi^0$}:} Consider, for instance, the decays $N \to \nu \pi^0$ and $\overline{N} \to \overline{\nu} \pi^0$ for a Dirac HNL, and the decay $N \to \nu \pi^0$ for a Majorana HNL. In principle, the distributions of $\pi^0$ lab-frame energies and angles relative to the beam direction are different between these two hypotheses; a large enough sample of decays could allow these distributions to be distinguished. For illustrative purposes, we explore this decay channel in detail here, although, as discussed in Section~\ref{HNL:subsec:Decays}, we expect this channel to have a large background from neutral-current events.

If the neutrinos, including $N$, are Majorana particles, the decays $N \to \nu \pi^0$ will produce an isotropic distribution of $\pi^0$s in the rest frame of $N$~\cite{Balantekin:2018ukw}. In contrast, for Dirac $N$, the decay $N \to \nu \pi^0$ is maximally anisotropic~\cite{Balantekin:2018ukw}. However, the DUNE beam is not a perfect environment for such searches -- since there will be some contamination of $\overline{N}$ particles in the beam. Even if $N$ is a Dirac fermion, the $\overline{N}$ decay to $\overline{\nu}\pi^0$, with the opposite anisotropy, will hinder the separation between the Majorana and Dirac fermion hypotheses.
\begin{figure}
    \centering
    \includegraphics[width=0.7\linewidth]{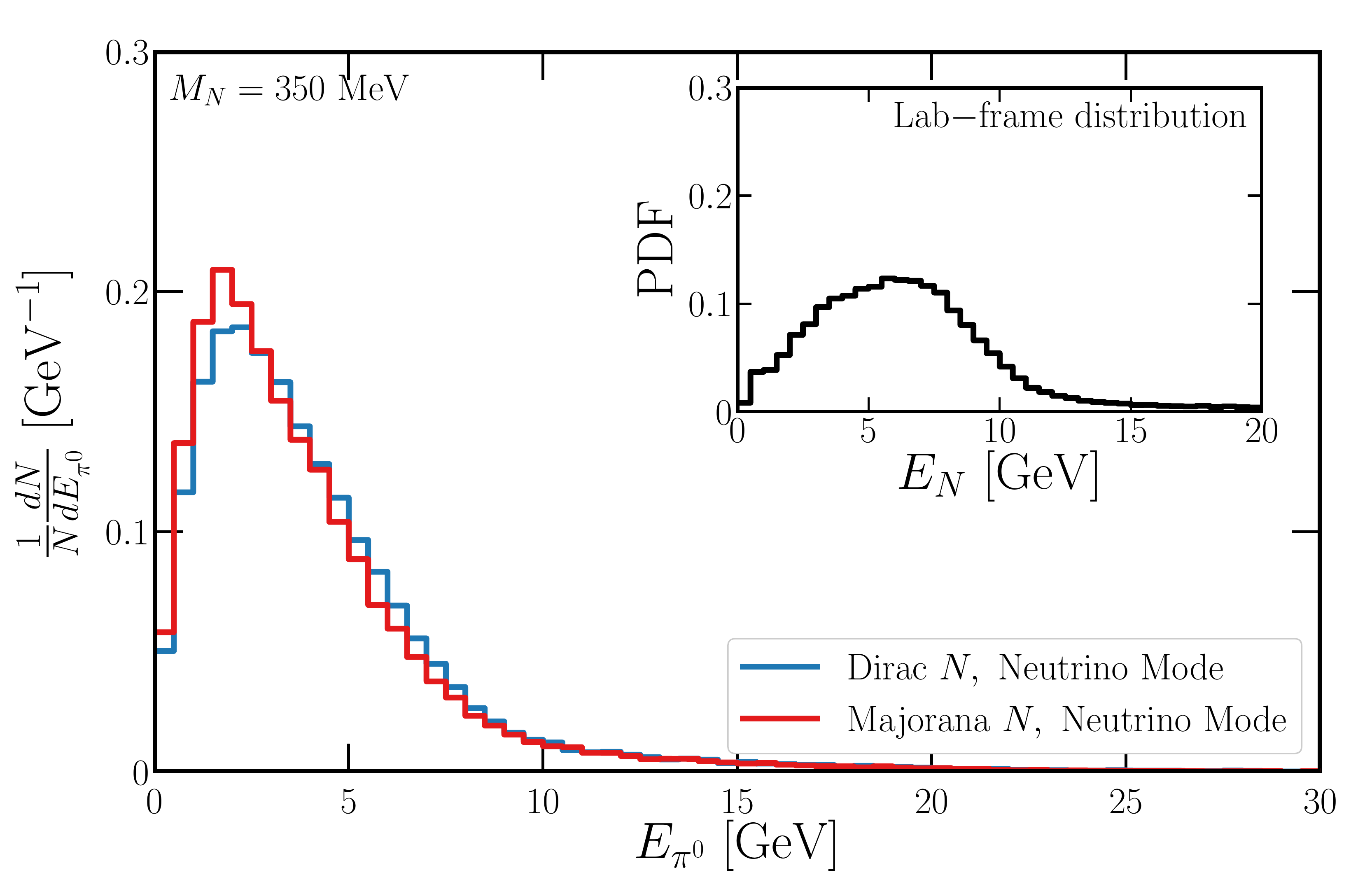}
    \caption{Expected distribution (area normalized) of HNL decays to $\nu \pi^0$ and $\overline{\nu} \pi^0$ as a function of lab-frame $\pi^0$ energy, assuming operation in neutrino mode at the DUNE MPD. For all cases we consider $M_N = 350$ MeV. The red line depicts this distribution for Majorana $N$ decays, in which the $N$ rest-frame decay is isotropic. The blue line depicts this distribution for Dirac $N$ and $\overline{N}$ decays, in which the rest-frame decay is maximally anisotropic. The inset shows (black) the lab-frame distribution of $N$ energies prior to their decay.}
    \label{fig:HNL:Pi0}
\end{figure}
For concreteness, we explore the kinematics of such decays assuming $M_N = 350$ MeV. We determine the flux (and polarization) of HNLs from $K^+$ and $K^-$ decays when the DUNE facility is operated in neutrino mode.\footnote{The results of this analysis in antineutrino mode are qualitatively equivalent.} When simulating the HNL decays, we assume they are either Majorana (i.e., the decay products are emitted isotropically in the rest frame) or Dirac (i.e., the decay products are emitted anisotropically in the rest frame) fermions, and so determine the lab-frame distribution of $\pi^0$ energies.

These resulting lab-frame $\pi^0$ energy distributions are depicted in Fig.~\ref{fig:HNL:Pi0} for $M_N = 350$ MeV; the Majorana (Dirac) hypothesis results are depicted in red (blue). The inset depicts the lab-frame distribution of $N$ energies as they travel to the detector. We note that the HNLs are highly boosted; most of them have energies between $5$ and $10$ GeV, corresponding to boost factors of ${\sim}10{-}20$. It is apparent in Fig.~\ref{fig:HNL:Pi0} that an extremely large sample of events (even neglecting background events) would be required to use this sample to distinguish between Dirac and Majorana HNLs.

The authors of Ref.~\cite{Balantekin:2018ukw} explored this decay channel as a possible means of distinguishing Dirac from Majorana HNLs in a toy scenario. Our analysis differs from theirs in two main ways. 
%First, we consider the contamination of $\overline{N}$ while operating in neutrino mode. \pjf{reword} 
First, we considered the decays of both $N$ and $\overline{N}$, instead of either one or the other. We did, however, determine that the impact of this wrong-lepton-number contamination is not very significant.
%However, we have generated the analogous version of Fig.~\ref{fig:HNL:Pi0} if we could ignore this contribution, and it looks nearly identical to Fig.~\ref{fig:HNL:Pi0}. 
%However, we have performed an analysis in which only one of these is considered at a time; the resulting figure looks nearly identical to Fig.~\ref{fig:HNL:Pi0}. 
Second, and more importantly, the toy analysis performed in Ref.~\cite{Balantekin:2018ukw} considered a flux of $N$ with boost factors between, roughly, $2{-}4$, significantly lower than what would occur for a $350$ MeV HNL at DUNE. %This lower boost factor causes the resulting $E_{\pi^0}$ distributions to be more distinguishable.
More strongly boosted HNLs yield predicted energy distributions that are substantially more difficult to distinguish.

We conclude from this that, if DUNE discovers an HNL that is produced predominantly from charged kaon decay, then such beam-neutrino environments are not ideal for studying the intrinsic nature of the HNL mass in the final state $\nu \pi^0$. Scenarios in which $N$ are generated from only one charge of kaon, and in which the $N$ are significantly less boosted, would be significantly more propitious. We leave the study of such scenarios to future work.

\textbf{Searches for \boldmath{$N\to \nu e^+ e^-$}:}  In Section~\ref{sec:HNL:Sensitivity}, we showed that the DUNE MPD is sensitive to a large region of currently unexplored parameter space by searching for the decay channel $N \to \nu e^+ e^-$. This is particularly true for the scenarios in which $N$ couples only to the muon (Fig.~\ref{fig:HNL:MuLimits}) or only to the tau (Fig.~\ref{fig:HNL:TauLimits}). Compared to the final state $\nu \pi^0$, this has significantly lower background. While it suffers from having unidentifiable lepton number, its kinematics are different for Dirac and Majorana $N$. Ref.~\cite{Ballett:2019bgd} explored this decay in more detail.
We anticipate that, because the relative angular distributions for Dirac/Majorana $N$ are not as discrepant (in the $N$ rest frame) as in the decay $N \to \nu \pi^0$, such Majorana vs. Dirac fermion differences would be difficult to measure in the DUNE environment. We leave a more detailed study of these three-body decays to future work.

%%%%%%%%%%%%%%%%%%%%%%%%%%%%%%%%%%%%%%%%%
%%%%%%%%%%%%%%%%%%%%%%%%%%%%%%%%%%%%%%%%%

\subsubsection{Final States with Only Charged Particles}\label{sec:HNL:DM:Charged}
In the previous subsection, we focused on decay modes in which the final-state lepton number is unobservable due to the presence of SM neutrinos. We now explore final states with only charged particles where, if particles \emph{and their charges} are identified, the lepton number will be determined. In a perfect world, assuming HNLs are Dirac fermions, an experiment operating in neutrino mode would produce HNLs with $L = +1$ directed toward a detector and any HNLs with $L = -1$ would be diverted away. Then, using measurements of charge/kinematics, one could deduce that the decaying HNLs have $L = +1$. Obviously, an experiment like DUNE is not perfect; even with the light neutrinos, a certain level of antineutrino contamination exists in neutrino mode, and vice versa. 

\begin{figure}
    \centering
    \includegraphics[width=\linewidth]{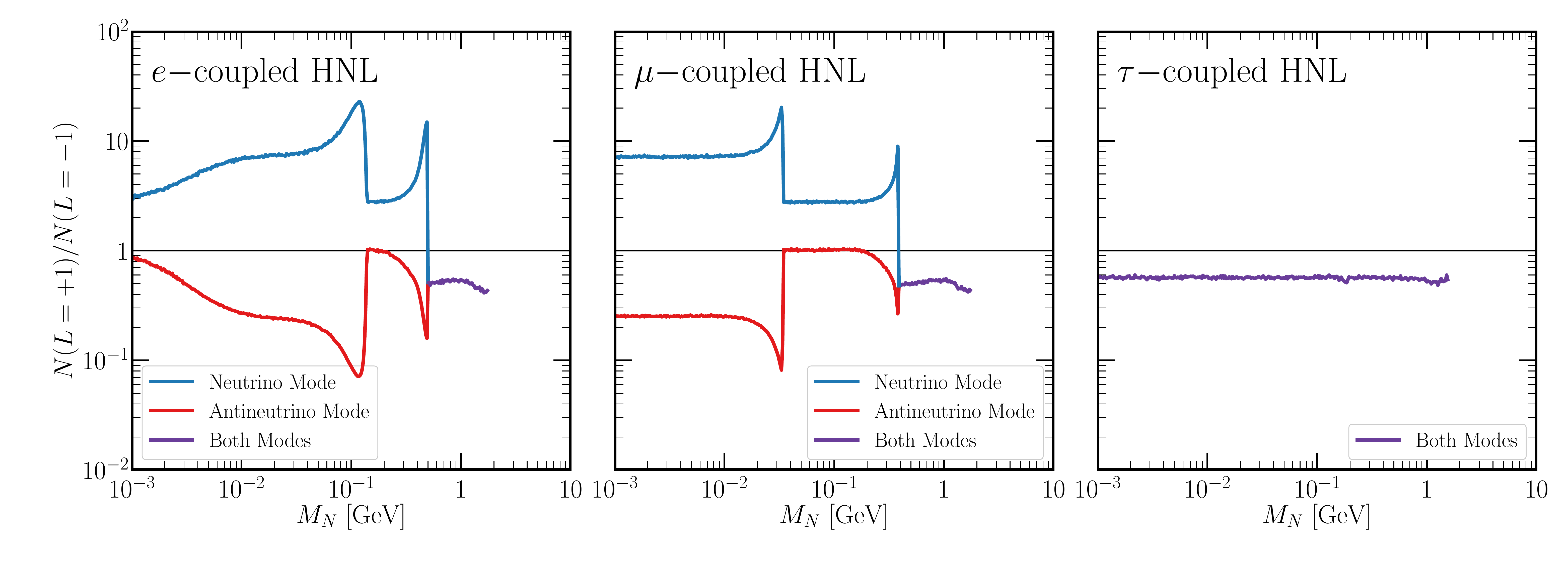}
    \caption{Expected ratio of heavy neutral lepton decays with lepton number $+1$ to those with lepton number $-1$, assuming the HNL are Dirac fermions, as a function of the mass $M_N$. The left panel is for $e$-coupled HNL, center is for $\mu$-coupled HNL, and right is for $\tau$-coupled HNL. For each panel, the blue (red) line corresponds to operation in (anti)neutrino mode. The results associated to the two operation modes overlap (purple lines) for $M_N \gtrsim 400$ MeV for the $e$- and $\mu$-coupled scenarios and for all masses in the $\tau$-coupled one.}
    \label{fig:HNL_Ratio}
\end{figure}

Assuming the HNLs are Dirac fermions, and using the production rates depicted in Fig.~\ref{fig:HNL:ElectronProduction}-\ref{fig:HNL:TauProduction}, we calculate the ratio of two fluxes as a function of $M_N$ (for each different coupling): the number of HNLs with $L = +1$ reaching the MPD to the number of HNLs with $L = -1$ also reaching the MPD. In a perfect world, this ratio is $\infty$ in neutrino mode and $0$ in antineutrino mode. Fig.~\ref{fig:HNL_Ratio} presents the ratios in neutrino mode (blue) and antineutrino mode (red) for $e$-(left), $\mu$-(center), and $\tau$-(right) coupled HNLs. When the distinction between neutrino and antineutrino modes is irrelevant (because the parent particles are unfocused), we use purple lines. If HNLs were Majorana fermions, then they would decay as if they have $L = +1$ and $L = -1$ in equal abundance --- the measured ratio would be $1$.

Several features in Fig.~\ref{fig:HNL_Ratio} are notable. First, the expected ratio typically deviates from $1$ more in neutrino mode than in antineutrino mode, making such a measurement more favorable. This is not because of the ability of the beam to focus one polarity of mesons over the other; it is simply because positively charged light mesons are produced more frequently in the beam than negatively charged ones (see Table~\ref{tab:ChargedMesonProduction} and Appendix~\ref{appendix:MesonProd}). This is not true for heavy ($D$/$D_s$) mesons, where negatively charged mesons are produced in more abundance. This explains why, for all couplings, for heavy enough $M_N$, both neutrino and antineutrino mode predict more $L=-1$ HNLs than $L=+1$ ones. Additionally, the ratio in neutrino mode is typically $\mathcal{O}(\mathrm{few})$. Even with perfect event-by-event identification of whether the HNL has $L = +1$ or $L = -1$, to determine whether the detected HNL is a Dirac (where the ratio is predicted by Fig.~\ref{fig:HNL_Ratio}) or Majorana fermion (where the ratio should be $1$) will require many events.

To demonstrate the power of measuring this ratio, consider a muon-coupled HNL with a mass $M_N = 250$ MeV. We calculate the expected number of $N \to \mu^\pm \pi^\mp$ events (at this mass, the branching ratio $N \to \mu^\pm \pi^\mp$ is approximately $20\%$ and this is the largest of the relevant channels) for five years of neutrino mode operation as a function of the mixing $|U_{\mu N}|^2$. For a given $|U_{\mu N}|^2$, we perform a number of pseudoexperiments where we assume that the HNL is a Majorana fermion which therefore decays to $\mu^- \pi^+$ and $\mu^+ \pi^-$ with equal probability. These event rates are defined as $N(L=+1)$ and $N(L=-1)$, respectively, and are drawn from a Poisson distribution. We then determine the 68.3\%, 95\%, and 99\% credible regions for the ratio $N(L=+1)/N(L=-1)$ after performing these pseudoexperiments.

\begin{figure}
    \centering
    \includegraphics[width=0.8\linewidth]{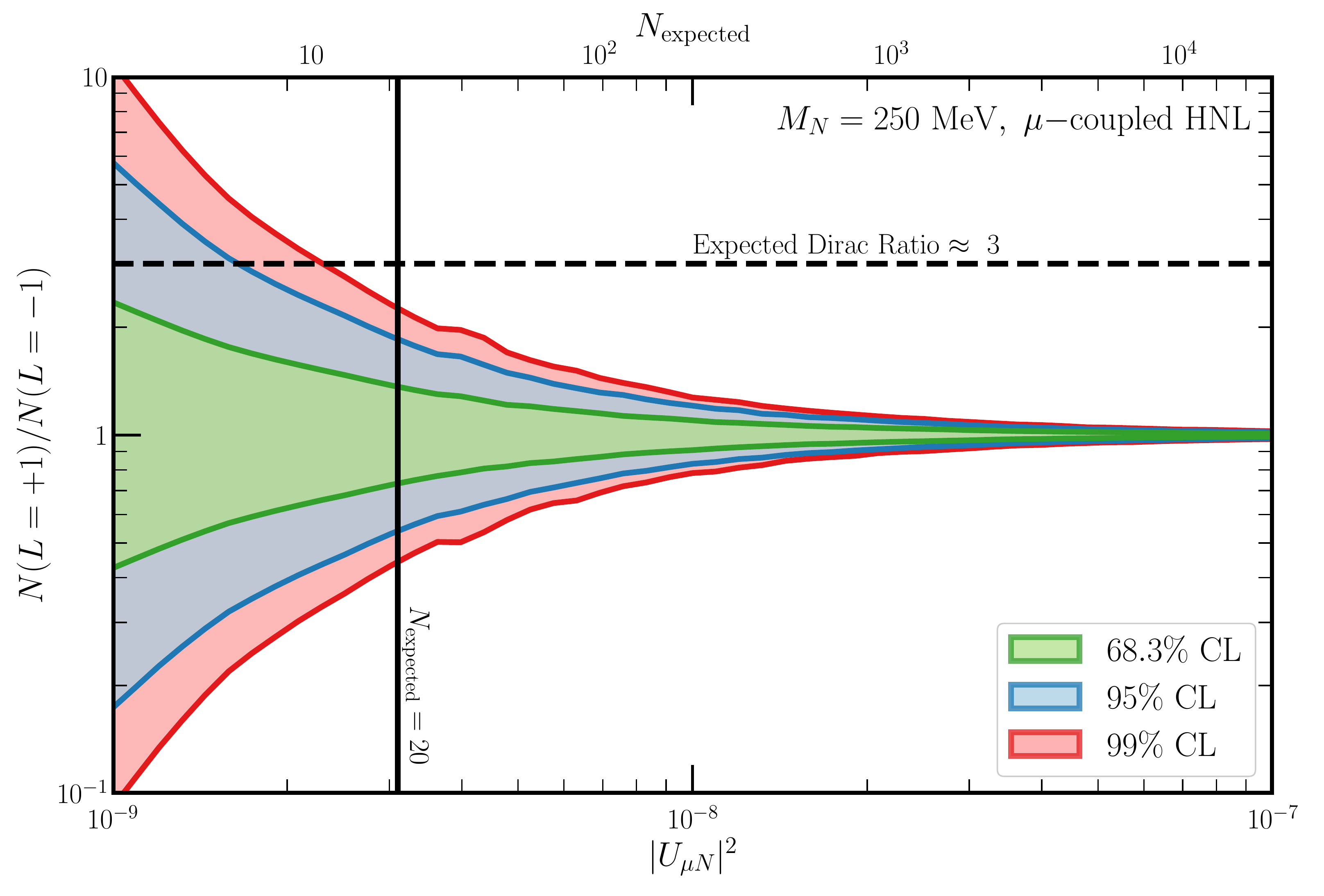}
    \caption{Expected ability to measure the ratio of HNL decays with lepton number $+1$ to those with lepton number $-1$ assuming perfect lepton-number-identification, as a function of the mixing $|U_{\mu N}|^2$ for a $M_N = 250$ MeV muon-coupled HNL. Here, we simulate data assuming $N$ is a Majorana fermion; if it were a Dirac fermion, the preferred ratio would be approximately $3$. Below $|U_{\mu N}|^2 \approx 10^{-9}$, fewer than one or two events are expected, and this procedure should not be used.  The vertical line indicates where the number of signal events is $20$. We have included only five years of data collection in neutrino mode.}
    \label{fig:HNL:RatioMeasurement}
\end{figure}

Figure~\ref{fig:HNL:RatioMeasurement} displays the result of this procedure assuming five years of data collection in neutrino mode. As $|U_{\mu N}|^2$ increases, so does the expected number of events, and it becomes easier to measure that the ratio is $1$, as predicted by the Majorana HNL hypothesis. If HNLs are Dirac fermions, and if we are operating in neutrino mode, then the expected ratio is approximately $3$ (see the center panel of Fig.~\ref{fig:HNL_Ratio}). The Dirac hypothesis can be robustly excluded (if every event can be properly identified as $L=+1$ or $L=-1)$ for $|U_{\mu N}|^2 \gtrsim 2\times 10^{-9}$. On the top axis of Fig.~\ref{fig:HNL:RatioMeasurement}, we give the number of expected signal events for a Majorana HNL given the corresponding value of $|U_{\mu N}|^2$.

Figure~\ref{fig:HNL:RatioMeasurement} demonstrates the ideal ability of the DUNE MPD to determine whether HNLs are Dirac or Majorana fermions if, on an event-by-event basis, the signal events can be perfectly identified as having lepton number $+1$ or $-1$ (by identifying the sign of the charged lepton in the final state). In the real world, a detector will misidentify these final states a finite fraction of the time. This is tantamount to redistributing the events between the samples, limiting the ability of the detector to separate the Dirac and Majorana hypotheses.

To study the extent to which the MPD could discriminate between the Dirac and Majorana hypotheses using these types of final states (without using kinematic information), we separately study the decays $N \to \mu^\mp \pi^\pm$, $N \to e^\mp \pi^\pm$, and $N\to \mu^\mp\rho^\pm$. For each final state, we calculate the expected number of decays to each charge state \emph{assuming HNLs are Majorana particles} as a function of the HNL mass and mixing.  We perform a frequentist analysis to address how distinct the Majorana-assumed data are from data generated under the Dirac hypothesis.\footnote{We also performed a Bayesian analysis and obtained qualitatively similar results.} Since such an analysis would take place only after an initial discovery of $N$, we optimistically assume that, with enough events, $M_N$ will be nearly perfectly measured. If $N$ decays completely visibly, then its mass may be measured from the invariant mass of the final state particles. While the detector is not perfect, we assume that it can attain $\mathcal{O}(10\ \mathrm{MeV})$ precision, corresponding to nearly perfect knowledge of $M_N$ as far as this analysis is concerned.

For the Majorana hypothesis, the expected numbers of $L = +1$ and $L = -1$ events are equal in both neutrino and antineutrino modes; we denote these expectations as $n_\pm^{\nu,M}$ and $n_\pm^{\bar{\nu},M}$.  For the Dirac hypothesis, there is a (mass-dependent) ratio between the expected numbers of $L = +1$ and $L = -1$ events in both running modes (see Fig.~\ref{fig:HNL_Ratio}).  We denote the numbers of events predicted by the Dirac hypothesis, as a function of the mixing $U_D=U_{\alpha N}$, as $n_\pm^{\nu,D}(U_D)$ and $n_\pm^{\bar{\nu},D}(U_D)$.  We use a likelihood ratio to compare the two hypotheses; in practice, we consider the difference in their log-likelihoods.  For observed numbers of events $N_\pm^{\nu}$ and predictions $n_\pm^\nu$, the log-likelihood is
\begin{eqnarray}
    \log {\mathcal{L}^\nu}(n_+^\nu,n_-^\nu)= &-&n_+^\nu + N_+^{\nu} \log \left(n_+^\nu\right)-\log\left(N_+^{\nu} !\right) \nonumber \\
    				     &-&n_-^\nu + N_-^{\nu} \log \left(n_-^\nu\right)-\log\left(N_-^{\nu} !\right)~,
\end{eqnarray}
and similarly for the antineutrino mode. For the Majorana hypothesis, the likelihood is clearly maximized for $n_\pm^{\nu,M} = N_\pm^{\nu}$, since this is our null hypothesis. For the Dirac hypothesis, we maximize the likelihood with respect to $U_D$ and compare the likelihood at this best-fit point, $\hat{U}_D$, to the Majorana hypothesis value.  Our test statistic for distinguishing between  the Dirac and Majorana fermion hypotheses is $-2\Delta \mathcal{L}$, with
\begin{equation}
\Delta \mathcal{L}\equiv  \log \mathcal{L}^\nu (n_\pm^{\nu,D}(\hat{U}_D)) +  \log \mathcal{L}^{\bar{\nu}}(n_\pm^{\bar{\nu},D}(\hat{U}_D)) -  \log \mathcal{L}^\nu(n_\pm^{\nu,M}) -  \log {\mathcal{L}^\nu}(n_\pm^{\bar{\nu},M})~. \label{eq:LnL}
\end{equation}
We consider the Majorana fermion hypothesis preferred over Dirac at $3\sigma$ confidence level
if $-2\Delta\mathcal{L}>9$.

Defining $r_\nu \equiv n_+^{\nu,D}/n_-^{\nu,D}$, we may re-express our test statistic, using Eq.~(\ref{eq:LnL}) as
\begin{equation}
-2\Delta\mathcal{L} = 2n_\pm^{\nu,M} \log{\left(\frac{\left(1+r_\nu\right)^2}{4r_\nu}\right)} + 2n_\pm^{\overline{\nu},M} \log{\left(\frac{\left(1+r_{\overline{\nu}}\right)^2}{4r_{\overline{\nu}}}\right)}. \label{eq:LnLAnalytic}
\end{equation}
The test statistic grows as $r_\nu$ and $r_{\overline{\nu}}$ deviate from $1$ -- it is easier to rule out the Dirac neutrino hypothesis when the expected ratio of positively charged and negatively charged leptons is more discrepant than what is expected for the Majorana neutrino hypothesis. As expected, the test statistic vanishes if $r_{\nu,\overline{\nu}} \to 1$, where there is no measurable difference between the two hypotheses.

So far, we have assumed that the charges of the final-state particles are always correctly identified.  To take into account that this may not be the case, we introduce an efficiency factor $a$ that corresponds to the fraction of decays that are correctly characterized in terms of particle species and charges. In practice, $a = 1$ corresponds to a perfect detector and $a = 0.5$ corresponds to pure guessing.  For the Majorana hypothesis, and thus the expected numbers of events, this has no effect.  In the Dirac interpretation, this will wash out any expected asymmetry.  To account for this imperfection, we make the replacements
\begin{eqnarray}
    n^{\nu,\mathrm{D}}_{+} \to a n_{+}^{\nu,\mathrm{D}} + (1-a)n_{-}^{\nu,\mathrm{D}}, \nonumber \\
    n^{\nu,\mathrm{D}}_{-} \to a n_{-}^{\nu,\mathrm{D}} + (1-a)n_{+}^{\nu,\mathrm{D}},
\end{eqnarray}
in the log-likelihood (likewise for antineutrino mode). This substitution modifies Eq.~(\ref{eq:LnLAnalytic}), which becomes
\begin{equation}
-2\Delta \mathcal{L} = 2 n_\pm^{\nu,M} \log{\left(\frac{\left(1+r_\nu\right)^2}{4\left(1 + a(r_\nu-1)\right)\left(a + (1-a)r_\nu\right)}\right)} + 2 n_\pm^{\overline{\nu},M} \log{\left(\frac{\left(1+r_{\overline{\nu}}\right)^2}{4\left(1 + a(r_{\overline{\nu}}-1)\right)\left(a + (1-a)r_{\overline{\nu}}\right)}\right)}.\label{eq:LnLAnalytic2}
\end{equation}

We present our results in terms of contours of the test statistic for a given assumption regarding the $M_N$ prior and the efficiency parameter $a$, and compare against the relevant parameter space from Figs.~\ref{fig:HNL:ELimits} and \ref{fig:HNL:MuLimits}. We focus on two cases: perfect identification ($a = 1.0$), corresponding roughly to the toy analysis presented in Fig.~\ref{fig:HNL:RatioMeasurement}; and imperfect identification ($a = 0.75$). Note that the worst-case scenario is not $a = 0$ but $a = 0.5$, in which one is effectively guessing the lepton number for each signal event. Our test statistic in Eq.~(\ref{eq:LnLAnalytic2}) vanishes for $a \to 0.5$.

\begin{figure}
    \centering
    \includegraphics[width=0.7\linewidth]{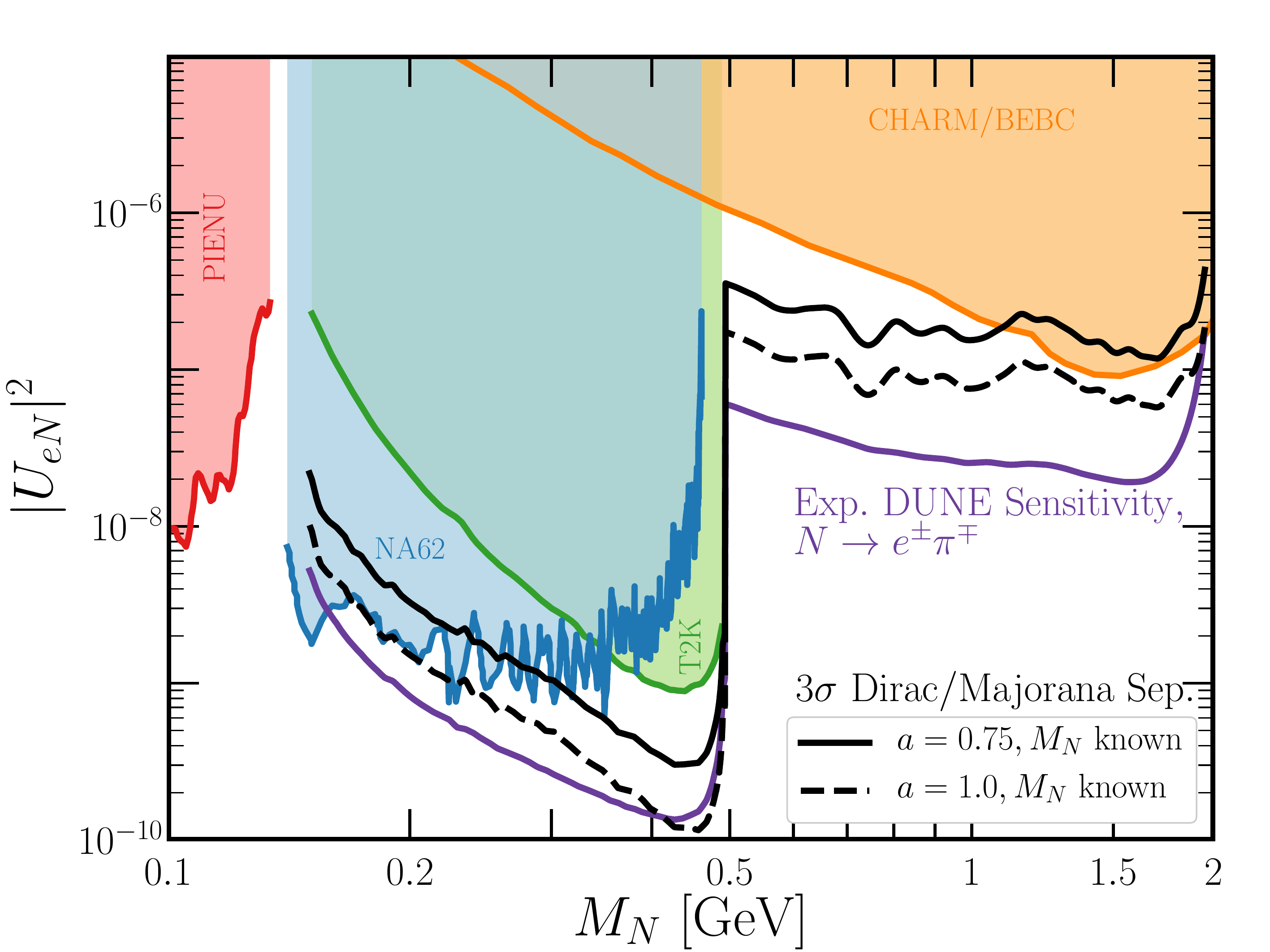}
    \caption{Expected Dirac/Majorana separation capability in the final state $N \to e^\pm \pi^\mp$. The purple curve is the expected DUNE sensitivity using the channel $N \to e^\pm \pi^\mp$. In contrast with Fig.~\ref{fig:HNL:ELimits}, this assumes that $N$ is a Majorana fermion. The dashed black curve is the 3$\sigma$ Dirac/Majorana distinction assuming perfect knowledge of $M_N$ and the efficiency factor $a = 1$. The solid black curve is $3\sigma$ Dirac/Majorana distinction assuming perfect knowledge of $M_N$ and efficiency factor $a = 0.75$.}
    \label{fig:HNL:DM:EPi}
\end{figure}
Figure~\ref{fig:HNL:DM:EPi} depicts the results of this analysis for the decay $N \to e^\pm \pi^\mp$ for an electron-coupled HNL. The purple curve shows our expected DUNE MPD sensitivity using the $N \to e^\pm \pi^\mp$ channel. In contrast to what was presented in Fig.~\ref{fig:HNL:ELimits}, we now assume (for consistency with this analysis) that $N$ is a Majorana fermion.\footnote{In practice, this means that we are sensitive to values of $|U_{\alpha N}|^4$ a factor of two (or $|U_{\alpha N}|^2$ a factor of $\sqrt{2}$) lower than in the Dirac fermion scenario.} In various colors, we show other existing constraints in the same parameter space (discussed above). The dashed, black curve shows the 3$\sigma$ contour from our analysis assuming a perfect efficiency of $a=1.0$; the solid, black curve shows the same for $a=0.75$. That the $3\sigma$ contours lie relatively close to the ten-event sensitivity is unsurprising. We note here that $a \approx 1$ should be realistic with the DUNE MPD -- there should be little confusion between identifying electron and pion tracks in the HPTPC.

Clearly, while more than ten events are required to make this distinction at moderate significance, having $\sim40$ events is sufficient; $\sim30$ decays of Dirac HNLs with $L=+1$ and $\sim10$ decays with $L=-1$ can look quite distinct from the Majorana HNL expectation of 20 for each. We note that the separation between the black and purple curves shrinks as $M_N$ increases towards the kaon-production threshold near $500$ MeV, owing to the increased ratio $N(L=+1)/N(L=-1)$. Lastly, it is particularly exciting that the parameter space occupied by these $3\sigma$ contours is as of yet largely unconstrained. Not only would the DUNE MPD have sensitivity to HNLs where other experiments have not, but it could have strong resolving power between Dirac and Majorana HNLs in this hitherto unexplored parameter space.

\begin{figure}
    \centering
    \includegraphics[width=0.7\linewidth]{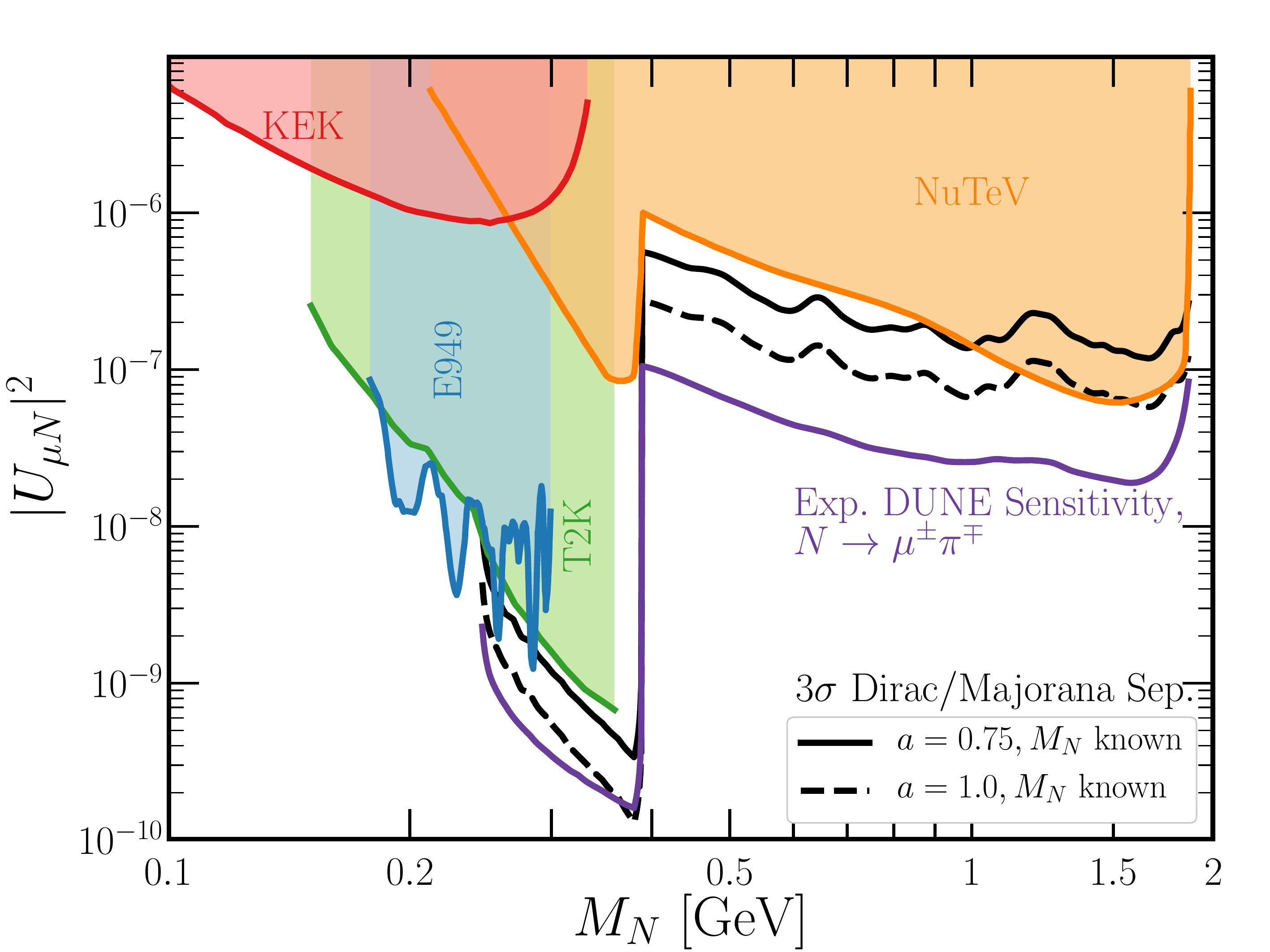}
    \caption{Expected Dirac/Majorana separation capability in the final state $N \to \mu^\pm \pi^\mp$. The purple curve is the expected sensitivity from the search $N \to \mu^\pm \pi^\mp$ assuming $N$ is a Majorana fermion (in contrast to the Dirac-neutrino assumption in Fig.~\ref{fig:HNL:MuLimits}). The dashed black curve is the 3$\sigma$ Dirac/Majorana distinction assuming perfect knowledge of $M_N$ and the efficiency factor $a = 1$. The solid black curve is $3\sigma$ Dirac/Majorana distinction assuming perfect knowledge of $M_N$ and efficiency factor $a = 0.75$.}
    \label{fig:HNL:DM:MuPi}
\end{figure}
Figure~\ref{fig:HNL:DM:MuPi} displays the results of this procedure for the final state $N \to \mu^\pm \pi^\mp$, assuming ten years of data collection, equal amounts of time in neutrino and antineutrino modes. The purple curve is the expected sensitivity curve for the DUNE MPD if $N$ is a Majorana fermion (again, assumed for consistency, in contrast to the curve presented in Fig.~\ref{fig:HNL:MuLimits}). Other existing constraints in the same parameter space are also included. As in Fig.~\ref{fig:HNL:DM:EPi}, the dashed (solid), black curve shows the 3$\sigma$ contour from our analysis assuming an efficiency of $a=1.0$ ($a=0.75$). Our analysis still indicates that ${\sim}40$ HNL decays are required in order to make a meaningful distinction, but more of the corresponding parameter space is ruled out by, in particular, T2K. For heavier, $D$-produced HNLs, we would require efficiency of at least $a = 0.75$ in order to discover the nature of $N$. In contrast to the $N\to e^\pm \pi^\mp$ search, $a \approx 1$ is significantly harder to attain because muon and pion tracks look nearly identical in the HPTPC. This result provides another reason for the need of a muon tagger for the DUNE MPD~\cite{NewGasSlides}.

We have also explored a similar study in the channel $N \to \mu^\pm \rho^\mp$. In Fig.~\ref{fig:HNL:MuLimits} we saw that this channel can improve on existing limits from CHARM and BEBC. However, we find that the regions of parameter space for which DUNE could discover an HNL and also determine its nature are already excluded. This is mostly because the predicted $\mu^+ \rho^-$ to $\mu^- \rho^+$ ratio in the Dirac-fermion case is close to 1 (see the middle panel of Fig.~\ref{fig:HNL_Ratio} where $M_N \approx 1$ GeV), meaning many events are required to differentiate between the two hypotheses.

Unfortunately, the case of a $\tau$-coupled HNL does not have a fully visible decay mode available for analysis at DUNE.  Study of the $\tau$-coupled HNL would require production of $N$ with mass $M_N>m_\tau$, which is not possible with the DUNE beam energy.  If $B$-mesons could be produced in abundance, then it would be possible to probe the Majorana versus Dirac hypotheses in the $\tau$-coupled case.

In principle, one could use additional kinematic information -- lab-frame energies, angular distributions, etc. -- to provide additional discriminating power between the Dirac and Majorana hypotheses. Specifically, the final states $N \to e^\pm \pi^\mp$ and $N \to \mu^\pm \pi^\mp$ share many of the same features regarding anisotropy as the final state $N \to \nu \pi^0$ (see Section~\ref{sec:HNL:DM:Uncharged}) when one combines the distributions of the charge-conjugated final states \cite{Balantekin:2018ukw}. This anisotropy leads to different predicted energy spectra for the lab-frame electron/muon/pion distributions. We have simulated these differences and found them to be small -- such information would only assist in the analyses of Figs.~\ref{fig:HNL:DM:EPi} and \ref{fig:HNL:DM:MuPi} if the number of signal events is already quite large.

%\afterpage{\clearpage}  
%------------------------------
%Conclusions Section
%------------------------------

\section{Conclusions}\setcounter{footnote}{0}

It is proposed that the DUNE near detector facility will include a magnetized multi-purpose detector consisting of a high-pressure gaseous argon TPC with a surrounding electromagnetic calorimeter. The components of the MPD allow for good particle identification, tracking, and momentum resolution, and the HPTPC enables measurements of particles down to low energies. The primary purpose of the MPD is to control the systematic uncertainties associated with neutrino oscillation measurements but these capabilities also lend it great power in searching for new light physics that is weakly coupled to the Standard Model.  The DUNE MPD is ideally suited to search for new particles that are light enough ($m\ltap 1$ GeV) to be copiously produced in meson decays and sufficiently weakly coupled that they have a long lifetime ($\gamma \beta\tau \sim 500$ m) and can reach the MPD.  In addition to the enhanced measurement properties of the MPD, the reduced target density means that the signals suffer from fewer beam-induced backgrounds than in the LArTPC.

New physics connected with neutrinos or dark matter is typically weakly coupled to the SM, and is long-lived.  If dark matter is GeV-scale in mass and thermally produced, then it must sit in a more complicated dark sector that contains additional light mediator states that can also be long-lived.  In this work, we have investigated the sensitivity of the MPD to sub-GeV dark mediators or heavy neutral leptons.  We have focused on models where these new states are coupled to SM degrees of freedom through one of the so-called renormalizable portals: the vector, scalar, or neutrino portals of Eq.~\eqref{eq:portals}.  Light particles with these couplings can be produced in the decays of mesons, predominantly neutral and charged pions and kaons, and then subsequently decay in the gaseous detector, often into charged final states.  In each case analyzed, we have also estimated the expected background rates and determined regions of parameter space where the new physics signal would be large enough to be seen above backgrounds.

In the case of kinetic mixing of a dark photon with the SM (Section \ref{sec:DarkPhoton}), we find that after ten years of data taking, the DUNE MPD will have sensitivity to presently untested regions of parameter space.  Due to different target-detector separations, DUNE is sensitive to a complementary region of parameter space to other future experiments like SHiP or SeaQuest.   For dark photon masses ranging from $200$ MeV to $1$ GeV and kinetic mixing parameter $10^{-8}\ltap \epsilon\ltap 10^{-7}$, the DUNE MPD will observe at least 10 decays of dark photons to $e^+e^-$, $\mu^+\mu^-$, or $\pi^+\pi^-$ final states, and, for some parts of parameter space, as many as 100 such decays.  The excellent measurement capabilities of the gaseous argon detector, coupled with the surrounding ECAL and magnet, allow the backgrounds from neutrino scattering events to be adequately suppressed.  

A related model that also has a sizable decay rate into charged leptons is a leptophilic gauge boson (Section \ref{sec:Leptophilic}).  Unlike the case of a kinetically-mixed $U(1)$, a leptophilic gauge boson has additional production through the leptonic decays of charged mesons.  There is a complicated interplay between which pair of lepton-flavor numbers is gauged and the mass of the vector which determines the expected number of observable decays in the DUNE near detector.  Over the accessible mass range, only the gauging of $L_\mu-L_\tau$ is expected to probe a region of parameter space not presently probed by terrestrial experiments.  For $L_\mu-L_\tau$ with gauge coupling between $10^{-6}$ and $10^{-4}$ and vector mass below $10$ MeV, DUNE could see 3-10 $V\rightarrow e^+ e^-$ events.  Should a leptonic resonance be discovered at DUNE, the ability to distinguish electrons from muons/pions will give excellent model discrimination.

The scalar portal coupling between the SM Higgs boson and a SM-singlet scalar $\varphi$ leads to scalar-Higgs mixing and allows for production of light scalars (Section \ref{sec:DarkHiggs}) through a loop decay of kaons.  Although the kinematics of production and decay are similar to the vector models, the scalar decays to the heaviest state available and thus, again, particle identification capabilities will be a great benefit for model discrimination.  For $50\ \mathrm{MeV}\ltap M_\varphi\ltap 400\ \mathrm{MeV}$, DUNE can make considerable improvements over existing bounds by searching for a resonance in the invariant mass distributions of charged leptons or pions.

The final portal we considered was heavy neutral leptons $N$ mixing with SM neutrinos through the neutrino portal (Section \ref{sec:HNL}).  The possible search channels for a heavy neutral lepton (HNL) include two- and three-body decays, and depend on whether $N$ is a Dirac fermion (and thus lepton number is conserved) or a Majorana fermion.  Considering HNLs with couplings to only one lepton flavor at a time, we find that the DUNE MPD has the potential to improve upon existing bounds by up to an order of magnitude, in the case of $e$ or $\tau$ coupling, and by up to two orders of magnitude for $\mu$ coupling.  The magnetic field acting on the MPD allows good charge identification and we find that, from the relative numbers of $N\rightarrow \ell^+ \pi^-$ and $N\rightarrow \ell^- \pi^+, \, (\ell=e,\mu)$ decays, it is possible to achieve $3\sigma$ discrimination between the Dirac and Majorana fermion hypotheses, after five years of data taken in neutrino mode.  We predict that learning the Dirac/Majorana nature of neutrinos from kinematics alone, as is necessary for final states involving SM neutrinos, will be more challenging.

All the portal models that we have studied have signals that can be separated from backgrounds by taking advantage of a combination of final-state particle energies, invariant mass cuts, pointing and particle identification.  The ability of DUNE to distinguish between model hypotheses, should an excess be found, or to rule out more parameter space, would be improved if pions and muons could be effectively distinguished.  Thus, we advocate for the inclusion of a muon tagger outside of the ECAL.  There are proposals to make the whole near detector system, including the MPD, movable, allowing measurements to be made both off- and on-axis.  The different kinematics of signal and background events, and between new physics models, means that taking data at multiple angles should allow for signal-background separation and model discrimination.  We have not investigated this potential here, leaving it for future work.  Overall, the inclusion of a gaseous detector in the DUNE near detector complex will provide superb sensitivity to a broad class of new physics models, enhancing the physics program of DUNE beyond neutrino oscillation measurements.

\subsection*{Acknowledgements}

We thank Gordan Krnjaic and Martin Bauer for clarification on different new physics searches contained in this work. We are extremely grateful to Laura Fields, Mary Bishai, and the entire DUNE Beam Interface Working Group for providing files regarding charged meson distributions in the DUNE beamline. The work of JMB is supported by DOE Office of Science awards~DE-SC0018327 and~DE-SC0020262, as well as NSF Grant~PHY-1630782 and by Heising-Simons Foundation Grant~2017-228. JMB further thanks the Fermilab Neutrino Physics Center for their hospitality during the completion of this work. The work of AdG is supported in part by DOE Office of Science award~\#DE-SC0010143. PJF, BJK, KJK, and JLR are supported by Fermi Research Alliance, LLC under Contract DE-AC02-07CH11359 with the U.S. Dept. of Energy.

%\afterpage{\clearpage}  

%\newpage
\appendix
\titleformat{\section}
{\normalfont\LARGE\bfseries}{Appendix~\thesection{:}}{1em}{}

%------------------------------
%Meson Production Appendix
%------------------------------

\section{Meson Production Details and Beam Energy Comparison}\label{appendix:MesonProd}\setcounter{footnote}{0}

In this appendix, we expand on the simulations discussed in Section~\ref{sec:Simulation}, providing more detail regarding the meson production rates extracted from {\sc Pythia8}. We also compare results for an 80 GeV proton beam with those for a 120 GeV proton beam, both of which are under consideration by the DUNE collaboration.

Table~\ref{tab:AppendixMesonProd} lists the number of mesons produced per proton-on-target under several assumptions, all using the \texttt{``SoftQCD:all''} flag within {\sc Pythia8} for the production. We provide the associated Particle ID for clarity. We perform four separate simulations; two each with 80 GeV protons and 120 GeV protons. Because the DUNE-LBNF target is expected to be graphite, we perform simulations assuming the protons are hitting either protons or neutrons, however the resulting output is similar in both cases. To obtain an expected number of mesons per proton-on-target, for our simulations, we average the $pp$ and $pn$ results, assuming an isoscalar target. This leads to the results in Tables~\ref{tab:ChargedMesonProduction} and~\ref{tab:NeutMesonProduction}.

\begin{table}[!htb]
    \centering
    \begin{tabular}{|c|c||c|c|c|c|} \hline
    Meson Type & Particle ID & 80 GeV $pp$ & 80 GeV $pn$ & $120$ GeV $pp$ & $120$ GeV $pn$ \\ \hline \hline
    $\pi^+$ & 211 & 2.5 & 2.2 & 2.8 & 2.5 \\ \hline
    $\pi^-$ & -211 & 1.9 & 2.2 & 2.2 & 2.6 \\ \hline
    $K^+$ & 321 & 0.21 & 0.19 & 0.24 & 0.23 \\ \hline
    $K^-$ & -321 & 0.12 & 0.12 & 0.15 & 0.16 \\ \hline
    $D^+$ & 411 & $1.1\times 10^{-6}$ & $1.4\times 10^{-6}$ & $3.6 \times 10^{-6}$ & $3.7 \times 10^{-6}$ \\ \hline
    $D^-$ & -411 & $2.3 \times 10^{-6}$ & $2.8\times 10^{-6}$ & $5.7 \times 10^{-6}$ & $6.2 \times 10^{-6}$ \\ \hline
    $D_s^+$ & 431 & $2.8 \times 10^{-7}$ & $4.4\times 10^{-7}$ & $1.1 \times 10^{-6}$ & $1.2 \times 10^{-6}$ \\ \hline
    $D_s^-$ & -431 & $4.3 \times 10^{-7}$ & $6.7\times 10^{-7}$ & $1.5 \times 10^{-6}$ & $1.7 \times 10^{-6}$ \\ \hline
    $\pi^0$ & 111 & 2.49 & 2.52 & 2.86 & 2.89 \\ \hline
    $\eta$ & 221 & $0.28$ & 0.28 & 0.32 & 0.33 \\ \hline
    $K_L^0$ & 130 & 0.15 & 0.16 & 0.18 & 0.19 \\ \hline
    $K_S^0$ & 310 & 0.15 & 0.16 & 0.18 & 0.19 \\ \hline
    \end{tabular}
    \caption{Average number of mesons produced per proton-on-target assuming 80 or 120 GeV protons striking protons or neutrons. See text for more detail.}
    \label{tab:AppendixMesonProd}
\end{table}

From Table~\ref{tab:AppendixMesonProd} we see that light meson production rates are around $10\%$ larger for a 120 GeV beam than for an 80 GeV beam, while $D/D_s$ meson production differs, roughly, by a factor of two. We also explored the energy distributions of outgoing mesons in these two samples and find them to be largely similar. Combining these two differences, we expect that the sensitivity to different types of new physics with an 80 GeV beam would be mildly less powerful than that with a 120 GeV beam but the differences will be minor.

Finally, we note that the DUNE collaboration is also exploring a ``high-energy'' beam configuration, in which the Standard Model neutrino flux consists of higher energy neutrinos. This configuration is obtained by changing the focusing magnets, not the proton beam energy or target composition, meaning that Table~\ref{tab:AppendixMesonProd} would be unchanged. All results we present regarding new physics particles being produced in the decay of neutral or short-lived mesons would be unchanged. We leave studies regarding the results of high-energy beam operation with new physics particles coming from long-lived, charged mesons (mostly heavy neutral leptons coming from $\pi^\pm$ and $K^\pm$ decays -- see Section~\ref{sec:HNL}) to future work.

%------------------------------
%Leptophilic Appendix
%------------------------------

\section{Derivation of Charged Meson Decays into Leptophilic Vector Bosons}\label{appendix:derivation}\setcounter{footnote}{0}

In Section~\ref{sec:Leptophilic}, many of our results for the sensitivity to the leptophilic vector bosons associated with the $L_\alpha - L_\beta$ current relied on production via charged meson decay, $\mathfrak{m^\pm} \to \ell^\pm \nu V$. In this appendix, we provide a derivation for the partial width of such a decay (see Ref.~\cite{Krnjaic:2019rsv} for a version of this result for the $K^\pm \to \mu^\pm \nu_\mu V$ decay channel).

\begin{figure}[!h]
\centering
\includegraphics[width=0.6\linewidth]{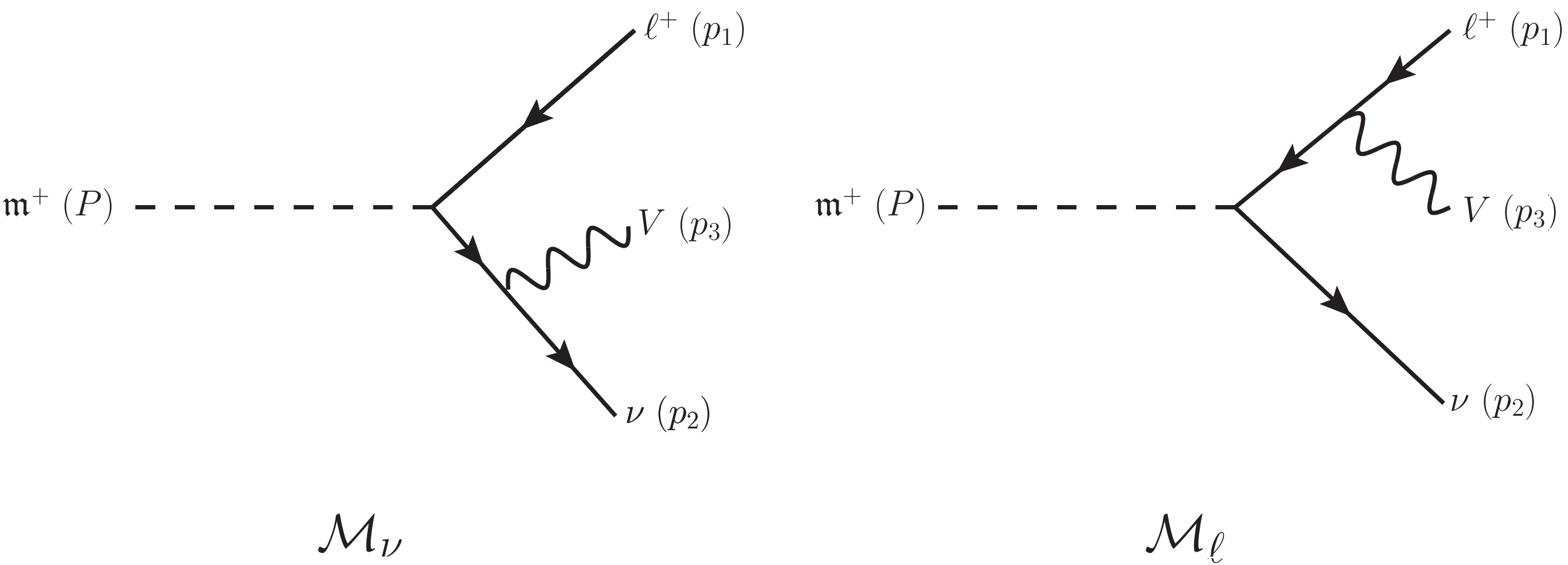}
\caption{Feynman diagrams contributing to the process $\mathfrak{m}^+ \to \ell^+ \nu V$. \label{fig:LeptophilicFD}}
\end{figure}

In Fig.~\ref{fig:LeptophilicFD} we show the two Feynman diagrams that contribute to this decay, in which $V$ is emitted by either the final-state neutrino $\nu_\alpha$ or the charged lepton $\ell^+$. We refer to the matrix elements of these two diagrams as $\mathcal{M}_\nu$ and $\mathcal{M}_\ell$, respectively. These matrix elements may be expressed as 
\begin{align}
    i\mathcal{M}_\nu &= \frac{2g_V\sqrt{2} G_F F_\mathfrak{m} V_{qq^\prime} }{M_V^2 + 2(p_2 \cdot p_3)}\left[ \overline{u}_{p_2} \gamma^\beta \left(\slashed{p_2} + \slashed{p_3} + m_\nu\right)\gamma^\alpha P_L v_{p_1}\right] P_\alpha \epsilon^*_\beta(p_3), \\
        i\mathcal{M}_\ell &= -\frac{2g_V\sqrt{2} G_F F_\mathfrak{m} V_{qq^\prime}}{M_V^2 + 2(p_1 \cdot p_3)} \left[ \overline{u}_{p_2} \gamma^\alpha P_L \left(\slashed{p_1} + \slashed{p_3} - m_\mu\right)\gamma^\beta v_{p_1}\right] P_\alpha \epsilon^*_\beta(p_3),
\end{align}
where $G_F = 1.16 \times 10^{-5}$ GeV$^2$ is the Fermi constant, $g_V$ is the new vector boson's gauge coupling, $F_\mathfrak{m}$ is the decay constant of the meson $\mathfrak{m}$, and $V_{q q^\prime}$ is the relevant quark mixing matrix element for this decay. We label the initial meson momentum as $P$ and the final state momenta as $p_1$ (charged lepton $\ell^+$), $p_2$ (neutrino $\nu_\alpha$), and $p_3$ (vector boson $V$). 

After squaring the total matrix element, we express kinematical factors in terms of invariants $s_{ij} \equiv (p_i + p_j)^2$, and eliminate $s_{13}$ by using the identity $s_{12} + s_{13} + s_{23} = m_\mathfrak{m}^2 + m_\ell^2 + m_\nu^2 + M_V^2$. Taking the limit $m_\nu \to 0$, the matrix element squared is
\begin{align}
    \left\lvert\mathcal{M}\right\rvert^2 &= \frac{16g_V^2 G_F^2 F_\mathfrak{m}^2 \left\lvert V_{qq^\prime}\right\rvert^2 m_\ell^2}{s_{23}^2 \left(m_\mathfrak{m}^2 + M_V^2 - s_{12} - s_{23}\right)^2} \times \nonumber \\
    &\left[ M_V^2 (m_\ell^2 - m_\mathfrak{m}^2) (m_\mathfrak{m}^2 + M_V^2 -s_{12})^2 - s_{23}^2\left(m_\mathfrak{m}^4 + 2m_\mathfrak{m}^2 M_V^2 + (M_V^2 - s_{12})(2m_\ell^2 + M_V^2 - s_{12})\right) \right. \nonumber \\
    &+\left. s_{23}(m_\mathfrak{m}^2 + M_V^2 - s_{12})(m_\mathfrak{m}^4 + 2m_\mathfrak{m}^2(M_V^2 - m_\ell^2) + 2m_\ell^4 + M_V^4 - 2s_{12}(m_\ell^2 + M_V^2) + 2m_\ell^2 M_V^2 + s_{12}^2) \right].\label{eq:msq}
\end{align}
As observed in Ref.~\cite{Krnjaic:2019rsv}, this process, like the two body decay $\mathfrak{m}^+ \to \ell^+ \nu$, is helicity-suppressed and vanishes when $m_\ell \to 0$. For scalar emission, that is not the case~\cite{Berryman:2018ogk,Kelly:2019wow,Krnjaic:2019rsv}.

To obtain the partial width for this decay, we must integrate over $s_{12}$ and $s_{23}$ in the kinematically allowed regions. It is convenient to define $m_\ell^2 \equiv w m_\mathfrak{m}^2$, $M_V^2 \equiv x m_\mathfrak{m}^2$, $s_{12} \equiv z m_\mathfrak{m}^2$, and $s_{23} \equiv y m_\mathfrak{m}^2$. We choose to integrate over $z$ first before $y$. In this parametrization, $y$ is integrated between $x$ and $(1 - \sqrt{w})^2$. For a given $y$, the integration region for $z$ is $z \in [z_-,z_+]$, where
\begin{equation}
    z_\pm = \frac{1}{2y} \left[ w(x+y) + (y-x) \left(1 - y \pm \sqrt{w^2 - 2w(1+y) + (1-y)^2}\right)\right].
\end{equation}
These ranges are determined by the Dalitz criteria, as described in Ref.~\cite{Tanabashi:2018oca}.

Hence, the partial width of $\mathfrak{m}^\pm \to \ell^\pm \nu V$ can be expressed as
\begin{equation}
    \Gamma(\mathfrak{m}^\pm \to \ell^\pm \nu V) = \frac{g_V^2 G_F^2 F_\mathfrak{m}^2 \left\lvert V_{qq^\prime}\right\rvert^2 m_\mathfrak{m} m_\ell^2}{16\pi^3} \int_{x}^{(1-\sqrt{w})^2} \int_{z_-}^{z_+} L^{(4)}(w,x,y,z) \mathrm{d}z\mathrm{d}y,
\end{equation}
where % $w \equiv m_\ell^2/m_\mathfrak{m}^2$ and $x \equiv M_V^2/m_\mathfrak{m}^2$ and
\begin{align}
    L^{(4)}(w,x,y,z) &= \frac{2-y(2-y)-2w(2-x-y)+2w^2}{y(1+x-y-z)} + \frac{(x+y)(w+y-1)}{y^2} + \frac{(w-1)(2w+x)}{(1+x-y-z)^2} - \frac{z-w}{y}
    %L^{(4)}(w,x,y,z) &= \frac{1}{y^2(y + z - x - 1)^2} \times \nonumber \\
    %&\left[ x\left( 2w^2 y - 2wy^2 + z^2(w + 3y-1) - 2z(2wy + w-(y-1)^2) + w - 2y^2 + 3y-1\right) \right. \nonumber \\
    %&+\left. y\left(-2w^2(z-1) + 2w(z(y+z)-1)-(z^2+1)(y+z-1)\right)\right. \nonumber \\
    %&+\left. x^3(w+y-1) + x^2\left(2w(y-z+1)-y(y+3z-3) + 2(z-1)\right)\right].
\end{align}

The dimensionless function $L^{(4)}(w,x,y,z)$ may be integrated over $z$ analytically, giving $L^{(3)}(w,x,y) \equiv \int_{z_-}^{z_+} L^{(4)}(w,x,y,z)\mathrm{d}z$. Defining $S \equiv (y-x)\sqrt{w^2 - 2w(1+y) + (1-y)^2}$,
\begin{align}
L^{(3)}(w,x,y) &= S\left[ \frac{(w+y-1)(x+3y)}{2y^3} + \frac{(1-w)(x+2w)}{(1+x)xy - x(x+(1-w)^2)-wy^2}\right] \nonumber \\
&+ \frac{2w(w+x-2) + 2(w-1) + y^2 + 2}{y} \log{\left(\frac{(x+y)(1-w)+xy-y^2+S}{(x+y)(1-w)+xy-y^2-S}\right)}.
\end{align}
The final width then is
\begin{equation}
\Gamma(\mathfrak{m}^\pm \to \ell^\pm \nu V) = \frac{g_V^2 G_F^2 F_\mathfrak{m}^2 \left\lvert V_{qq^\prime}\right\rvert^2 m_\mathfrak{m} m_\ell^2}{16\pi^3} \int_{x}^{\left(1-\sqrt{w}\right)^2} L^{(3)}(w,x,y) \mathrm{d}y.
\end{equation}

\begin{figure}
    \centering
    \includegraphics[width=0.7\linewidth]{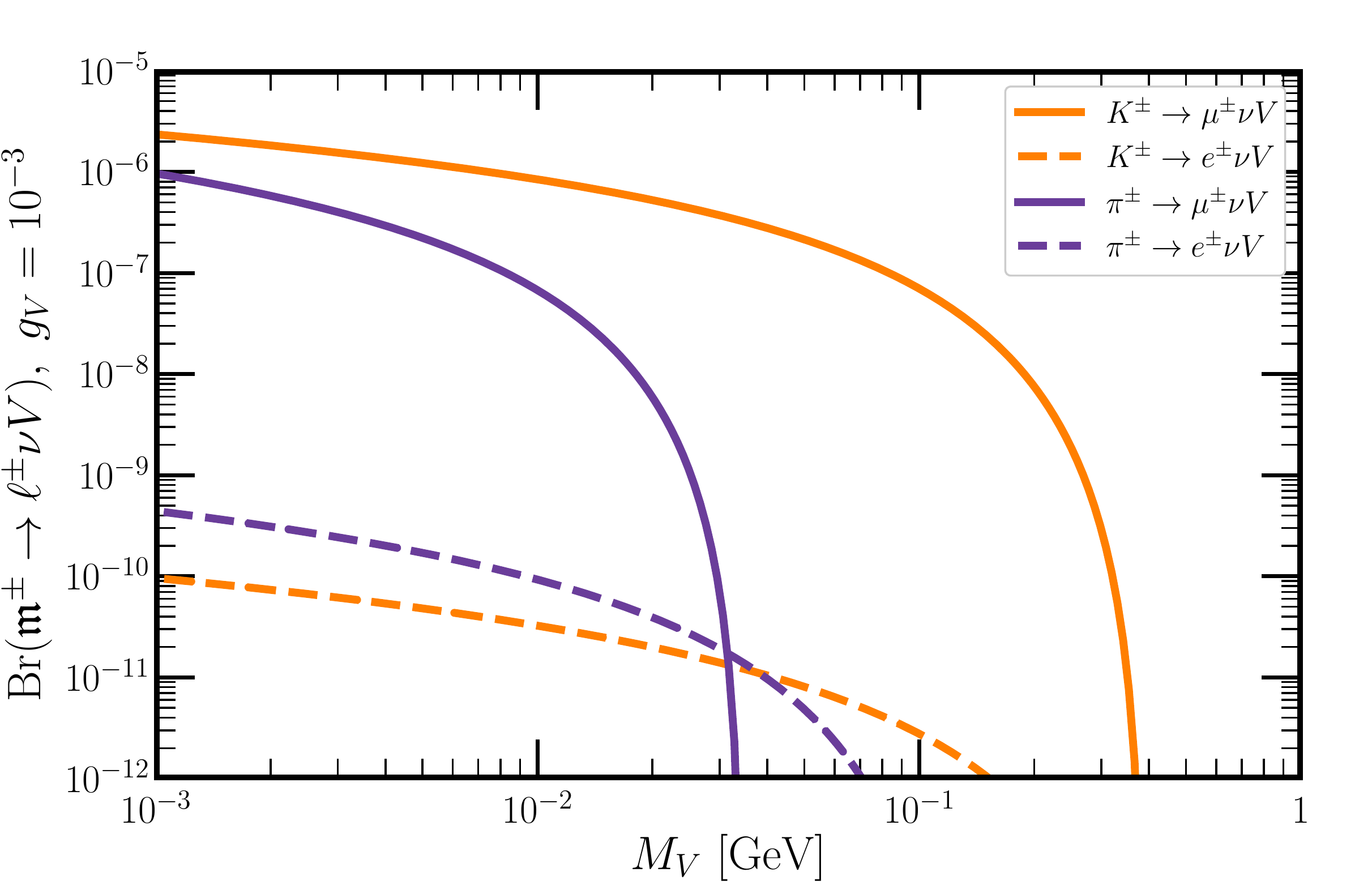}
    \caption{Branching fractions for gauge boson $V$ production for $g_V = 10^{-3}$ via the decays $\mathfrak{m}^\pm \to \ell^\pm \nu V$, with orange lines representing kaon decays and purple lines representing pion decays. Solid lines are for $\ell = \mu$ and dashed are for $\ell = e$.}
    \label{fig:Lmutau:BranchingFractions}
\end{figure}
We use this result to calculate the branching fractions of interest: $\mathrm{Br}(K^\pm \to \mu^\pm \nu V)$, $\mathrm{Br}(K^\pm \to e^\pm \nu V)$, $\mathrm{Br}(\pi^\pm \to \mu^\pm \nu V)$, $\mathrm{Br}(\pi^\pm \to e^\pm \nu V)$. The relevant constants for this calculation are $F_K |V_{us}| = 35.09$ MeV, $F_\pi |V_{ud}| = 127.13$ MeV, $m_K = 493.68$ MeV, $m_\pi = 139.57$ MeV, $m_\mu = 105$ MeV, $m_e = 511$ keV~\cite{Tanabashi:2018oca}. The branching fractions of these decays, which scale as $g_V^2$ for small $g_V$, are depicted in Fig.~\ref{fig:Lmutau:BranchingFractions}.

\bibliographystyle{JHEP}
\bibliography{GasRefs}

\end{document}